\documentclass[usenatbib]{mn2e}

\usepackage{graphicx}
\usepackage{amsmath}
\usepackage{amssymb}
\usepackage{url}
\usepackage{multirow}
\usepackage{ctable}

\usepackage[compatibility=false]{caption}
\usepackage[labelformat=empty]{subfig}

\usepackage[rounding]{rccol}
\rcDecimalSign{.}

\newcommand{\gaa}{\text{\AA}}
\newcommand{\nan}{\multicolumn{1}{c}{---}}
\newcommand{\nansep}[1]{\multicolumn{1}{#1}{---}}
\newcommand{\cpar}[1]{\left( #1 \right)}
\newcommand{\spar}[1]{\left[ #1 \right]}

\newcommand{\best}{{\tt BEST}}
\newcommand{\llll}{{\tt LINE}}

\newcommand{\oiii}{\mbox{[O\,{\sc iii}]}}
\newcommand{\civ}{\mbox{C\,{\sc iv}}}
\newcommand{\mgii}{\mbox{Mg\,{\sc ii}}}

\def\aj{AJ}  % Astronomical Journal
\def\apj{ApJ}% Astrophysical Journal
\def\apjl{ApJ}% Astrophysical Journal, Letters
\def\apjs{ApJS}% Astrophysical Journal, Supplement
\def\aap{A\&A}% Astronomy and Astrophysics
\def\mnras{MNRAS}% Monthly Notices of the RAS
\def\pasp{PASP}% Publications of the ASP
\def\ssr{Space~Sci.~Rev.}% Space Science Reviews
\def\nat{Nature}% Nature
\title[Mass of RL-NLS1]{Black hole mass estimate for a sample of
  radio--loud Narrow--line Seyfert 1 galaxies}

\author[G. Calderone et al.]
  {G. Calderone$^{1}$\thanks{E-mail: {\tt giorgio.calderone@mib.infn.it}},
  G. Ghisellini$^{2}$,
  M. Colpi$^{1,3}$,
  M. Dotti$^{1,3}$\\
  $^{1}$Universit\`a di Milano - Bicocca, Dip. di Fisica G. Occhialini, 
    Piazza della Scienza 3, I-20126 Milano, Italy\\
  $^{2}$INAF Osservatorio Astronomico di Brera, Via E. Bianchi 46, I-23807 Merate (LC), Italy\\ 
  $^{3}$INFN Sezione Universit\`a di Milano - Bicocca, Piazza della Scienza 3, I-20126 Milano, Italy\\
}

\begin{document}

\pagerange{\pageref{firstpage}--\pageref{lastpage}} \pubyear{2011}

\maketitle
\label{firstpage}

\begin{abstract}
We discuss the relationship between a standard \citet{1973-ssad}
accretion disk model and the Big Blue Bump (BBB) observed in Type 1
AGN, and propose a new method to estimate black hole masses.  We apply
this method to a sample of 23 radio--loud Narrow--line Seyfert 1
(RL--NLS1) galaxies, using data from WISE (Wide-–field Infrared Survey
Explorer), SDSS (Sloan Digital Sky Survey) and GALEX.  Our black hole
mass estimates are at least a factor $\sim$6 above previous results
based on single epoch virial methods, while the Eddington ratios are
correspondingly lower.  Hence, the black hole masses of RL--NLS1
galaxies are typically above $10^8 M_{\sun}$, in agreement with the
typical black hole mass of blazars.
\end{abstract}


%---------------------------------------------------------------------
\section{Introduction}
\label{sec-intro}
The Spectral Energy Distribution (SED) of Active Galactic Nuclei (AGN)
spans several orders of magnitude in frequency and results from the
superposition of radiation emitted by different components.

In radio--quiet Type 1 sources, characterized by the presence of broad
emission lines in their optical spectrum, the most luminous components
are the ``Big Blue Bump'' (BBB, between $\sim$1 $\mu$m and $\sim$3 nm,
or $\log(\nu / {\rm Hz}) \sim $14.5--17) and the ``infrared bump'' (IR
bump, between $\sim$1 mm and $\sim$1 $\mu$m or $\log(\nu / {\rm Hz})
\sim $11.5--14.5).  The former is the most prominent feature in the
SED \citep{1989-sanders-torusReproBBB, 1994-Elvis-atlasQuasar,
  2006-richards-meanSED}, while the latter accounts for 20--40\% of
the bolometric AGN luminosity.  The BBB is thought to be thermal
radiation from the accretion disk, while the IR bump is thermal
radiation emitted from a dusty torus located a $\sim$1 pc from the
black hole \citep{1989-sanders-torusReproBBB}.  Superimposed to the
BBB there is often a minor component named ``Small Blue Bump'' (SBB,
extending from 2200\gaa{} to 4000\gaa{}) which is likely the blending
of several iron lines and hydrogen Balmer continuum
\citep{1985-Wills-SBB, 2001-vanden-composite}.  This scheme roughly
describes the SED of AGN over at least 5 orders of magnitude in
bolometric luminosity \citep{1989-sanders-torusReproBBB,
  1994-Elvis-atlasQuasar, 2006-richards-meanSED}.  It also applies to
powerful blazars, although in these cases two more components are
needed to describe the entire SED: the ``synchrotron hump'' (extending
from radio to IR/optical wavelengths) and the ``Compton hump''
(extending from X--rays to TeV energies) which may overwhelm the torus
and the BBB radiation.  These further components characterize
radio--loud sources whose jet is closely aligned to the line of sight,
and are due to the synchrotron and inverse Compton processes,
respectively.

The common energy production process in AGN is believed to be
accretion onto a super--massive black hole ($M \sim 10^{6-10}
M_{\sun}$), through a disk whose observational properties depend
(among other parameters) on the black hole mass and accretion rate.
This interpretation led several authors to use models of geometrically
thin, optically thick accretion disks \citep[][hereafter AD
  model]{1973-ssad} to fit the optical/UV SED of AGN in order to
determine the black hole mass and the accretion rate
\citep[e.g.][]{1978-Shields-3c273disc, 1982-Malkan-BlackBodyInBBB,
  1983-Malkan-EvidenceForAccretionDisc, 1995-Zheng-mrk335-ssad,
  1989-Sun-FittingAccretionDiscModels}.  The AD fitting method allowed
to estimate such quantities for those active nuclei which are too
distant ($z \gtrsim 0.1$) or too bright for other direct methods, such
as resolved stellar/gas dynamics, to be applied
\citep{2005-Ferrarese-reviewSMBH}.  However, as discussed extensively
in \citet{1999-Koratkar-Review-AccDisc} and references therein, such
simple models provide only rough fits to the observed data.  Among the
major issues with this interpretation, we point out a few ones: the
broad--band continuum slopes $\alpha_{\nu}$ (with $F_{\nu} \propto
\nu^{\alpha_{\nu}}$) at optical--NUV wavelengths found in literature
\citep[e.g.][]{1979-Neigebauer-slope-QSO, 2001-vanden-composite,
  2007-Davis-UVContinuum-ModelsSlopes,
  2007-Bonning-AccretionTemperaturesColors} are incompatible with the
slope $\alpha_{\nu} = 1/3$, expected from the AD model; the spectrum
from a simple accretion disk does not reproduce the observed power law
extending at X--rays and the soft X--ray excess
\citep[e.g.][]{1986-Pounds-XRayExcess,
  1994-Nandra-XRaySpectraOfSeyfert, 2005-Fabian-XRaySpectrumIronLine};
the gross properties of the spectrum of radio--quiet AGN appear to
scale with the luminosity \citep{1989-sanders-torusReproBBB,
  1993-Walter-AGN-BBB-univshape}, but does not shift in frequency
\citep{2011-Laor-CommentOnEfficiency}.  The latter issue indicates
that the BBB spectrum peaks always at, or near, the same frequency.
However, this observation is hard to reconcile with reasonably broad
distributions of black hole masses, accretion rates, inclinations and
radiative efficiencies.  Given these difficulties, the AD fitting
method is not widely employed as a black hole mass estimation method.
Rather, it is sometimes used to indirectly infer other parameters such
as the accretion efficiency
(\citealt{2011-Davis-EfficiencyFromSSADFitting,
  2011-Laor-CommentOnEfficiency}, but see
\citealt{2012-Raimundo-MeasureEfficiency}) or to explain specific
quasar properties \citep{2011-Laor-ColdDiscLinelessQuasar}, while the
black hole mass is usually estimated using reverberation mapping
calibrated scaling relations \citep[the so--called ``single epoch
  virial method'', or SE virial, ][]{1993-peterson-rm,
  2004-Peterson-AnalysisRM-DB, 2004-Onken-CalibrationRM,
  2006-vp-masscalib, 2009-bentz-calib-bhmass}.

The SE virial method is currently employed to estimate black hole
masses in large catalogs \citep[e.g.][]{2011-shen-catdr7} for its
simple applicability.  However, the resulting black hole mass
estimates are known to be affected by uncertainties of a factor of
$\sim$3 \citep{2012-Park-calibSEVir}.  In addition, there may be
systematic uncertainties related to the Broad Line Region (BLR)
geometry and inclination; the role of radiation pressure; the modeling
of emission line profiles; the contribution from other components
(e.g. host galaxy); intrinsic differences between different AGN; and
whether gravity dominates the motions of the BLR clouds
\citep{2001-Krolik-ErrorsInRMBasedMasses, 2008-Marconi, 2008-Decarli,
  2011-decarli-blrgeom, 2011-Peterson-MassesOfNLS1}.  Black hole
masses estimated in this way are subjected to a number of assumptions
and may therefore be just order of magnitude estimates
(e.g. \citealt{2011-Croom-DoFWHMCarryInfoOnMass}, but see
\citealt{2012-Assef-ImportanceOfFWHMInMassEstimate}).

There has been claims about the existence of a correlation between the
SE virial mass and the radio luminosity, either absolute
\citep{1998-Franceschini-RelationMbhRadiolum} or normalized to the
optical luminosity (i.e. the radio loudness parameter,
\citealt{2000-Laor-thresholdInMbhForJet}).  It was further noticed
that a black hole mass greater than $\sim 3 \times 10^8 M_{\sun}$
would be required in order to develop a relativistic jet as observed
in powerful blazars.  Later, these findings have been revisited with
the availability of larger
samples. \citet{2002-Woo-MbhIndependentRadioLoudness} found that both
the radio--quiet and radio--loud AGN span the same range of black hole
mass, and that there is no evidence for a strong correlation between
the radio--loudness and black hole mass.  Recently, the issue on the
black hole mass threshold has been revisited by
\citet{2011-chiaberge-rp}: by taking into account the radiation
pressure on the BLR clouds in computing the SE virial masses they
found that a black hole mass $\gtrsim 10^8$ is required to produce a
radio--loud AGN.

In this paper we revisit the AD spectrum fitting method and show (\S
\ref{sec-ssad-bbb}) that the AD model provides a rather satisfactory
description of the Type 1 AGN SED in the majority of cases (at least
at optical/NUV wavelengths) once the contributions from other emitting
components (such as host galaxy and/or jet) have been properly taken
into account.  Therefore, the AD modeling method is a viable and
independent way to estimate black hole masses.  This is particularly
interesting for a class of AGN sources for which black hole masses are
suspected to be systematically underestimated by the SE virial method:
the Narrow Line Seyfert 1 (NLS1) galaxies \citep{2008-Marconi,
  2008-Decarli, 2011-Peterson-MassesOfNLS1}.

NLS1 sources are characterized by the relatively small values of the
full width at half maximum (FWHM) of the ``broad'' component of the
H$\beta$ emission line (FWHM(H$\beta$) $<$ 2000 km s$^{-1}$), by the
presence of strong blended iron lines, and of a prominent soft X--ray
excess \citep{1985-Osterbrock-spectra_of_nls1,
  2000-Pogge-review_nls1}.  By estimating the virial black hole mass
using the H$\beta$ emission line, and \oiii{} width as a surrogate for
the bulge stellar velocity dispersion, \citet{2004-Grupe-Msigma-NLS1}
and \citep{2005-Mathur-Msigma-HighEdd} claimed that NLS1 lie
systematically below the $M$--$\sigma_*$ of non--active and active
Broad Line galaxies (BLS1).  This indicates that the black hole masses
of NLS1 are systematically smaller than the black hole masses of BLS1
for a given value of $\sigma_*$.  The same considerations apply when
considering objects of the same luminosity: NLS1 appear to accrete at
a higher Eddington ratio with respect to BLS1, with some objects
exceeding the Eddington luminosity
\citep{2006-Zhou-comprehensivestudy}.

Recently, a few NLS1 sources have been confirmed to be part of a new
class of $\gamma$--ray emitting sources, as detected by {\it
  Fermi}/LAT \citep{2009-Abdo-discovery_pmnj0948,
  2009-Abdo-rlnls1_newclass, 2011-foschini-procnls1}, besides blazars
and radio--galaxies.  Variability of the $\gamma$--ray emission
\citep{2011-calderone-variab} allows to exclude a starburst origin of
the $\gamma$--rays and confirms the presence of powerful relativistic
jets, such as those found in typical blazars.  The emerging picture is
that $\gamma$--NLS1 sources are very similar to powerful blazars,
except for their small widths of broad lines, and consequently their
small SE virial black hole masses ($10^{6-8} M_{\sun}$,
\citealt{2008-Yuan-population_rlnls1_with_blazar_prop}).  Hence these
sources are the best candidates to settle the question on whether very
massive black holes ($\gtrsim 10^8 M_{\sun}$) power relativistic jets.
If the SE virial mass estimates will be confirmed this would imply
that a large mass is not required to produce a radio--loud AGN.  On
the other hand, if masses turn out to be systematically
under--estimated in NLS1, we would then conclude that extragalactic
jetted sources preferentially live in large black hole mass systems.

Furthermore we may find that NLS1 too lie on the ``canonical''
$M$--$\sigma_*$ relation of broad line sources and accrete close to
(albeit below) the Eddington rate.  Possible explanations for the
observed small widths of the broad lines in NLS1 include: the virial
mass scaling relations may need to be modified in order to account for
radiation pressure effects \citep{2008-Marconi, 2011-chiaberge-rp};
the BLR may have a disk--like geometry and be oriented almost face-on,
so that the Doppler shifted line velocity, projected along the line of
sight, turns out to be small \citep{2008-Decarli}; a combination of
these effects \citep{2011-Peterson-MassesOfNLS1}.

In this work we show that the broad--band composite SEDs of AGN are
roughly compatible with a simple, non--relativistic AD model (\S
\ref{sec-ssad-bbb}), and discuss a method to estimate the total disk
luminosity using either the continuum (\S \ref{sec-cont2ld}) or the
line luminosities (\S \ref{sec-line2ld}) as proxy. Then, we perform a
spectroscopic analysis of the SDSS spectra of 23 radio--loud NLS1 (\S
\ref{sec-sample}) in order to disentangle the host galaxy and/or jet
contribution from the AGN continuum, and estimate line luminosities
(\S \ref{sec-gasf}).  In \S \ref{sec-method} we discuss a new method
to estimate the black hole mass and accretion rate, using AD spectrum
modeling, and apply this method on the afore mentioned sample. We
discuss our results in \S \ref{sec-results} and \ref{sec-discussion},
and draw our conclusion in \S \ref{sec-conclusion}.  The observational
properties of the \citet{1973-ssad} AD model are summarized in the
appendix.

Throughout the paper, we assume a $\Lambda$CDM cosmology with H$_0$ =
71 km s$^{-1}$ Mpc$^{-1}$, $\Omega_{\rm m}$ = 0.27, $\Omega_\Lambda$ =
0.73.

\subsection{Notation}
\label{sec-notation}
In what follows we will consider a non--relativistic,\footnote{General
  relativistic correction are negligible for the purpose of our work,
  see \S \ref{sec-GRcorrections}} steady state, geometrically thin,
optically thick accretion disk \citep{1973-ssad}, extending from
$R_{\rm in} = 6 R_{\rm g}$ to $R_{\rm out} = 2 \times 10^3 R_{\rm g}$,
where $R_{\rm g} = G M / c^2$ is the gravitational radius of the black
hole. The integrated disk luminosity is $L_{\rm d} = \int L_{\nu} d
\nu = \eta \dot{M} c^2$, with $\eta \sim 0.1$ (radiative efficiency).
The corresponding Eddington ratio is $\ell = L_{\rm d} / L_{\rm Edd}$
with $L_{\rm Edd} = 1.3 \times 10^{47} (M / 10^9 M_{\sun})$ erg
s$^{-1}$.

The relation between the disk luminosity and its ``isotropic
equivalent'' counterpart is $L_{\rm d}^{\rm iso} = \langle 2 \cos
\theta \rangle L_{\rm d}$ (Eq. \ref{eq-ad-iso-corr}), where $\theta$
is the angle between the normal to the disk and the line of sight.
For Type 1 AGN we take $\langle 2 \cos \theta \rangle = 1.7$
(Eq. \ref{eq-ad-iso-corr2}).

We will refer to the peak frequency in the $\nu L_{\nu}$
representation as $\nu_{\rm p}$, and to the luminosity of the peak as
$\nu_{\rm p} L_{\nu_{\rm p}}$.  These quantities scale with the
physical parameters as follows (Eq. \ref{eq-ad-scaling} and
\ref{eq-ad-scaling-edd}):
\begin{equation}
  \nonumber
  \begin{aligned}
    \nu_{\rm p} \propto M^{-1/2} \dot{M}^{1/4}\\
    \nu_{\rm p} L_{\nu_{\rm p}}  \propto \dot{M}.
  \end{aligned}
\end{equation}In particular we notice that $\nu_{\rm p} L_{\nu_{\rm p}}
\sim 0.5 L_{\rm d}$ (Eq. \ref{eq-ad-scaling-q}).  The location of the
peak (i.e. its luminosity and frequency) determines uniquely the black
hole mass and accretion rate.  Details about the observational
properties of the AD spectrum are given in appendix \ref{sec-ad-ss73}.

All spectral slopes $\alpha_{\nu}$ are defined as $F_{\nu}
\propto \nu^{\alpha_{\nu}}$.

\section{Accretion disk spectrum in AGN spectra}
\label{sec-ssad-bbb}
\begin{figure*}
  \includegraphics[width=\textwidth]{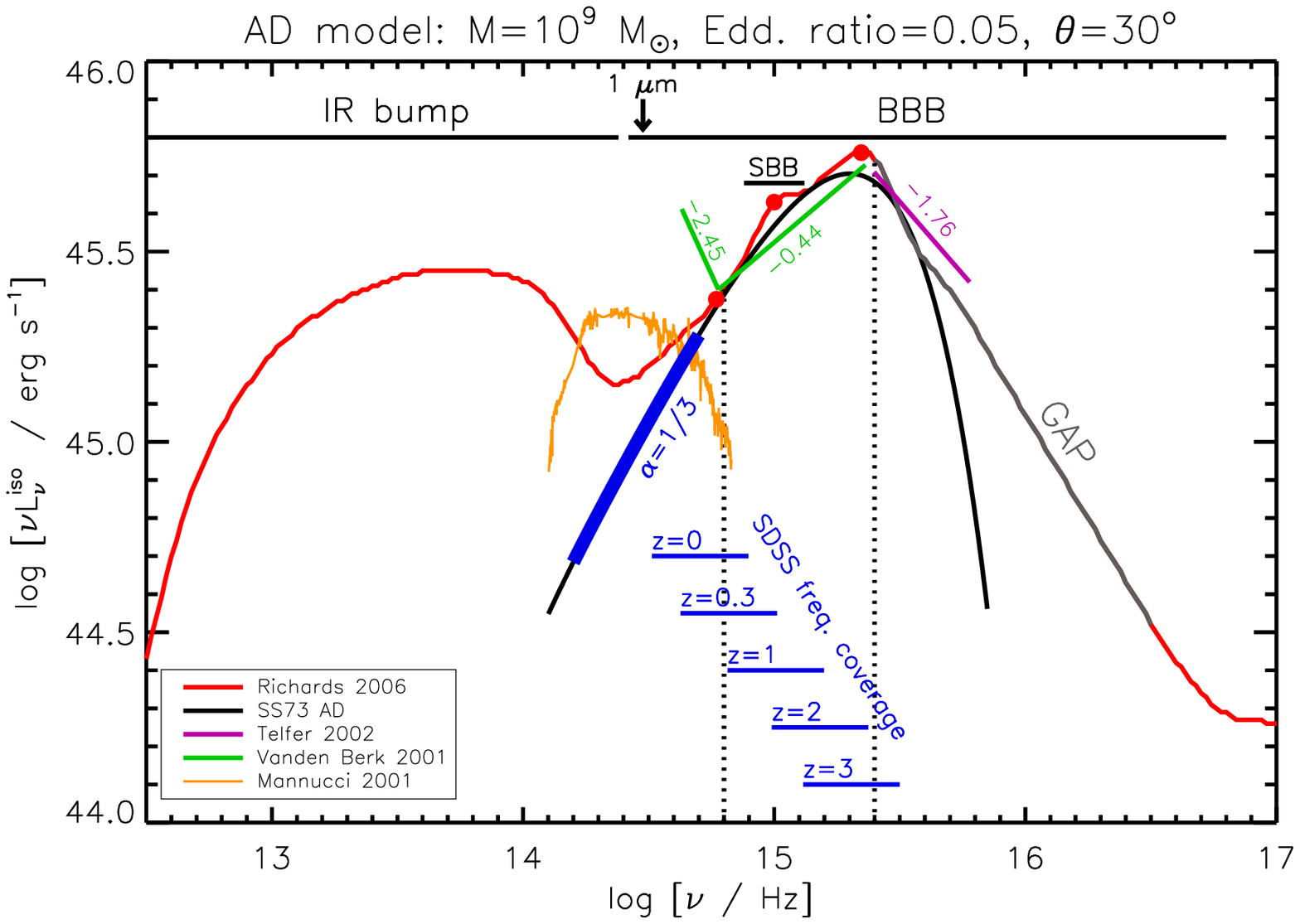}
  \caption{Comparison of the composite Type 1 AGN SED (red line) from
    \citet{2006-richards-meanSED} and an AD model (\S
    \ref{sec-notation}) for $\log(M/M_{\sun})$=9, $\ell$=0.05 and
    $\theta=30^\circ$ (black line).  Also shown are: a spiral galaxy
    template as given in
    \citet[][orange]{2001-Mannucci-galaxytemplate}, normalized to have
    a bolometric luminosity of 10$^{45.5}$ erg s$^{-1}$; the location
    of the Small Blue Bump \citep[SBB, ][]{1985-Wills-SBB,
      2001-vanden-composite}; three reference frequencies
    corresponding to 5100\gaa{}, 3000\gaa{} and 1350\gaa{} (red filled
    circles), commonly used in calculation of bolometric luminosity;
    the average spectral slopes found in literature, as measured on
    composite spectra at near IR, optical/UV
    \citep[][green]{2001-vanden-composite} and far UV wavelengths
    \citep[][purple]{2002-telfer-UVPropOfQSO}; the rest frame
    frequency range covered by SDSS, for values of $z=$0, 0.3, 1, 2
    and 3 (thin blue lines).  Thick blue line highlights the portion
    of the AD spectrum characterized by the slope $\alpha_\nu=1/3$.
    The rest frame frequency range inside which the AD model
    reproduces the shape of the AGN composite SED ($\log(\nu / {\rm
      Hz})$=14.8--15.5) is shown with dotted black lines.}
  \label{fig-template-ssad}
\end{figure*}

\citet{2006-richards-meanSED} built an average Type 1 QSO SED using
data from 259 (mainly radio--quiet) sources, observed with instruments
ranging from radio wavelengths to X--rays.  Individual SEDs have been
interpolated between available bands.  An average SED is then computed
as a geometric mean of individual ones, and is shown in
Fig. \ref{fig-template-ssad}, (red line).
% ------------------------------------------------- 
% ------------------------------------------------- 
Also shown in Fig. \ref{fig-template-ssad} are: a spiral galaxy
template as given in \citet[][orange]{2001-Mannucci-galaxytemplate},
normalized to have a bolometric luminosity of 10$^{45.5}$ erg
s$^{-1}$; the location of the Small Blue Bump \citep[SBB,
][]{1985-Wills-SBB, 2001-vanden-composite}; three reference
frequencies corresponding to 5100\gaa{}, 3000\gaa{} and 1350\gaa{}
(red filled circles), commonly used in calculation of bolometric
luminosity (see below); the average spectral slopes found in
literature, as measured on composite spectra at near IR, optical/UV
\citep[][green]{2001-vanden-composite} and far UV wavelengths
\citep[][purple]{2002-telfer-UVPropOfQSO}; the rest frame frequency
range covered by SDSS, for values of $z=$0, 0.3, 1, 2 and 3 (thin blue
lines).  Finally, we show the AD spectrum that best fits the composite
Type 1 SED at optical/UV wavelengths (black line).  The parameter for
the AD model are: $\log(M/M_{\sun})$=9, $\ell$=0.05 and
$\theta=30^\circ$.

The agreement between the AD model and the composite SED is rather
good, therefore the association between the BBB and thermal emission
from simple AD model is justified, at least in the interval
1000--5000\gaa{}, or $\log(\nu / {\rm Hz})$=14.8--15.5 (black dotted
vertical lines).  A few discrepancies between the AD model and the
composite Type 1 QSO SED arise:
\begin{itemize}
\item at $\log(\nu / {\rm Hz}) <$14.7 a further component emerges in
  the spectrum, which may be either the host galaxy, the emission from
  a dusty torus or some other component;
\item at $\log(\nu / {\rm Hz}) \sim$15 a Small Blue Bump (SBB) is
  present, likely due to a blending of iron lines and Hydrogen Balmer
  continuum;
\item at $\log(\nu / {\rm Hz}) \gtrsim$15.6 other physical components
  contribute to the flux (e.g. a corona).
\end{itemize}
Note that, in this interpretation of the BBB, the portion of the AD
spectrum characterized by the $\alpha_\nu=1/3$ slope (thick blue line)
is hidden by the host galaxy and the torus components, and cannot be
revealed directly with observations (although in some case it may be
detected in polarized light,
\citealt{2008-Kishimoto-DiskInPolarizedLight}).  The average slopes at
optical/UV and far UV (green and purple lines) are roughly consistent
with the slopes near the peak of the AD spectrum.  We notice however
that fixed spectral features (such as the SBB) may affect the
estimation of spectral slopes.  Furthermore, the value of the slope
likely depends on the width of the wavelength range inside which it is
defined.  Therefore it is not always possible to infer the presence of
an AD spectrum by just checking the spectral slopes at optical/UV
wavelengths.  By contrast, at near IR the average slope (green line)
is inconsistent with an AD spectrum, but this is likely due to the
host galaxy component.

We conclude that the AD model provides a reasonable description of the
gross properties of Type 1 AGN SED at optical/NUV wavelengths, and the
similarity between the predicted spectrum and the average BBB is
rather strong.  Under this assumption it is possible to infer the
black hole mass and the accretion rate by comparing the observed SED
with the AD spectrum, as discussed in \S \ref{sec-method}.  Our black
hole mass estimation method requires an estimate of the disk
luminosity ($L_{\rm d}$), as discussed in the following two sections.

\subsection{Continuum luminosity as a proxy to disk luminosity}
\label{sec-cont2ld}

The broad--band similarity among AGN spectra allows to use the
continuum luminosity at a given wavelength as a proxy for the
bolometric luminosity, that is $L_{\rm bol} = {\rm C_{\rm bol}} \times
\lambda L_{\lambda}$.  In order to explore this relationship
\citet{2006-richards-meanSED} measured the bolometric luminosity for
each spectrum (defined to be the integral isotropic luminosity between
100$\mu$m and 10 keV) and derived a bolometric correction (C$_{\rm
  bol}$) based on the continuum luminosity at 3000\gaa{}, 5100\gaa{}
and $3 \mu {\rm m}$.  The resulting values are: C$_{\rm
  bol}(3000{\gaa{}}) = 5.62 \pm 1.14$, C$_{\rm bol}(5100{\gaa{}}) =
10.33 \pm 2.08$, C$_{\rm bol}(3 \mu {\rm m})=9.12 \pm 2.62$.  The
distribution of C$_{\rm bol}$ values is relatively narrow, with a
relative dispersion of the order of $\sim$20\%.  Note however, that in
particular cases the C$_{\rm bol}$ value can differ by as much as 50\%
from the mean value.  \citet[][hereafter S11
  catalog]{2011-shen-catdr7} have slightly re--calibrated the C$_{\rm
  bol}$ values, and extended the analysis to 1350\gaa{}, in order to
compute bolometric luminosities for all the sources in their sample.
Their values are: C$_{\rm bol}(5100{\gaa{}}) = 9.26$, C$_{\rm
  bol}(3000{\gaa{}}) = 5.15$ and C$_{\rm bol}(1350{\gaa{}}) = 3.81$.

In order to calibrate analogous relations to estimate the disk
luminosity $L_{\rm d}^{\rm iso}$ we numerically estimate the
bolometric luminosity of the composite SED in
\citet{2006-richards-meanSED}, and compare it with the disk luminosity
for the AD model shown in Fig. \ref{fig-template-ssad}.  The resulting
relation is:
% ------------------------------------------------- 
\begin{equation}
  \label{eq-lbol2ld}
  L_{\rm d}^{\rm iso} \sim \frac{1}{2} L_{\rm bol}
\end{equation}
% ------------------------------------------------- 
Then, we compare $L_{\rm d}^{\rm iso}$ with the luminosities at
$5100{\gaa{}}$, $3000{\gaa{}}$ and $1350{\gaa{}}$ wavelengths, as measured
on the composite SED:
% ------------------------------------------------- 
\begin{equation}
\label{eq-cont2disclum}
\begin{aligned}
  L_{\rm d}^{\rm iso} &\sim 4.4\  \nu L_\nu (5100{\gaa{}})\\
  L_{\rm d}^{\rm iso} &\sim 2.4\  \nu L_\nu (3000{\gaa{}})\\
  L_{\rm d}^{\rm iso} &\sim 1.8\  \nu L_\nu (1350{\gaa{}})
\end{aligned}
\end{equation}
% -------------------------------------------------
The locations of these wavelengths are shown with red filled circles
in Fig. \ref{fig-template-ssad}.  Considering the uncertainties
($\sim$20\%) of C$_{\rm bol}$ we conclude that our relations
(Eq. \ref{eq-lbol2ld} and \ref {eq-cont2disclum}) are compatible with
those of S11.  Eq. \ref{eq-cont2disclum} provides a reliable estimate
of $ L_{\rm d}^{\rm iso}$ as long as the source continuum is not
dominated by other emitting components such as host galaxy starlight
or synchrotron radiation from a relativistic jet (for radio--loud
sources).  In these cases we need alternative luminosity estimators,
as discussed in the following section.

\subsection{Line luminosities as a proxy to disk luminosity}
\label{sec-line2ld}
\begin{figure}
  \includegraphics[width=.48\textwidth]{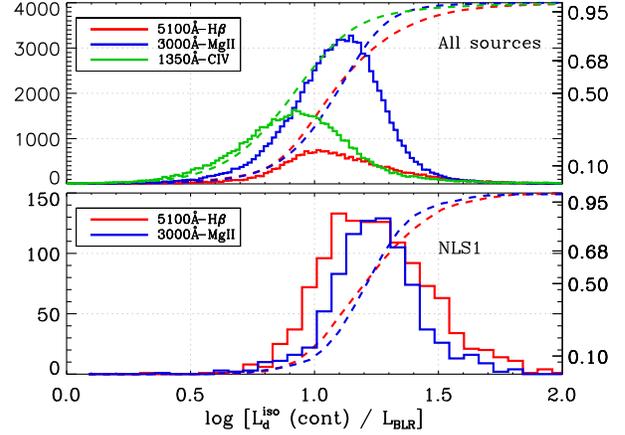}
  \caption{Upper panel: distribution of the $\kappa$ parameter
    (Eq. \ref{eq-kappa-def} for the three combinations
    5100\gaa{}--H$\beta$, 3000\gaa{}--\mgii{} and 1350\gaa{}--\civ{}.
    Both the continuum and line luminosity estimates are those
    reported in the S11 catalog. The number of sources in each
    subsample are 22644, 85514 and 52157 respectively (note that a
    single source typically belongs to two such subsample).  Lower
    panel: the same as upper panel, for the subsample of Narrow--Line
    Seyfert 1 sources common to both the S11 and
    \citet{2006-Zhou-comprehensivestudy} catalog.}
  \label{fig-lineVsCont} 
\end{figure}

Relations similar to Eq. \ref{eq-cont2disclum} can be obtained by
using line luminosities.  Line ratios are known to be approximately
constant among AGN \citep{1991-francis-composite,
  2001-vanden-composite}: by setting the Ly$\alpha$ luminosity to 100,
relative luminosities of H$\beta$, \mgii{} and \civ{} (both narrow and
broad components) lines are 22, 34 and 63, respectively, while the
total line luminosity is 555.8 \citep{1991-francis-composite,
  1997-Celotti-LblrVsJetPower}.  Therefore it is possible to have a
rough estimate of the luminosity of all emission lines by measuring
the luminosity of a single line.  Also, according to the
photo--ionization model, the line--emitting gas is ionized by the
accretion disk continuum radiation.  Therefore we expect (to a first
approximation) the disk to line luminosity ratio to be a constant:
$L_{\rm d}^{\rm iso} = \kappa\, L_{\rm line}$.  This provides a way to
estimate the disk luminosity using a single (or a few) line luminosity
estimates.  In order to calibrate the $\kappa$ parameter we consider
all sources in the S11 catalog having both a continuum and line
luminosity estimate for at least one of the combinations:
5100\gaa{}--H$\beta$, 3000\gaa{}--\mgii{} and 1350\gaa{}--\civ{}.  The
number of sources in each subsample are 22644, 85514 and 52157
respectively (note that a single source typically belongs to two such
subsamples).  For each source we estimate $L_{\rm line}$ using the
broad and narrow line luminosities given in S11 and the coefficients
given in \citet{1991-francis-composite} and
\citet{1997-Celotti-LblrVsJetPower}.  Then we compute the $\kappa$
parameter as follows:
\begin{equation}
  \label{eq-kappa-def}
  \kappa = 
  \frac{ L_{\rm d}^{\rm iso} ({\rm Eq. \ref{eq-cont2disclum}}) }
       {L_{\rm line}}
\end{equation}
where the disk luminosity $L_{\rm d}^{\rm iso}$ is computed using the
continuum luminosity given in S11, and Eq. \ref{eq-cont2disclum}.  The
distributions of $\kappa$ for the three combinations are approximately
log--normal (Fig. \ref{fig-lineVsCont}, upper panel) with median
values:
\begin{equation}
  \label{eq-kappa}
  \begin{aligned}
    \log \kappa &(5100 \gaa{} - {\rm H}\beta) = 1.08 \pm 0.28 \\
    \log \kappa &(3000 \gaa{} - {\rm \mgii{}}  ) = 1.10 \pm 0.21\\
    \log \kappa &(1350 \gaa{} - {\rm \civ{}}   ) = 0.92 \pm 0.28
  \end{aligned}
\end{equation}
The widths of the $\kappa$ distributions in Fig. \ref{fig-lineVsCont}
show that the disk luminosity $L_{\rm d}^{\rm iso}$ computed using the
continuum and the line intensities differs by $\lesssim$0.3 dex,
i.e. a factor $\lesssim$2.  Hence, the relationships between continuum
and line luminosities seem quite robust.
% ------------------------------------------------- 
% ------------------------------------------------- 
A possible explanation for the larger dispersion in the
5100\gaa{}--H$\beta$ case, with respect to the 3000\gaa{}--\mgii{}
one, may be that the continuum luminosity at 5100\gaa{} is
contaminated by the host galaxy.

Both the continuum and the line luminosities in S11 are affected by
uncertainties, therefore the distributions shown in
Fig. \ref{fig-lineVsCont} are likely broadened by measurement errors.
Thus, the intrinsic dispersion is expected to be smaller than 0.3 dex
(a factor of $\sim$2) for H$\beta$ and \civ{}, and 0.2 dex (a factor
of $\sim$1.6) for \mgii{}.  This is yet another evidence that SEDs in
most AGN show some degree of universality: the constancy of the
continuum to line luminosity ratio at optical wavelengths implies a
constant optical continuum to ionizing UV luminosity ratio.

Since the samples in the S11 catalog are dominated by radio--quiet
sources (the great majority are undetected in the FIRST survey), we
repeat our analysis on radio--loud sources, i.e. those sources for
which the radio--loudness parameter\footnote{The radio--loudness
  parameter provides an indication of whether the AGN SED is dominated
  by radiation at radio frequencies or optical band.  Historically, it
  is defined as the ratio of 5 GHz to optical B--band luminosity
  \citep{1989-Kellermann-def_radio_loudness}.  The values we used here
  are those given in S11, defined as the ratio of flux densities at 6
  cm and 2500\gaa{} (rest frame).}  is greater than 100.
Interestingly, the $\kappa$ parameters for the radio--loud sub--sample
of S11 differ by at most $\sim$5\% from the values quoted above.

By using the values given in Eq. \ref{eq-kappa} and the coefficients
to compute the line luminosity $L_{\rm line}$ discussed above, we are
able to estimate the total disk luminosity as follows:
% ------------------------------------------------- 
\begin{equation}
\label{eq-line2disclum}
\begin{aligned}
  L_{\rm d}^{\rm iso} &= 12   \, L ({\rm H} \beta) \, \frac{555.8}{22} = 303  \,   L ({\rm H} \beta)\\
  L_{\rm d}^{\rm iso} &= 12.5 \, L ({\rm \mgii{}}) \, \frac{555.8}{34} = 204  \,   L ({\rm \mgii{}})   \\
  L_{\rm d}^{\rm iso} &= 8.4  \, L ({\rm \civ{}})  \, \frac{555.8}{63} =  74.1\,   L ({\rm \civ{}}).
\end{aligned}
\end{equation}
% -------------------------------------------------

We repeat the above analysis on the subsample of NLS1 sources, that
are the focus of our study.  In particular, we consider the sources
common to both the \citet{2006-Zhou-comprehensivestudy} and the S11
catalog (1210 sources).  The distributions\footnote{The
  1350\gaa{}--\civ{} case is missing since the SDSS wavelength
  coverage does not allow to observe both the \civ{} line and the
  H$\beta$ line (required to classify the source as a NLS1).} of the
$\kappa$ parameter are still log--normal, and are shown in the lower
panel of Fig. \ref{fig-lineVsCont}.  Median values are now $\sim$0.15
dex (i.e. a factor $\sim$1.4) greater:
\begin{equation}
  \label{eq-kappa-nls1}
  \begin{aligned}
    \log \kappa &(5100\gaa{} - {\rm H}\beta) = 1.23 \pm 0.23\\
    \log \kappa &(3000\gaa{} - {\rm \mgii{}})   = 1.24 \pm 0.21.
  \end{aligned}
\end{equation}
The uncertainties are of the same order of magnitude.  The class of
NLS1 sources is therefore characterized by both a smaller width and a
smaller luminosity of lines.  The resulting disk luminosities are:
% ------------------------------------------------- 
\begin{equation}
\label{eq-line2disclum-rlnls1}
\begin{aligned}
  L_{\rm d}^{\rm iso}  &= 424\,  L ({\rm H} \beta)\\
  L_{\rm d}^{\rm iso}  &= 286\,  L ({\rm \mgii{}}).
\end{aligned}
\end{equation}
% ------------------------------------------------- 

In order to estimate the accretion disk luminosity using a single
spectrum we can use either Eq. \ref{eq-cont2disclum} (whose
uncertainties are $\sim$20\%) or Eq. \ref{eq-line2disclum-rlnls1}
(whose uncertainties are a factor $\sim$2).  In cases where the
observed continuum radiation is dominated by components other than AD,
e.g. synchrotron emission from the jet or host galaxy starlight,
Eq. \ref{eq-cont2disclum} would overestimate the disk luminosity.
Therefore Eq. \ref{eq-line2disclum-rlnls1} is our preferred choice to
estimate $L_{\rm d}^{\rm iso}$.

\section{The sample}
\label{sec-sample}
The aim of this work is to estimate the black hole masses of the
sample of 23 Radio--Loud, Narrow--Line Seyfert 1 sources (RL--NLS1)
given in \citet[][ hereafter
  Y08]{2008-Yuan-population_rlnls1_with_blazar_prop}.  We identify
each source with a sequential index (\#1, \#2, etc...), following the
same order as in Tab. 1 of Y08.

All sources have been spectroscopically observed in the SDSS, and 21
over 23 sources are also in the S11 catalog.  The IR photometry at
3.4$\mu$m, 4.6$\mu$m, 11.6$\mu$m, 22.1$\mu$m from WISE
\citep[Wide--field Infrared Survey Explorer, ][]{2010-Wright-wise} is
available for all sources.  Finally, 21 over 23 sources have
photometric measurements by GALEX \citep{2005-Martin-Galex}, either in
the Medium Imaging Survey (MIS), or the All sky Imaging Survey (AIS).
We noticed that when multiple GALEX observations were available, we
find significant variability in a few cases (\#3, \#8, \#18) possibly
due to the jet component.  In these cases we chose preferably the MIS
photometry with lower luminosity.

The redshifts are in the range $z=$0.1--0.8, therefore the continuum
in the SDSS spectra will likely trace the AD component (\S
\ref{sec-ssad-bbb}).  The FWHM(H$\beta$) are less than 2200 km
s$^{-1}$, as required by the definition of NLS1 given in
\citet{2006-Zhou-comprehensivestudy}.  The SE virial black hole masses
are in the range $\log(M / M_{\sun})=$6--8, while the Eddington ratio
are $\ell=$0.5--3
\citep{2008-Yuan-population_rlnls1_with_blazar_prop}.  The radio
morphology is compact, unresolved on $5^{\prime\prime}$ scale, and the
radio loudness \citep{1989-Kellermann-def_radio_loudness} is $>$100
for all sources.

The overall observational properties are very similar to that of
blazars \citep{2008-Yuan-population_rlnls1_with_blazar_prop}, and the
$\gamma$--ray emission from these sources has been predicted, and
later detected in 7 RL-NLS1 sources \citep{2009-Abdo-rlnls1_newclass,
  2011-calderone-variab, 2011-foschini-procnls1}, 4 of which are in
the Y08 sample.  However, these sources show unusually small widths of
broad emission lines, and consequently small SE virial black hole mass
estimates, when compared to typical blazars.

In order to apply our black hole mass estimation method (\S
\ref{sec-method}) to the sources in the sample we need to perform a
spectroscopic analysis of the SDSS data.  In particular we need to
disentangle the host galaxy and/or jet contribution from the AGN
continuum, and estimate the emission line luminosities.  This
procedure is described in the following section.

\subsection{Spectral analysis}
\label{sec-gasf}
We used the spectra from the Sloan Digital Sky Survey
\citep[SDSS,][]{2000-york-sdss-summary}, data release 7
\citep[DR7,][]{2009-sdss-dr7}.  We dropped spectral bins marked by at
least one of the following mask flags:\footnote{See
  \url{http://www.sdss.org/dr7/dm/flatFiles/spSpec.html}.}
\url{SP_MASK_FULLREJECT}, \url{SP_MASK_LOWFLAT},
\url{SP_MASK_SCATLIGHT}, \url{SP_MASK_BRIGHTSKY},
\url{SP_MASK_NODATA}, \url{SP_MASK_COMBINEREJ},
\url{SP_MASK_BADSKYCHI}.  Also, we dropped 100 bins at the beginning
and end of each spectrum, in order to eventually avoid artifacts from
instrument or pipeline.

Each spectrum has been de-–reddened using the Galactic extinction
values estimated from dust IR emission maps in
\citet{1998-Schlegel-galaxy-irmap}, and the extinction law reported in
\citet{1989-cardelli-extinction} and \citet{1994-odonnell-updateCCM}.
We are currently neglecting any intrinsic reddening in the rest--frame
of the source.  Then we transformed the spectra to the rest frame by
assuming isotropic emission (i.e. multiplying the flux by 4$\pi D_{\rm
  L}^2$).  The redshift estimates are provided by the SDSS pipeline.
Finally, we rebinned each spectrum by a factor of 3 in order to
improve the signal to noise ratio, resulting in a spectral resolution
of $\lambda / \delta \lambda \sim$1450
(corresponding to $\sim$200 km s$^{-1}$).

The model used to fit the spectra consists of five components:
\begin{itemize}
\item a smoothly broken power law to account for the AGN continuum
  (``AGN continuum'' component) ;

\item a spiral\footnote{The results of the spectral fitting procedure
  do not change significantly by considering the elliptical galaxy
  template from \citet{2001-Mannucci-galaxytemplate}.} host galaxy
  template spectrum from \citet{2001-Mannucci-galaxytemplate} and a
  power law to (eventually) account for the synchrotron emission from
  the jet (``galaxy'' and ``jet'' components respectively).  The
  galaxy component has a single free parameter (the overall
  normalization).  The parameters for the jet component are estimated
  using data from WISE \citep[Wide--field Infrared Survey Explorer,
  ][]{2010-Wright-wise}.  In particular we use the photometry in the
  two bands at the longest wavelengths (11$\mu$m and 22$\mu$m) to
  estimate the luminosity and the slope of the power law.\footnote{In
    analyzing the source SDSS J094857.32+002225.5 (\#5) we also
    applied an exponential cutoff at $\log [\nu / {\rm Hz}] = 14$
    \citep{2009-Abdo-mw_monitor_pmnj0948}.}  If the resulting slope is
  greater than --1 we extrapolate the power law to optical wavelengths
  and subtract the contribution from the SDSS spectrum.  Otherwise we
  do not consider any jet component.  Parameter of the jet component
  are fixed during the fitting process;

\item the iron templates from \citet{2001-vestergaard-UV-iron} (at UV
  wavelengths) and from \citet{2004-veron-spectra-izw1} (at optical
  wavelengths);

\item a Gaussian profile for each emission line listed in
  Tab. \ref{tab-knownline}.  The FWHM of narrow lines are forced to be
  in the range 200--1000 km s$^{-1}$, while that of broad lines are
  forced in the range 1000--3000 km s$^{-1}$.  Furthermore, the FWHM
  and velocity offset of the H$\beta$ narrow component is tied to the
  width and offset of \oiii{} $\lambda$4959 and \oiii{} $\lambda$5007.
% -------------------------------------------------------
\begin{table}
  \caption{List of emission lines used in modeling SDSS spectra. Third
    column (Type) indicates if a broad (B), a narrow (N) or both
    components are used in the fit.}
  \label{tab-knownline}
  \begin{tabular}{lcc|lcc}
    \toprule
        {Line}             &
        {Wave [\gaa{}]}    &
        {Type}             &
        {Line}             &
        {Wave [\gaa{}]}    &
        {Type}          \\
        \midrule
        C II         &    2326  &    B       &  N I          &    5199  &     N\\
        Mg II        &    2798  &    BN      &  He I         &    5876  &    BN\\
        Ne V         &    3426  &     N      &  Fe VII       &    6087  &     N\\
        O II         &    3727  &     N      &  O I          &    6300  &     N\\
        Ne III       &    3869  &     N      &  Fe X         &    6375  &     N\\
        H $\delta$   &    4101  &    B       &  N II         &    6548  &     N\\
        H $\gamma$   &    4340  &    BN      &  H $\alpha$   &    6563  &    BN\\
        O III        &    4363  &     N      &  N II         &    6583  &     N\\
        He II        &    4686  &    BN      &  S II         &    6716  &     N\\
        H $\beta$    &    4861  &    BN      &  S II         &    6731  &     N\\
        O III        &    4959  &     N      &  Ar III       &    7136  &     N\\
        O III        &    5007  &     N                                        \\
        \bottomrule
  \end{tabular}
\end{table}
% -------------------------------------------------------

\item a maximum of 10 additional Gaussian line profiles which are not
  ``a priori'' associated to any specific transition.  These
  components are necessary to account for (e.g.) the iron blended
  emission lines in the range 3100--3500\gaa{} (not covered by the
  above--cited iron templates), or line asymmetries.  The FWHM of the
  additional lines are forced to be in the range 1000--3000 km
  s$^{-1}$, except for lines in the range 3100--3500\gaa{} for which
  the upper limit is $10^4$ km s$^{-1}$.  A posteriori, we check
  whether the wavelength range identified by the full width at half
  maximum of these additional lines contains any of the transition
  lines listed in Tab. \ref{tab-knownline}.  In this case we associate
  the two components, and numerically compute the line luminosity on
  the composite line profile.
\end{itemize}
% -------------------------------------------------------

Results of the spectral fitting are shown in Tab. \ref{tab-results}
and Fig. \ref{fig-spec}.
% -------------------------------------------------------
\begin{table*}
    \caption{Results of the spectral fitting for the 23 RL--NLS1
      sources in \citet{2008-Yuan-population_rlnls1_with_blazar_prop}
      catalog.  Columns are: (1) source numeric identifier; (2) SDSS
      name of the source; (3) redshift; (4) luminosity and error of
      the H$\beta$ emission line (both the broad and narrow
      components); (5) luminosity and error of the \mgii{} emission
      line (both the broad and narrow components); (6) wavelength
      $\lambda_0$ and (7) luminosity $\lambda_0 L_{\lambda_0}$ used to
      constrain the \llll{} model (see \S \ref{sec-method-auto}); (8)
      jet component (extrapolated from WISE data to wavelength
      $\lambda_0$) to AGN continuum luminosity ratio.}
    \label{tab-results}
    \begin{tabular}{r c R{1}{3} R{2}{2}@{ $\pm$}R{1}{2}  R{2}{2}@{ $\pm$}R{1}{2}  R{4}{0} R{2}{2} R{1}{2}}
      \toprule
         \multicolumn{1}{c}{\#}
      &  \multicolumn{1}{c}{SDSS Name}
      &  \multicolumn{1}{c}{$z$}
      &  \multicolumn{2}{c}{\parbox[c]{2cm}  {\begin{equation}\nonumber   \log \frac{L({\rm H}\beta)}              {\rm erg\ s^{-1}}  \end{equation}}}
      &  \multicolumn{2}{c}{\parbox[c]{2cm}  {\begin{equation}\nonumber   \log \frac{L({\rm \mgii{}})}             {\rm erg\ s^{-1}}  \end{equation}}}
      &  \multicolumn{1}{c}{\parbox[c]{1cm}  {\begin{equation}\nonumber   \log \frac{\lambda_0}                    {\gaa{}}           \end{equation}}}
      &  \multicolumn{1}{c}{\parbox[c]{1.5cm}{\begin{equation}\nonumber   \log \frac{\lambda_0 L_{\lambda_0}}      {\rm erg\ s^{-1}}  \end{equation}}}
      &  \multicolumn{1}{c}{\parbox[c]{1.5cm}{\begin{equation}\nonumber        \frac{L_{\rm J,0}}{L_{\lambda_0}}                      \end{equation}}}\\
         \midrule
          1 & J081432.11+560956.6 &      0.509000 &       42.9633 &    0.00990504 &       42.9889 &     0.0161646 &       3169.65 &       45.1058 &          \nan\\
          2 & J084957.98+510829.0 &      0.583000 &       42.2925 &      0.118548 &       42.5188 &     0.0355501 &       3039.09 &       43.5794 &       5.66765\\
          3 & J085001.17+462600.5 &      0.523000 &       42.4825 &     0.0429985 &       42.5078 &     0.0310060 &       3156.82 &       44.6096 &      0.128548\\
          4 & J090227.16+044309.6 &      0.532000 &       42.6128 &     0.0291780 &       42.9280 &     0.0168167 &       3127.10 &       44.6656 &     0.0711497\\
          5 & J094857.32+002225.5 &      0.584000 &       42.8066 &     0.0255847 &       42.8504 &     0.0288248 &       3038.92 &       45.1527 &   0.00004    \\
          6 & J095317.09+283601.5 &      0.657000 &       42.5406 &     0.0328668 &       42.8422 &     0.0205168 &       2893.03 &       44.8835 &      0.102857\\
          7 & J103123.73+423439.3 &      0.376000 &       42.3066 &     0.0177352 &       \multicolumn{2}{c}{---} &       3486.03 &       44.1230 &          \nan\\
          8 & J103727.45+003635.6 &      0.595000 &       42.5157 &     0.0497300 &       42.3083 &     0.0449329 &       3020.11 &       44.7337 &          \nan\\
          9 & J104732.68+472532.1 &      0.798000 &       43.0992 &     0.0291574 &       43.0388 &     0.0393281 &       2673.53 &       45.1531 &      0.119078\\
         10 & J111005.03+365336.3 &      0.630000 &       42.3647 &     0.0438111 &       42.5921 &     0.0257101 &       2944.00 &       44.2556 &    0.00767284\\
         11 & J113824.54+365327.1 &      0.356000 &       42.2630 &     0.0148032 &       \multicolumn{2}{c}{---} &       3547.03 &       43.8736 &      0.641377\\
         12 & J114654.28+323652.3 &      0.465000 &       42.6832 &    0.00982629 &       42.6469 &     0.0256486 &       3285.20 &       44.7465 &          \nan\\
         13 & J123852.12+394227.8 &      0.622000 &       42.4095 &     0.0476098 &       42.4781 &     0.0326746 &       2960.56 &       44.5907 &     0.0154246\\
         14 & J124634.65+023809.0 &      0.362000 &       42.4318 &     0.0199813 &       \multicolumn{2}{c}{---} &       3524.91 &       44.6613 &     0.0498457\\
         15 & J130522.75+511640.3 &      0.785000 &       43.7869 &     0.0108274 &       43.4719 &     0.0118704 &       2692.23 &       45.7367 &      0.474174\\
         16 & J143509.49+313147.8 &      0.501000 &       42.4296 &     0.0213833 &       42.6506 &     0.0273201 &       3202.91 &       44.4637 &      0.723604\\
         17 & J144318.56+472556.7 &      0.703000 &       42.8172 &     0.0411853 &       43.2358 &     0.0178257 &       2826.08 &       45.4169 &      0.149988\\
         18 & J150506.48+032630.8 &      0.408000 &       41.8958 &     0.0299009 &       42.5522 &     0.0403410 &       3407.00 &       44.3932 &      0.138913\\
         19 & J154817.92+351128.0 &      0.478000 &       42.8394 &    0.00985655 &       42.9291 &     0.0214729 &       3248.82 &       45.0540 &     0.0592851\\
         20 & J163323.58+471859.0 &      0.116000 &       41.6562 &     0.0248987 &       \multicolumn{2}{c}{---} &       4303.39 &       43.7399 &      0.433409\\
         21 & J163401.94+480940.2 &      0.494000 &       42.4459 &     0.0219728 &       42.0241 &     0.0223438 &       3213.10 &       44.5574 &      0.136507\\
         22 & J164442.53+261913.2 &      0.144000 &       41.8557 &     0.0114359 &       \multicolumn{2}{c}{---} &       4200.00 &       43.9497 &      0.248787\\
         23 & J172206.03+565451.6 &      0.425000 &       42.5522 &    0.00960540 &       42.6412 &     0.0261559 &       3370.04 &       44.6811 &     0.0443078\\
      \bottomrule
    \end{tabular}
\end{table*}
% -------------------------------------------------------

\section{Black hole mass estimation method}
\label{sec-method}
The AGN continuum in the rest frame wavelength range 1000--5000\gaa{}
(or $\log(\nu / {\rm Hz})$=14.8--15.5), if interpreted as radiation
emitted from a \citet{1973-ssad} accretion disk (\S
\ref{sec-ssad-bbb}), allows to constrain an AD model, and to infer the
black hole mass.  Once we assume proper values for the inner radius of
the disk $R_{\rm in}$ and the viewing angle $\theta$ (\S
\ref{sec-method-hyp}), the luminosity and frequency of the peak of the
AD spectrum uniquely identify a value of the black hole mass.  In the
following sections we will discuss two methods to locate the peak of
the AD spectrum, and infer the black hole mass and accretion rate.  An
example of the application of both methods to a specific case will be
discussed in \S \ref{sec-example}.

\subsection{Hypotheses}
\label{sec-method-hyp}
The methods relies on the following hypotheses, which need to be
independently verified:
\begin{enumerate}
\item accretion in AGN occurs through steady--state, geometrically
  thin, optically thick, non--relativistic accretion disks.  The
  emitted spectrum is well described by an AD model \citep{1973-ssad};

\item once the galaxy and/or jet contribution has been subtracted, the
  continuum radiation in the range $\log(\nu / {\rm Hz})$=14.8--15.5
  (\S \ref{sec-ssad-bbb}) is emitted directly from the accretion disk,
  i.e. it has not been reprocessed by intervening material, nor it is
  emitted by some other component;

\item the spatial extention of the disk is $R_{\rm in} = 6 R_{\rm g}$,
  corresponding to a radiative efficiency $\eta \sim 0.1$.  The outer
  radius of the disk $R_{\rm out} = 2\times 10^3 R_{\rm g}$ is not
  critical, since at frequencies much smaller than $\nu_{\rm p}$ the
  AD spectrum will always be hidden by other emitting components. The
  assumption for $R_{\rm in}$, on the other hand, is more critical,
  since our black hole mass estimates show a linear dependence on this
  value (case (v) of \S \ref{sec-ad-peakshift});

\item the relation between disk luminosity and its ``isotropic
  equivalent'' counterpart is $L_{\nu}^{\rm iso} = \langle 2 \cos
  \theta \rangle L_\nu$ (Eq. \ref{eq-ad-iso-corr2}).  Since we are
  interested in Type 1 AGN the viewing angle is in the range 0--45 deg
  \citep[i.e. the aperture of the obscuring
    torus,][]{2012-calderone-torus}.  The averaged de--projection
  factor is thus $\langle 2 \cos \theta \rangle \sim 1.7$
  (Eq. \ref{eq-ad-mean-cost}, \ref{eq-ad-iso-corr2}), corresponding to
  a viewing angle of $\sim$30 deg.
\end{enumerate}
The AD model has four parameters: $M$, $\dot{M}$, $R_{\rm in}$ and
$\cos \theta$ (\S \ref{sec-obsprop}).  With the assumptions discussed
above, the remaining unknown parameters are the black hole mass $M$
and the accretion rate $\dot{M}$.

\subsection{Procedure}
\label{sec-method-gen}
Usually the localization of the peak of the AD spectrum is not
accessible by using a single instrument, requiring optical/UV
multiwavelength observations.  When these observations are available
it is possible to constrain the AD model, and estimate the frequency
$\nu_{\rm p}$ and luminosity $\nu_{\rm p} L_{\nu_{\rm p}}^{\rm iso}$
of the peak.  The latter can then be used to infer the total disk
luminosity $L_{\rm d}^{\rm iso}$ (Eq. \ref{eq-ad-scaling-q}).
Finally, the black hole mass and the accretion rate can be estimated
as follows:
\begin{equation}
  \begin{aligned}
    \label{eq-ssad-compute_M}
    \frac{M}{10^9 M_{\sun}} &= 
    1.44 \cpar{ \frac{\nu_{\rm p}}{10^{15}\ {\rm Hz}} }^{-2}
    \cpar{
      \frac{L_{\rm d}^{\rm iso}}{\langle 2 \cos \theta \rangle \times 10^{45}\ {\rm erg\ s^{-1}}}
    }^{1/2}\\
    \frac{\dot{M}}{M_{\sun}\ {\rm yr}^{-1}} &= 
    0.21 \cpar{
      \frac{L_{\rm d}^{\rm iso}}{\langle 2 \cos \theta \rangle \times 10^{45}\ {\rm erg\ s^{-1}}} 
    }
  \end{aligned}
\end{equation}
The uncertainties on these results can be estimated by propagating the
uncertainties in the $\nu_{\rm p}$ and $\nu_{\rm p} L_{\nu_{\rm
    p}}^{\rm iso}$ parameters in the above equations.  Hence, whenever
the data allow to constrain the location of the peak of the AD
spectrum, the accuracy of the black hole mass estimate is determined
only by the accuracy of the data points
\citep[e.g.][]{2012-Sbarrato-blazar_z5.3}.  When UV observation are
not available (or not reliable) the location of the peak cannot be
constrained, and we must resort to an alternative method.

\subsubsection{The \llll{} procedure}
\label{sec-method-auto}
\begin{figure}
  \includegraphics[width=.5\textwidth]{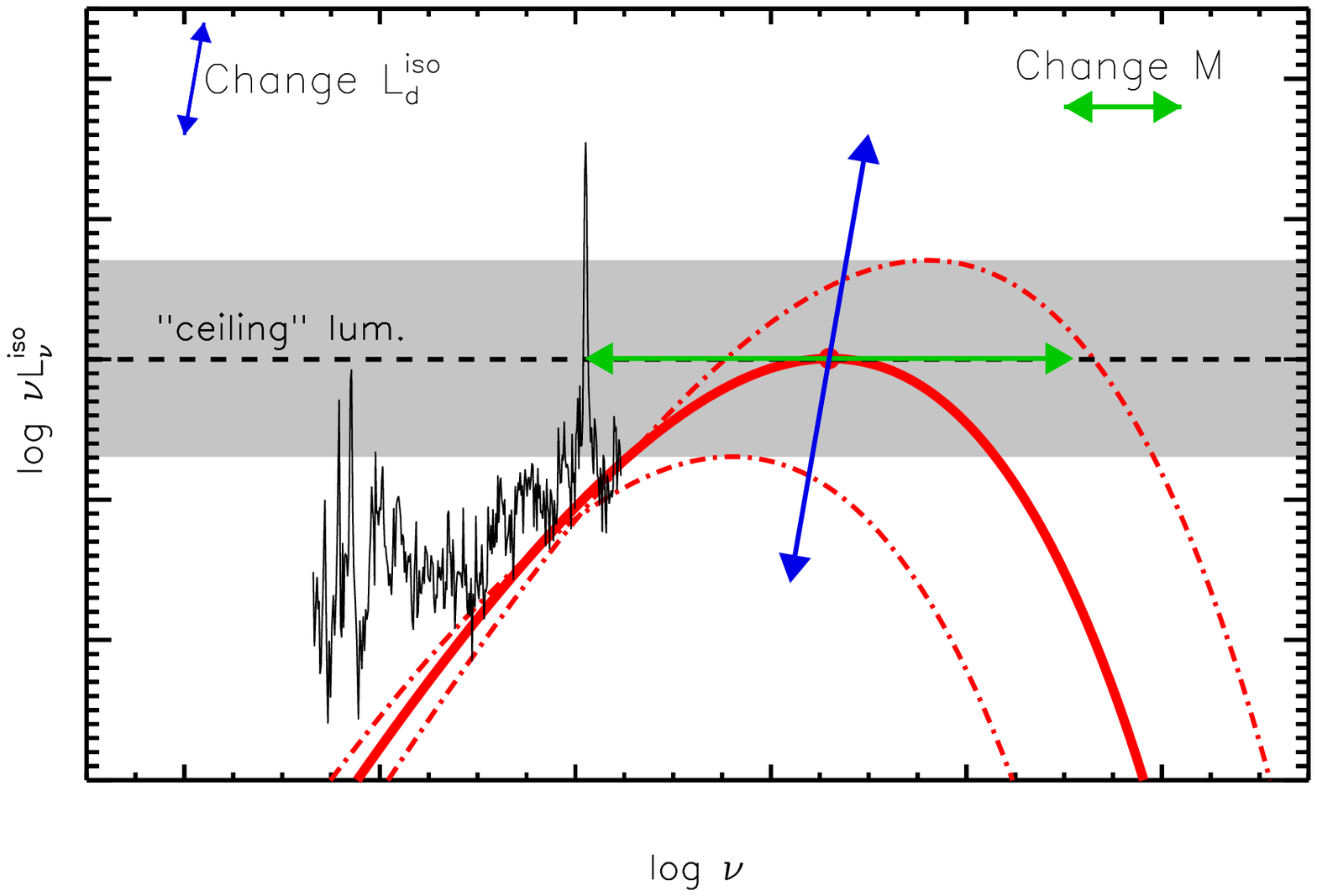}
  \caption{The \llll{} black hole mass estimation procedure: the SDSS
    source spectrum (black line) is analyzed with our fitting
    procedure (\S \ref{sec-gasf}) in order to estimate the broad line
    luminosities.  Then we use Eq. \ref{eq-line2disclum-rlnls1} to
    estimate $L_{\rm d}^{\rm iso}$.  This is equivalent to estimate a
    value for the luminosity of the peak $\nu_{\rm p} L_{\nu_{\rm
        p}}^{\rm iso}$ (Eq. \ref{eq-ad-scaling-q}), i.e. to fix a
    ``ceiling'' in the $\nu L_\nu$ representation (black dashed line).
    Then we use the SDSS spectrum to constrain the peak frequency
    $\nu_{\rm p}$, which is related to the black hole mass.  In
    particular, we shift the AD spectrum horizontally (green arrow),
    until the AD spectrum reproduces the AGN continuum identified in
    \S \ref{sec-gasf}.  The resulting AD model (red solid line)
    provides an estimate for $\nu_{\rm p}$, to be used (along with
    $L_{\rm d}^{\rm iso}$) in Eq. \ref{eq-ssad-compute_M} to infer the
    black hole mass and accretion rate.  The uncertainty in the disk
    luminosity $L_{\rm d}^{\rm iso}$ (a factor 2, \S
    \ref{sec-line2ld}) is shown as a grey shade.  In order to evaluate
    a confidence interval for our estimate of $M$ we repeat the whole
    process requiring the peak luminosity of the AD spectrum to lie
    respectively at the top and the bottom of the grey stripe.  The
    resulting AD models (dot--dashed red lines) provide respectively
    the lower and upper limits of the confidence interval on the black
    hole mass.}
  \label{fig-example1}
\end{figure}

Here we propose a new method for the AD modeling which relies on broad
line luminosities to estimate the total disk luminosity.
Fig. \ref{fig-example1} illustrates the method.
% ------------------------------------------------- 
% -------------------------------------------------
We use Eq. \ref{eq-line2disclum-rlnls1} to estimate $L_{\rm d}^{\rm
  iso}$.  When both the H{$\beta$} and \mgii{} line luminosities were
provided by our spectral fitting (\S \ref{sec-gasf}) we considered the
average of the resulting disk luminosities.  This enables us to
estimate a value for the luminosity of the peak $\nu_{\rm p}
L_{\nu_{\rm p}}^{\rm iso}$ (Eq. \ref{eq-ad-scaling-q}), i.e. to fix a
``ceiling'' in the $\nu L_\nu$ representation (black dashed line in
Fig. \ref{fig-example1}): the peak of the AD spectrum must lie on this
line.  Then we use observations from a single instrument (SDSS) to
constrain the peak frequency $\nu_{\rm p}$, which is related to the
black hole mass.  In particular, we shift the AD spectrum horizontally
(green arrow), until the AD spectrum reproduces the AGN continuum
identified in \S \ref{sec-gasf}.  Note that the model to be compared
with data in Fig. \ref{fig-example1} is an ``isotropic equivalent'' AD
spectrum (Eq. \ref{eq-ad-iso-corr}).  The resulting AD model (red
solid line) provides an estimate for $\nu_{\rm p}$, to be used (along
with $L_{\rm d}^{\rm iso}$) in Eq. \ref{eq-ssad-compute_M} to infer
the black hole mass and accretion rate.  Finally, we compute the
Eddington ratio $\ell$ using the luminosity of the disk:\footnote{We
  are neglecting the contribution from the torus in computing the
  Eddington luminosity since it is reprocessed radiation from the
  disk.  If we had used the bolometric luminosity $L_{\rm bol}$,
  instead of $L_{\rm d}$, the resulting value would be overestimated
  by a factor 3.4 on average (Eq. \ref{eq-lbol2ld} and
  \ref{eq-ad-iso-corr2}).} $\ell = L_{\rm d} / L_{\rm Edd}$.

The main source of uncertainty in the process is the uncertainty in
the disk luminosity $L_{\rm d}^{\rm iso}$ (a factor 2, \S
\ref{sec-line2ld}).  This uncertainty is shown as a grey shade in
Fig. \ref{fig-example1}.  In order to evaluate a confidence interval
for our estimates of $M$ we repeat the whole process requiring the
peak luminosity of the AD spectrum to lie respectively at the top and
the bottom of the grey stripe.  The resulting AD models (dot--dashed
red lines) provide respectively the lower and upper limits of the
confidence interval on the black hole mass, which typically is $\pm
0.5$ dex.  In some case these limiting AD models are too distant from
the data to provide a meaningful description of the AGN continuum.
This occurs typically for the low luminosity solution, corresponding
to the upper limit in black hole mass (e.g. \S \ref{sec-example}).  In
these cases a visual inspection would reduce the thickness of the grey
stripe, and hence the uncertainty on the black hole mass.

Further sources of uncertainties are the assumption on the radiative
efficiency $\eta \sim 0.1$ and on the viewing angle $\theta \sim 30$
deg (\S \ref{sec-method-hyp}).  The uncertainty due to the former can
be estimated by considering that our black hole mass estimate is $M
\propto \eta$ (case (v) of \S \ref{sec-ad-peakshift}), and that the
actual value of $\eta$ is expected to range from $\sim$6\% (for
non--rotating black hole) to at most $\sim$30\% \citep[for a spin
  parameter $a = J c / G M^2 = 0.998$,][]{1974-thorne-BHspin0.998}.
Therefore the uncertainty on the black hole mass due to the
uncertainty on $\eta$ (and ultimately on the black hole spin) is
+0.5/-0.2 dex.  If we consider the possibility that the black hole can
be maximally counter--rotating (with respect to the direction of
accretion) then the uncertainty on the black hole mass due to the
uncertainty on $\eta$ becomes $\pm 0.5$ dex.  The uncertainty due to
the assumption on the viewing angle can be estimated by propagating
the error in Eq. \ref{eq-ssad-compute_M}.  Typically, this is
negligible compared to the uncertainties discussed above, being at
most 0.04 dex (a factor $\sim$1.1), provided $\theta_{\rm max} <$45
deg.  The uncertainties due to $L_{\rm d}^{\rm iso}$ and $\eta$ are
likely uncorrelated, therefore, the maximum expected uncertainty for
the black hole mass estimate is $\pm 0.7$ dex.

The \llll{} procedure can be implemented without any fitting
procedure, provided we have an estimate for the broad line
luminosities and the AGN continuum (\S \ref{sec-gasf}).  The search
for the peak frequency can be implemented by identifying a wavelength
$\lambda_0$ and the corresponding luminosity of the AGN continuum
$\lambda_0 L_{\lambda_0}$, and requiring the AD spectrum to match this
luminosity at the same wavelength.  A comparison between the resulting
AD model and the AGN continuum can be performed ``a posteriori'' in
order to assess the reliability of the black hole mass estimate (\S
\ref{sec-method-best}).  For SDSS spectra of sources with $z<0.8$ the
value of $\lambda_0$ has been chosen empirically as follows:
\begin{equation}
  \nonumber
  \lambda_0 = \lambda_{\rm min}
  \cpar{ \frac{\lambda_{\rm max}} {\lambda_{\rm min}}}^{0.25}
\end{equation}
where $\lambda_{\rm min}$ and $\lambda_{\rm max}$ are the minimum and
maximum rest--frame wavelengths of the SDSS spectrum.  This value is
sufficiently close to the short wavelength edge in order to minimize
the contamination from other continuum components (either galaxy or
jet); $\lambda_0$ is also sufficiently far from the shortest available
wavelength, at which the estimated luminosity may be unreliable due to
noise and/or edge artifacts.  Following these prescriptions the
\llll{} method can be efficiently implemented as an automated
procedure on large samples.

\subsubsection{The \best{} procedure}
\label{sec-method-best}
In order to assess the reliability of the \llll{} procedure we proceed
with a visual localization of the AD spectrum peak using the SDSS and
GALEX observations.  Photometry from GALEX has been de--reddened
following the same procedure as for the optical SDSS spectra (\S
\ref{sec-gasf}).  In addition, when a jet component is considered in
the spectral fitting, we compute the jet--subtracted GALEX photometry.
We can not exclude that further absorption took place either in the
AGN environment or the intervening medium, therefore we consider the
photometry as lower limits to the actual rest--frame luminosity.  Note
that SDSS and GALEX data are not simultaneous, therefore it may happen
that these data sets trace the source in two different state of
emission, e.g. a disk or a jet dominated state
\citep{2012-Calderone-0954}.

For each source we require the slope of the AD spectrum to match as
close as possible the slope in the AGN continuum (\S \ref{sec-gasf}),
and to lie above the (jet--subtracted) GALEX photometry.  A few
exceptions to these rules will be considered in \S
\ref{sec-discussion}.  Since this is a manually tuned AD model we give
no error associated to the corresponding black hole mass.

\subsection{Example of application of the methods}
\label{sec-example}
\begin{figure*}
  \includegraphics[width=\textwidth]{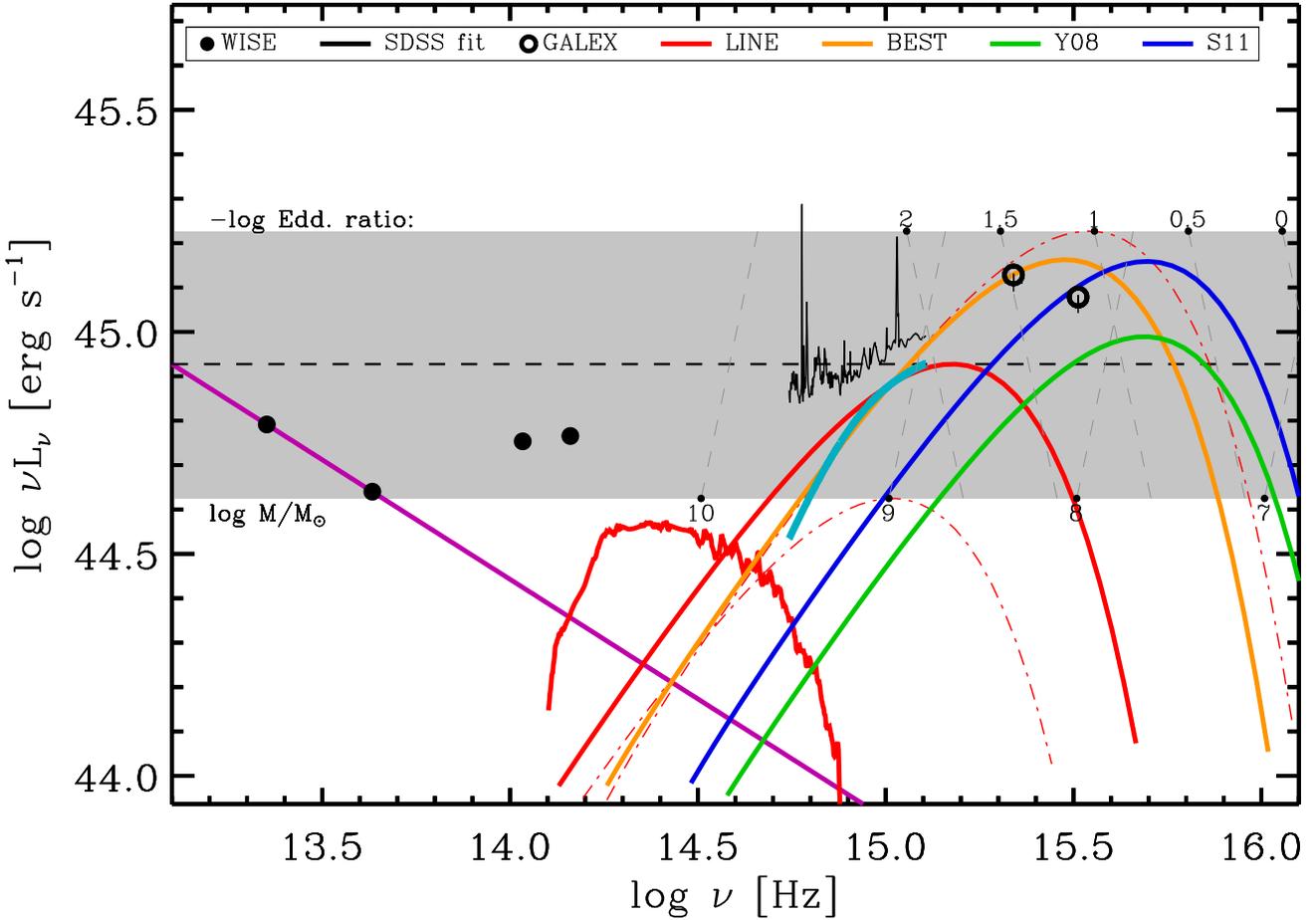}
  \caption{Application of the black hole mass estimation methods to
    the source SDSS J09531.7.09+283601.5 (\#6).  The WISE photometry
    is shown with black filled circles.  The spectral fit (\S
    \ref{sec-gasf}) is shown as a black line, while the AGN continuum
    component is shown as cyan line.  The jet power law extrapolation
    from IR data is shown as a purple line.  The GALEX photometry and
    their jet subtracted counterparts are shown as open circles and
    ``+'' symbols respectively.  The disk luminosity $L_{\rm d}^{\rm
      iso}$ with its uncertainty of a factor 2 is shown as a grey
    stripe: the peak of the AD model must lie within this region.  The
    grey dashed grid shows the location of peaks for AD models with
    values of black hole mass and Eddington ratio shown respectively
    below and above the grey stripe.  The \llll{} (\S
    \ref{sec-method-auto}) and \best{} (\S \ref{sec-method-best}) AD
    models are shown with a red and orange solid line respectively.
    In order to evaluate the uncertainty on the \llll{} black hole
    mass we repeat the procedure by requiring the AD model to peak at
    the top and the bottom of the grey stripe.  The resulting AD
    models are shown with dot--dashed red lines.}
  \label{fig-example}
\end{figure*}

As an example we discuss the case of SDSS J09531.7.09+283601.5 (\#6),
in Fig. \ref{fig-example}.  The WISE photometry is shown with black
filled circles.  The spectral fit (\S \ref{sec-gasf}) is shown as a
black line, while the AGN continuum component is shown as cyan line.
The jet power law extrapolation from IR data is shown as a purple
line.  The GALEX photometry and their jet subtracted counterparts are
shown as open circles and ``+'' symbols respectively.
% ------------------------------------------------- 
% ------------------------------------------------- 
The disk luminosity $L_{\rm d}^{\rm iso}$ with its uncertainty of a
factor 2 is shown as a grey stripe: the peak of the AD model must lie
within this region.  The grey dashed grid shows the location of peaks
for AD models with values of black hole mass and Eddington ratio shown
respectively below and above the grey stripe.

By applying our black hole mass estimation methods we identify the
\llll{} and \best{} AD models, shown with a red and orange solid line
respectively.  Both AD models provide a rather good representation of
the AGN continuum.  The \best{} AD model, however, needs a slightly
higher luminosity than the \llll{} model in order to lie above the
(jet--subtracted) GALEX photometry.  Note that the observed spectrum
(black line) has a significantly lower spectral slope (i.e. it is
``redder'') than the AGN continuum, because of the host galaxy and jet
contributions.  Having considered these components in the spectral
analysis allows us to reveal the real AGN continuum (cyan line) whose
slope agrees with our AD spectrum.

In order to evaluate the uncertainty on the \llll{} black hole mass we
repeat the procedure by requiring the AD model to peak at the top and
the bottom of the grey stripe.  The resulting AD models (shown with
dot--dashed red lines) are found to bracket the real case: the lower
one cannot account for the AGN continuum, while the higher one is
significantly above the GALEX photometry.  This situation often
occurred during the analysis of the sources (\S \ref{sec-discussion}),
therefore our black hole mass uncertainties are rather conservative.

Our AD models can be compared with those corresponding to the SE
virial masses and bolometric luminosities reported in the Y08 and S11
catalogs (green and blue lines).  We consider the disk luminosity as
computed using Eq. \ref{eq-lbol2ld}.  Note that our peak luminosities
are very similar to those of Y08 and S11, since this is the condition
we required (on average) to calibrate
Eq. \ref{eq-line2disclum-rlnls1}.  However, these models do not
provide a good description of the AGN continuum because their peak
frequencies lie $\sim$0.25 dex above our estimates of $\nu_{\rm p}$,
therefore our black hole mass estimates are 0.5 dex (a factor $\sim$3,
Eq. \ref{eq-ad-scaling}) greater than the virial ones.  The possible
reasons to explain such differences will be discussed in \S
\ref{sec-mass-discrepancy}.

\section{Results}
\label{sec-results}
We analyzed the data from the 23 RL--NLS1 sources of the Y08 catalog.
The spectral analysis (\S \ref{sec-gasf}) of each individual source is
shown in Fig. \ref{fig-spec}.  The results are summarized in
Tab. \ref{tab-results}.  The fitting models are in good agreement with
data with reduced $\chi^2$ in the range 1.16--1.86.  Also, the jet
contribution at optical/NUV wavelengths is typically negligible,
except for the \#2, \#11, \#15, \#16, \#20 and \#22 sources.

The results of our black hole mass estimation methods (\S
\ref{sec-method}) are shown graphically in Fig. \ref{fig-mass}
(adopting the same notation as in Fig. \ref{fig-example}).  The
results are summarized in Tab. \ref{tab-results2}.
% -------------------------------------------------------
\begin{table*}
  \caption{Results of our black hole mass estimation method. Columns
    are: (1) source numeric identifier; (2) flag to indicate if the AD
    ``signature'' (i.e. the slope $\alpha_\nu > -1$ a t optical
    wavelengths, see \S \ref{sec-discussion}) is missing. (3) peak
    frequency of the AD model, (4) black hole mass estimate (with its
    uncertainties) and (5) Eddington ratio for the AD model identified
    by our automatic procedure (\llll{} model); (6), (7), (8)
    corresponding quantities for the \best{} model; single epoch (SE)
    virial black hole mass estimate given in the (9) Y08 and (10) S11
    catalogs.}
  \label{tab-results2}
  \begin{tabular}{rc   |R{2}{1} R{1}{1} @{ (+} R{1}{1} @{,\ } R{1}{1} @{)}  R{1}{3}     |R{2}{1} R{1}{1} R{1}{3}      |R{1}{1} |R{1}{1}}
    \toprule
       \multicolumn{1}{c}  {}
    &  \multicolumn{1}{c}  {}
    &  \multicolumn{5}{|c} {\bf Method}
    &  \multicolumn{3}{|c} {\bf Best}
    &  \multicolumn{1}{|c} {\bf Y08}
    &  \multicolumn{1}{|c} {\bf S11}\\
     \midrule
       \multicolumn{1}{c}  {\#}
    &  \multicolumn{1}{c}  {Bad}
    &  \multicolumn{1}{|c} {\parbox[c]{1.3cm}  {\begin{equation}\nonumber   \log \frac{\nu_{\rm p}}       {\rm Hz}        \end{equation}}}
    &  \multicolumn{3}{c}  {\parbox[c]{1.7cm}  {\begin{equation}\nonumber   \log \frac{M}                 {M_{\sun}}      \end{equation}}}
    &  \multicolumn{1}{c}  {\parbox[c]{1.3cm}  {\begin{equation}\nonumber   \ell                                          \end{equation}}}
    &  \multicolumn{1}{|c} {\parbox[c]{1.3cm}  {\begin{equation}\nonumber   \log \frac{\nu_{\rm p}}       {\rm Hz}        \end{equation}}}
    &  \multicolumn{1}{c}  {\parbox[c]{1.3cm}  {\begin{equation}\nonumber   \log \frac{M}                 {M_{\sun}}      \end{equation}}}
    &  \multicolumn{1}{c}  {\parbox[c]{1.3cm}  {\begin{equation}\nonumber   \ell                                          \end{equation}}}
    &  \multicolumn{1}{|c} {\parbox[c]{1.3cm}  {\begin{equation}\nonumber   \log \frac{M}                 {M_{\sun}}      \end{equation}}}
    &  \multicolumn{1}{|c} {\parbox[c]{1.3cm}  {\begin{equation}\nonumber   \log \frac{M}                 {M_{\sun}}      \end{equation}}}\\
    \midrule
     1 &   &       15.2373 &       8.81809 &      0.364945 &     -0.505803 &     0.0216231 &       15.4806 &       8.43314 &     0.0842501 &       8.00000 &        8.0558154\\
     2 & * &          \nan &          \nan &          \nan &          \nan &   \nansep{c|} &          \nan &          \nan &   \nansep{c|} &       7.40000 &        7.9859325\\
     3 &   &       15.2629 &       8.52645 &      0.409763 &     -0.486595 &     0.0139844 &       15.1060 &       8.78896 &    0.00616250 &       7.20000 &        7.5498205\\
     4 &   &       15.5062 &       8.17890 &      0.589035 &     -0.390557 &     0.0586484 &       15.3045 &       8.48622 &     0.0187281 &       7.70000 &        8.0382013\\
     5 &   &       14.9972 &       9.22328 &       0.00000 &     -0.627451 &    0.00603557 &       15.0836 &       9.10084 &     0.0100663 &       7.50000 &        7.7770281\\
     6 &   &       15.1765 &       8.79639 &      0.166467 &     -0.582633 &     0.0117683 &       15.4774 &       8.30980 &     0.0622173 &       7.80000 &        7.8740957\\
     7 & * &          \nan &          \nan &          \nan &          \nan &   \nansep{c|} &          \nan &          \nan &   \nansep{c|} &       7.30000 &        7.5574819\\
     8 &   &       15.0004 &       9.02581 &       0.00000 &     -0.704282 &    0.00395413 &       14.9684 &       9.11142 &    0.00360821 &       7.30000 &        8.4509951\\
     9 & * &          \nan &          \nan &          \nan &          \nan &   \nansep{c|} &          \nan &          \nan &   \nansep{c|} &       8.10000 &        8.2354631\\
    10 & * &          \nan &          \nan &          \nan &          \nan &   \nansep{c|} &          \nan &          \nan &   \nansep{c|} &       7.10000 &        9.0186519\\
    11 & * &          \nan &          \nan &          \nan &          \nan &   \nansep{c|} &          \nan &          \nan &   \nansep{c|} &       7.10000 &        7.6185349\\
    12 &   &       15.2917 &       8.55467 &      0.505802 &     -0.460984 &     0.0196547 &       15.1156 &       8.84919 &    0.00757739 &       7.80000 &        7.8602964\\
    13 &   &       15.2373 &       8.54753 &      0.307323 &     -0.518608 &     0.0117519 &       15.4454 &       8.22100 &     0.0369584 &       6.80000 &        7.5221105\\
    14 &   &       15.1605 &       8.73749 &      0.307323 &     -0.525010 &    0.00892808 &       15.2053 &       8.66066 &     0.0114301 &       7.30000 &        7.6665477\\
    15 &   &       15.4134 &       8.82355 &      0.569828 &     -0.441777 &      0.110020 &       15.2277 &       9.20592 &     0.0492932 &       8.50000 &        8.5163740\\
    16 &   &       15.4614 &       8.14848 &      0.717087 &     -0.403362 &     0.0365709 &       15.1317 &       8.79330 &    0.00768087 &       7.50000 &        7.6433210\\
    17 &   &       15.0292 &       9.26265 &       0.00000 &     -0.499399 &    0.00895479 &       15.5670 &       8.44392 &      0.192660 &       7.80000 &        8.2813772\\
    18 &   &       15.2309 &       8.48385 &      0.416166 &     -0.486594 &    0.00945093 &       15.3621 &       8.29494 &     0.0205953 &       6.60000 &        7.6026572\\
    19 &   &       15.1605 &       8.92275 &      0.230493 &     -0.550620 &     0.0136226 &       15.2885 &       8.71146 &     0.0273683 &       7.90000 &        8.0108602\\
    20 & * &          \nan &          \nan &          \nan &          \nan &   \nansep{c|} &          \nan &          \nan &   \nansep{c|} &       6.30000 &             \nan\\
    21 &   &       14.9972 &       8.97077 &       0.00000 &     -0.768307 &    0.00338304 &       15.0356 &       8.90111 &    0.00414574 &       7.40000 &        7.8857021\\
    22 &   &       15.2533 &       8.26224 &      0.646659 &     -0.428971 &    0.00707894 &       15.2757 &       8.23023 &    0.00791555 &       6.90000 &             \nan\\
    23 &   &       15.2661 &       8.57042 &      0.473789 &     -0.467387 &     0.0158144 &       15.5478 &       8.14144 &     0.0789348 &       7.40000 &        7.6434172\\
    \bottomrule
  \end{tabular}
\end{table*}
% -------------------------------------------------------
The AD models identified by the \llll{} procedure provide a rather
good description of the AGN continuum in 17 over 23 cases (indicated
with a blank in the second column of Tab. \ref{tab-results2}).  The
remaining 6 sources cannot be modeled with an AD spectrum, and are
considered ``bad cases'' (indicated with a ``*'' symbol in the second
column of Tab. \ref{tab-results2}, see \S \ref{sec-discussion} for a
discussion of these sources).  These sources will not be considered in
the following analysis.

\begin{figure}
  \includegraphics[width=.5\textwidth]{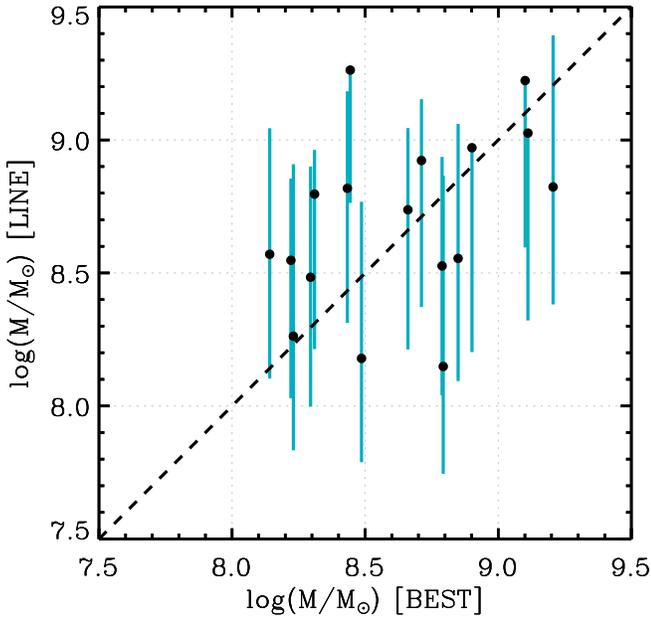}
  \caption{Comparison between black hole mass estimates obtained using
    the \llll{} and \best{} procedures.  Since the \best{} estimate is
    a manually tuned AD model we give no error associated to the
    corresponding black hole mass.}
  \label{fig-cmp-mass2}
\end{figure}

The comparison between the black hole mass estimates for the \llll{}
and \best{} AD models are shown in Fig. \ref{fig-cmp-mass2}.  
% -------------------------------------------------
% -------------------------------------------------
The mean value for the ratio of the two mass estimates is:
\begin{equation}
  \label{eq-ratio-best-auto}
  \left \langle \log \frac{M [{\rm \llll{}}]}{M [{\rm \best{}}]} \right \rangle =
  0.07 \pm    0.37
\end{equation}
The two black hole mass estimates are therefore compatible, within the
uncertainties associated to the \llll{} procedure (\S
\ref{sec-method-auto}).

In Fig. \ref{fig-cmp-mass} we show the comparison between the black
hole masses from the AD models (\llll{} in upper panels, \best{} in
lower panels) and the black hole masses from SE virial method, as
given in the Y08 (left panels) and S11 (right panels) catalogs.  The
uncertainty associated to SE virial mass is assumed to be 0.5 dex.  In
Fig. \ref{fig-cmp-mass-h} we show the histogram of the ratio of our
black hole mass estimates to the SE virial ones from the Y08 (left
panel) and S11 (right panel) catalogs.  The mean values for the
ratio of the mass estimates are:
\begin{equation}
  \label{eq-ratio-ad-vir}
  \begin{aligned}
    \left \langle \log \frac{M [{\rm \llll{}}]}{M [{\rm Y08    }]} \right \rangle = 1.2 \pm 0.5  \,\,\,\, &&
    \left \langle \log \frac{M [{\rm \llll{}}]}{M [{\rm S11    }]} \right \rangle = 0.8 \pm 0.3  \\
    \left \langle \log \frac{M [{\rm \best{}}]}{M [{\rm Y08    }]} \right \rangle = 1.1 \pm 0.4  \,\,\,\, &&
    \left \langle \log \frac{M [{\rm \best{}}]}{M [{\rm S11    }]} \right \rangle = 0.8 \pm 0.3  \\
  \end{aligned}
\end{equation}
The mean values are of the same order (or even greater) than the
maximum uncertainty associated to the \llll{} black hole mass estimate
(0.7 dex, \S \ref{sec-method-auto}), therefore our black hole mass estimates
are not compatible with the SE virial ones.

\section{Discussion}
\label{sec-discussion}
As discussed in \S \ref{sec-ssad-bbb}, the characteristic disk
spectral slope $\alpha_\nu = 1/3$ cannot be directly observed in AGN
SED.  However for values of the black hole mass $\log (M / M_{\sun})
\gtrsim 8$ and Eddington ratio $\ell \lesssim 1$, the peak of the AD
spectrum is (in principle) observable.  Indeed, for 17 over 23 sources
considered here, the SDSS continuum show an increasing trend in the
$\nu L_\nu$ representation (slope $\alpha_\nu >-1$) at $\log (\nu/{\rm
  Hz}) \gtrsim 14.8$, where the accretion disk is expected to dominate
over other emitting components.  Assuming that the observed BBB is
actually radiation emitted directly from the accretion disk, the slope
$\alpha_\nu >-1$ at $\log (\nu/{\rm Hz}) \gtrsim 14.8$ becomes the
``signature'' of the presence of the AD component.  In the considered
sample (\S \ref{sec-sample}) 17 sources over 23 show such signature
(the remaining six ``bad'' cases will be discussed below).  Our
spectral fitting procedure (\S \ref{sec-gasf}) reveals an emission
component that is well described by an AD model, although this has not
been included ``a priori'' in our fitting model.  Therefore,
observational data are in agreement with hypothesis (i), as discussed
in \S \ref{sec-method-hyp}.  Furthermore, the absorption column
densities $N_{\rm H}$, as estimated from X--ray spectral fitting, are
compatible with the Galactic values
\citep{2008-Yuan-population_rlnls1_with_blazar_prop,
  2010-Grupe-SEDOfXRaySelectedAGN}.  Hence, we expect the radiation we
observe not to be re--processed by any intervening medium, as assumed
in hypothesis (ii) (\S \ref{sec-method-hyp}).  Furthermore, we expect
the contribution from other emitting components, such as host galaxy
or jet, to be negligible at frequencies where our AD models are
constrained ($\log (\nu/{\rm Hz}) \gtrsim 14.8$).  In particular, the
slopes in the AGN continuum component ($\alpha_\nu > -1$) is
incompatible with the ones inferred from the galaxy template of
\citet{2001-Mannucci-galaxytemplate}.  Also, the jet component is
expected to decay at frequencies above a cutoff frequency of $\log
(\nu/{\rm Hz}) \lesssim 15$, as in typical Flat Spectrum
radio--quasars or powerful blazars \citep{2010-Ghisellini-blazarprop}.

The AD models identified by the \llll{} procedure (\S
\ref{sec-method-auto}) for the 17 ``good'' sources provide a rather
good description of the AGN continuum (Fig. \ref{fig-mass}).  In
particular, the AGN continuum slopes in the frequency range covered by
SDSS are in good agreement with the ones from \llll{} AD models (red
solid lines).  Also, the two limiting solutions (dot--dashed red
lines) likely bracket the real case, providing a robust estimate of
our uncertainties.  The average uncertainty on the black hole mass
estimates are of the order of $\pm 0.5$ dex (Tab. \ref{tab-results2}).
By taking into account the uncertainties due to hypotheses (iii) and
(iv) (radiative efficiency $\eta \sim 0.1$ and viewing angle $\theta
\sim 30$ deg) we obtain a maximum uncertainty of $\sim$0.7 dex (\S
\ref{sec-method-auto}).

In order to further assess the reliability of the \llll{} black hole
mass estimates we considered the \best{} AD models, identified by
visually tuning the $L_{\rm d}^{\rm iso}$ and $\nu_{\rm p}$ parameters
in order to achieve the best possible match between the AGN continuum
identified in \S \ref{sec-gasf} and the GALEX photometry.  In a few
cases we had to relax these requirements, as discussed below:
\begin{itemize}
\item the assumption that we can reliably estimate the jet
  contribution at optical/UV wavelengths by extrapolating a power law
  from the WISE photometry (\S \ref{sec-gasf}) may not be correct.
  For the \#15 and \#16 sources this assumption does not apply since
  the power law extrapolation (purple line) lies above the WISE
  photometry at shorter wavelengths (note that the error bars are
  smaller than the symbol in the plot).  Source \#5 would also falls
  in this class if the cutoff of synchrotron radiation at $\sim
  10^{14}$ Hz (\S \ref{sec-gasf}) is neglected, since jet
  extrapolation would intercept optical SDSS data.  In order to
  identify the \best{} AD model for the \#15 and \#16 sources we used
  the continuum observed in SDSS data (black solid lines in
  Fig. \ref{fig-mass}) rather than the jet--subtracted one (cyan
  lines) as requirement at optical wavelengths.  For \#16 we obtained
  a good agreement between the \best{} model and GALEX photometry.
  Lack of such agreement for \#15 will be discussed below.  A similar
  situation (jet component overestimated at optical/NUV wavelengths)
  possibly occurs also for source \#22.  In order to be conservative,
  for this source we retained the original constraints to identify the
  \best{} model.  For the other sources the jet extrapolation is
  marginal at optical/NUV wavelengths (Tab. \ref{tab-results}), hence
  the assumption discussed here has a negligible effect.

\item for the \#5 and \#15 sources the GALEX photometry does not
  follow the extrapolation from the SDSS slope.  Therefore a single AD
  model is not compatible with both the SDSS and GALEX observations.
  This may be a consequence of source variability, since the SDSS and
  GALEX data are not simultaneous.  Indeed, we found significant
  variability in GALEX photometry in a few sources (\S
  \ref{sec-gasf}).  In particular, source \#5 is known to be a
  variable source \citep{2009-Abdo-discovery_pmnj0948,
    2009-Abdo-mw_monitor_pmnj0948, 2010-Foschini-outburst}.  In these
  cases the \best{} model is computed relaxing the requirement of
  taking GALEX photometry into account, and using only the SDSS data
  as guidelines.

\item for the \#8 and \#21 sources the SDSS and GALEX data appear to
  trace the peak of the AD spectrum.  For these sources we neglected
  the jet component.  Note that the \best{} AD model for these sources
  provide a robust estimate of the black hole mass, since the peak of
  the AD spectrum has been directly observed.
\end{itemize}
The \best{} AD models are in good agreement with the \llll{} ones.  In
particular, note that the peak in \best{} AD spectra lie inside the
grey stripes for all sources except \#17.  Hence, the
Eq. \ref{eq-line2disclum-rlnls1} are well calibrated.  The resulting
\best{} black hole mass estimates are compatible the \llll{} ones
(Eq. \ref{eq-ratio-best-auto}, Fig. \ref{fig-cmp-mass2}).  Also, the
scatter in Fig. \ref{fig-cmp-mass2} (0.4 dex) is compatible with the
uncertainty on the \llll{} estimates due to our ignorance on $L_{\rm
  d}^{\rm iso}$ (0.5 dex, \S \ref{sec-method-auto}).  This provides
further support for the reliability of the \llll{} black hole mass
estimates.  We conclude that, under the assumptions discussed in \S
\ref{sec-method-hyp} our \llll{} procedure provides a reliable
estimate of the black hole mass, within the quoted uncertainties.

In six cases the \llll{} method do not provide an acceptable
description of data (sources marked with a ``*'' symbol in the second
column of Tab. \ref{tab-results2}.  For these sources the observed
SDSS continuum does not show an increasing trend (in the $\nu L_\nu$
representation ) at $\log (\nu/{\rm Hz}) \gtrsim 14.8$.  In two cases
the SDSS continuum are dominated by the jet and/or host galaxy
emission (\#2 and \#11), and the AD spectrum is not directly visible.
In four cases (\#7, \#9, \#10 and \#20) the observed SDSS continuum
appears ``flat'' in the $\nu L_\nu$ representation, with no hints for
a change of slope.  Although the jet--subtracted continuum (cyan line)
suggests the presence of an AD spectrum, this decomposition strongly
depends on the assumption that the extrapolation of the jet component
from IR data is also valid at optical wavelengths.  In order to be
conservative, we mark these sources as ``bad'', and neglect them in
our analysis.

\subsection{Comparison with SE virial mass estimates}
\label{sec-mass-discrepancy}
\begin{figure*}
  \includegraphics[width=\textwidth]{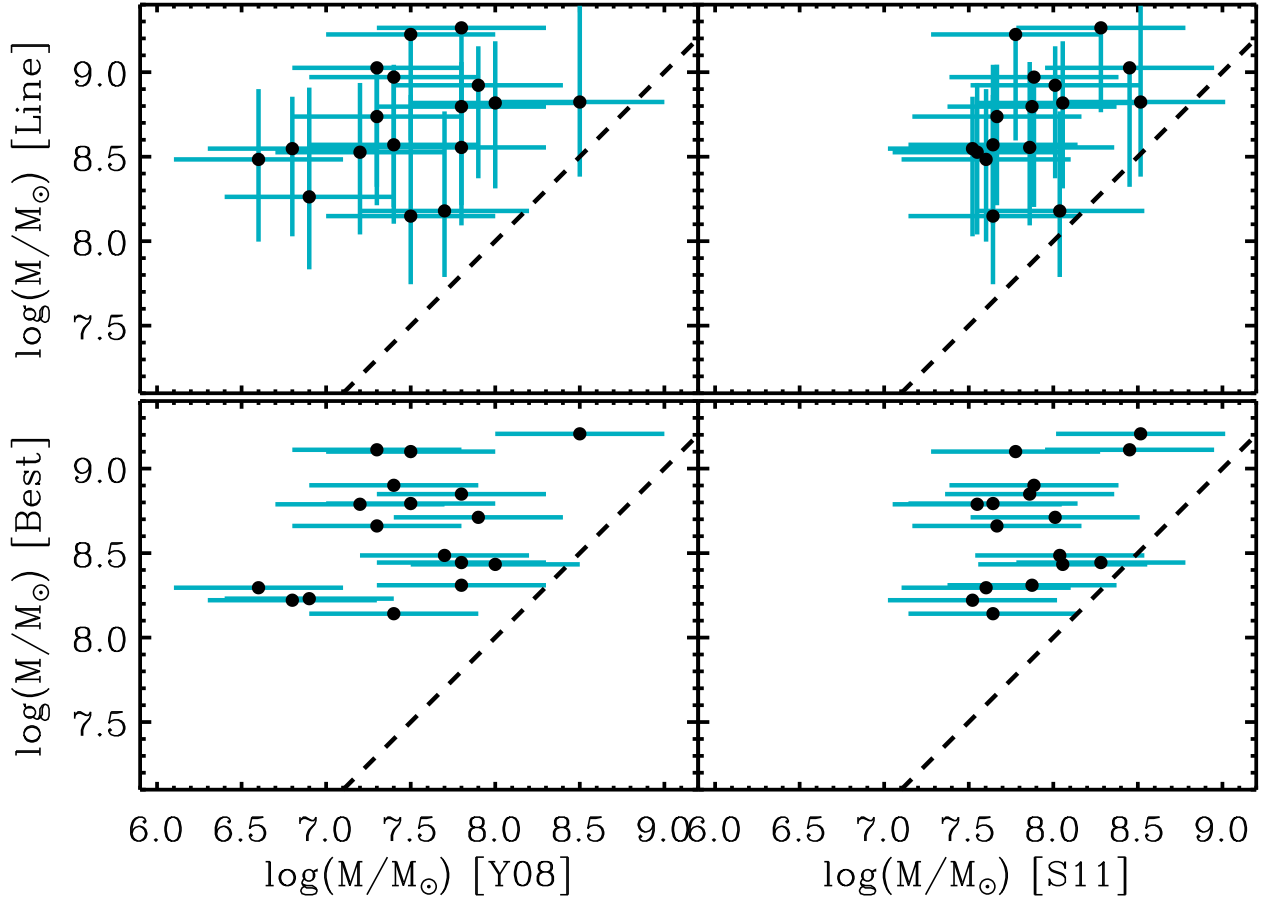}
  \caption{Comparison between black hole mass estimates of AD models
    identified by our procedures (\llll{} models, upper panels,
    \best{} models lower panels) and single epoch (SE) virial masses
    as given in the Y08 (left panels) and S11 (right panels) catalogs.
    The uncertainty associated to SE virial mass is 0.5 dex.}
  \label{fig-cmp-mass}
\end{figure*}

\begin{figure*}
  \includegraphics[width=\textwidth]{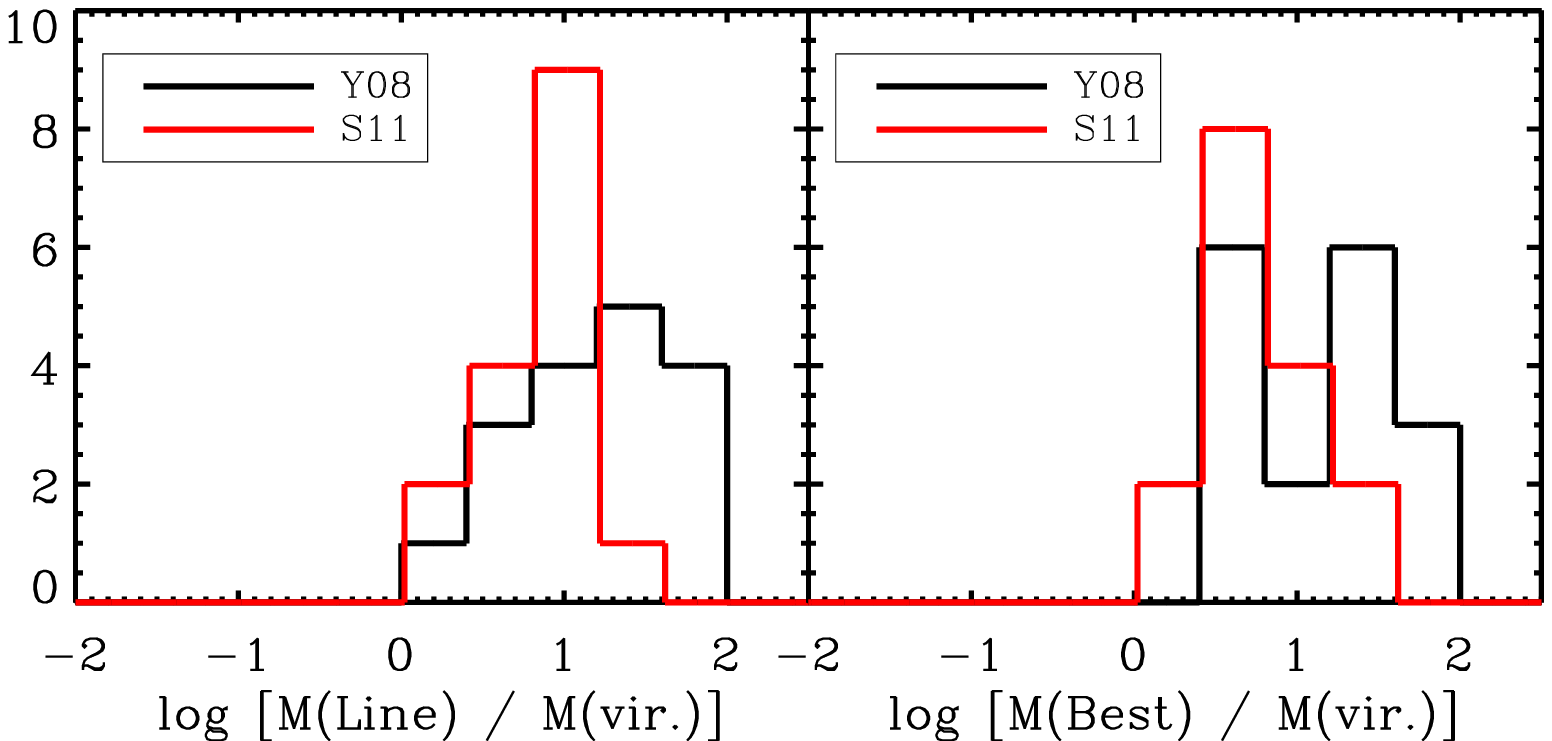}
  \caption{Histogram of our black hole mass estimates to the SE virial
    ones from the Y08 (left panel) and S11 (right panel) catalogs.
    Our estimates are greater than SE virial ones by $\sim$1 dex for
    Y08 and $\sim$0.8 for S11 (Eq. \ref{eq-ratio-ad-vir}).}
  \label{fig-cmp-mass-h}
\end{figure*}

The comparison with the SE virial mass reveals a systematic
discrepancy between our mass estimate and those from the Y08 and S11
catalogs (Fig. \ref{fig-cmp-mass} and \ref{fig-cmp-mass-h}).
% -------------------------------------------------
% ------------------------------------------------- 
% -------------------------------------------------
Although the discrepancy ($\gtrsim $0.7 dex,
Eq. \ref{eq-ratio-ad-vir}) is of the same order of the maximum
uncertainty associated to the \llll{} procedure (\S
\ref{sec-method-auto}), it appears systematic.  Therefore our black
hole mass estimates are not compatible with the virial ones given in
the Y08 and S11 catalogs.  On average, our black hole masses turn out
to be a factor $\sim$6 (0.8 dex) greater than virial ones.  A possible
explanation for this black hole mass discrepancy may involve a
radiative efficiency which is a factor $\sim$6 lower than assumed.
However, a value of $\eta \sim 0.02$ is lower than the minimum
efficiency expected for accretion onto a maximally counter--rotating
black hole ($\sim 0.03$).  Notice that the AD spectrum suggests that
the accretion disk is still in the ``radiatively efficient'' regime:
half the gravitational energy gained by matter at each radius is
locally emitted as radiation, i.e. it is not advected into the hole.

The mass discrepancy is not due to an inaccurate estimation of the jet
contribution at optical/NUV wavelengths.  If the actual jet
contribution is lower than estimated, the corresponding AGN continuum
luminosity ($\lambda_0 L_{\lambda_0}$ (Tab. \ref{tab-results}) would
correspondingly be higher.  The ``ceiling'' luminosity level, on the
other hand, are not affected by the presence of the jet, since it
relies on line luminosities.  In order to reproduce an higher
$\lambda_0 L_{\lambda_0}$, retaining the same peak luminosity, the
\llll{} AD model must shift to lower frequencies.  Therefore, if the
actual jet contribution is lower than estimated, we would have
obtained greater \llll{} black hole mass estimates, and greater
discrepancy with SE virial masses.  Furthermore, the mass discrepancy
is not due to having neglected the general relativistic corrections in
the AD model.  As discussed in \S \ref{sec-GRcorrections}, the AD
model used throughout this work mimics the more sophisticated general
relativistic one with $\eta_{\rm gr} \sim 0.1$, as long as frequencies
below the peak are concerned.

We speculate that a possible explanation for the mass discrepancy is a
selection effect in calibrating the SE virial method.  The virial
method relies on the calibration of both a BLR size--continuum
luminosity relation and of a virial factor
\citep{2012-Park-calibSEVir}.  However, the sample used to perform the
calibration consists of a few dozens of sources: the ones that has
been reverberation mapped.  As a consequence, the calibration of the
method may be biased by selection effects.  In particular, the method
may provide significantly underestimated black hole masses if the BLR
has a flat disk--like geometry, and it is seen almost face--on
\citep{2011-decarli-blrgeom}.  If these conditions apply, then the
discrepancy between our mass estimates and the virial ones would be
greater (on average) for AGN showing the smallest widths of broad
emission lines, i.e. the class of NLS1 sources \citep{2008-Decarli}.
The black hole mass estimates provided by our method, on the other
hand, are scarcely affected by the viewing angle and the geometry of
the BLR (\S \ref{sec-method-auto}).

The Eddington ratios are in the range $\ell=$0.04--0.2
(Tab. \ref{tab-results2}), significantly below the values reported in
Y08.  This discrepancy is due both to our greater black hole mass
estimates (a factor $\sim$6) and to the fact that we used the disk
luminosity $L_{\rm d}$ (instead of $L_{\rm bol}$) to compute the
Eddington ratio (a factor $\sim$3.4, Eq. \ref{eq-lbol2ld} and
\ref{eq-ad-iso-corr2}).  Hence, our values of Eddington ratio a factor
$\sim 20$ smaller than SE virial ones on average.  With such small
values of $\ell$ the role of radiation pressure in determining the SE
virial masses \citep{2008-Marconi, 2011-chiaberge-rp} is expected to
be small, not sufficient to explain the black hole mass discrepancy.

\subsubsection{The ``temperature'' argument}
\label{sec-caldofreddo}
If the assumptions discussed in \S \ref{sec-method-hyp} apply, then
all the independently estimated black hole masses (even the SE virial
ones) should produce an AD spectrum compatible with the observed data.
This provides a simple way to compare our results with those reported
in Y08 and S11.  The black hole mass discrepancy arises because the
peak frequencies of the \llll{} and \best{} AD spectra are
significantly lower then the peak frequencies of the Y08 and S11 ones,
while the peak luminosities are compatible.  If the Y08 and S11 AD
models (blue and green lines in Fig. \ref{fig-mass}) were the correct
ones, there must be a physical process able to shift photon to lower
frequencies (i.e. to lower ``temperatures'') in order to account for
the observed data.  However, such a process cannot exist on a
thermodynamic basis, since black body spectra (which build up the AD
model) already have the lowest temperature corresponding to a given
luminosity and emitting surface.  Since the luminosities are the same,
the only way to reduce the temperature is to increase the emitting
surface, that is by increasing the black hole mass.

\section{Conclusions}
\label{sec-conclusion}
In this work we analyzed the relationship between the Big Blue Bump
(BBB) observed in the SED of Type 1 AGN and a \citet{1973-ssad}
accretion disk model (AD model).  The characteristic disk spectral
slope $\alpha_\nu = 1/3$ cannot be directly observed in the AGN SED
because of the contributions from other emitting components such as
the host galaxy, the torus or (for radio--loud sources) the jet (\S
\ref{sec-ssad-bbb}).  Once these contributions are taken into account,
the observations are compatible with the presence of an emitting
component which is well described by an AD model.  In particular, the
peak of such component can be observed directly in the frequency range
$\log(\nu / {\rm Hz}) =$ 14.8--15.5.  By comparing the average Type 1
AGN SED from \citet{2006-richards-meanSED} with the AD model we
calibrate the relations to estimate the total disk luminosity using
the continuum line luminosities (at 1350\gaa{}, 3000\gaa{} and
5100\gaa{}) as proxy (\S \ref{sec-cont2ld}).  Furthermore, by using
the emission line templates from \citet{1991-francis-composite}, we
calibrate analogous relations based on the line luminosities of
H$\beta$, \mgii{} and \civ{} (\S \ref{sec-line2ld}).  The latter
provide more reliable disk luminosity estimates when the continuum is
not dominated by the AD spectrum.

The interpretation of the BBB as being due to the thermal emission
from an AD allows to link the AGN observed properties to the
properties of the super massive black hole.  In particular, the
luminosity and frequency at the peak of the AD spectrum uniquely
identifies a value for the black hole mass and the accretion rate (\S
\ref{sec-method}).  However, the direct observation of the peak of the
AD spectrum requires broad--band multiwavelength observations.  In
order to estimate the black hole mass also for sources with a poor
observational coverage, we propose a new method which relies only on
spectral observations at optical wavelengths.  In particular, we
constrain the luminosity of the peak by using the line luminosities.
Then we constrain the frequency of the peak by requiring the AD model
continuum and slope to reproduce the observed AGN continuum beneath
the emission lines (\S \ref{sec-method-auto}).  The maximum
uncertainty on our black hole mass estimates is $\sim 0.7$ dex (on
average).  This uncertainty is greatly reduced if the disk luminosity
can be accurately determined, namely when the peak of the AD spectrum
is visible within the frequency range of the data.

We applied our method to the sample of 23 radio--loud Narrow line
Seyfert 1 galaxies (RL--NLS1) of
\citet{2008-Yuan-population_rlnls1_with_blazar_prop}.  The method
provides reliable black hole mass estimates for 17 sources over 23, if
our interpretation of the BBB is correct (\S \ref{sec-results},
\ref{sec-discussion}).  The remaining six sources are either dominated
by synchrotron radiation from the jet, or do not show ``hints'' for
the presence of an AD--like emitting component.  The resulting black
hole mass estimates are a factor $\sim 6$ (on average) greater than
the corresponding single epoch virial mass estimates, and the
Eddington ratios are a factor $\sim 20$ below.  The discrepancy
between our black hole mass estimates and the single epoch virial ones
may be due to selection effects occurred in the calibration of the
BLR--continuum relation and of the virial factor.

The black hole masses estimated in this paper for the sample of
RL--NLS1 are in the interval $\log (M / M_{\sun}) = $ 8--9, and the
Eddington ratios are $\ell = $0.04--0.2.  Therefore, very radio--loud
NLS1 appear not to be extreme in terms of black hole masses and
Eddington ratios.  Their black hole masses are similar to those of
blazars.  We find no evidence for jetted sources with mass below $10^8
M_{\sun}$.

\bigskip

\noindent 
{\bf ACKNOWLEDGMENTS}\\
We acknowledge M. Vestergaard and B.J. Wilkes for having provided the
UV iron template.  We thank R. Decarli for fruitful discussion.

\appendix

\section{Accretion disk model}

\subsection{Shakura\&Sunyaev Accretion Disc (AD)}
\label{sec-ad-ss73}
Here we review the properties of the AD model for a geometrically
thin, optically thick accretion disk \citep[][hereafter SS73
  model]{1973-ssad} adopted in our analysis of the SED.

The amount of gravitational energy released from each annulus of the
disk is given by
\begin{equation}
\label{eq-ad-flux}
  F(R) = \frac{3}{8 \pi}
  \cpar{\frac{R}{R_{\rm g}}}^{-3}
  \spar{1 - 
    \cpar{\frac{R}{R_{\rm in}}}^{-1/2}
  }
  \frac{\dot{M} c^2}{R_{\rm g}^{2}}
\end{equation}
where $R$ is the distance from the black hole, $R_{\rm g} = G M / c^2$
is the gravitational radius of the black hole, $R_{\rm in}$ is the
inner radius of the disk.  By introducing the adimensional parameters:
\begin{equation}
  \label{eq-ad-adimpar}
  x = \frac{R}{R_{\rm in}}
  \hspace{1cm}
  \eta = \frac{R_{\rm g}}{2 R_{\rm in}}
\end{equation}
we rewrite the emitted flux as $F(R) = \tilde{F}(x)\ {\cal P}$, with
\begin{equation}
  \label{eq-ad-flux-norm}
    \tilde{F}(x) = \frac{3}{\pi} x^{-3} (1 - x^{-1/2})
    \hspace{1cm}
    {\cal P} = \eta^3 \frac{\dot{M} c^2}{R_{\rm g}^{2}}
\end{equation}
where all physical quantities are cast into the ${\cal P}$ parameter,
while $\tilde{F}(x)$ accounts for dimensionless flux distribution.
The total disk luminosity is given by
\begin{equation}
  \begin{aligned}
    L_{\rm d} = 2 \times \int_{R_{\rm in}}^{R_{\rm out}} 2 \pi
    R\ F(R)\ {\rm d}R\\ 
    = \spar{3 \int_{1}^{x_{\rm out}}
      x^{-2} (1 - x^{-1/2})\ {\rm d}x} 
    \eta \dot{M} c^2
  \end{aligned}
\end{equation}
where $x_{\rm out} = R_{\rm out}/R_{\rm in}$ is the normalized outer
radius of the disk.  The quantity in squared parentheses is equal to 1
(provided $R_{\rm out} \gg R_{\rm in}$), therefore the parameter
$\eta$ as defined above is the radiative efficiency of the disk.  By
assuming $R_{\rm in} = 6 R_{\rm g}$ (appropriate for a non--rotating
black hole) we obtain $\eta \sim 0.1$.

The maximum amount of energy flux is (by differentiating
Eq. \ref{eq-ad-flux-norm}):
\begin{equation}
  \label{eq-ad-maxflux}
  {\rm MAX}[F(R)] = F(R_{\rm max}) = 
  \frac{64}{\pi} \cpar{\frac{3}{7}}^7
  {\cal P}
\end{equation}
and it is emitted at a radius $R_{\rm max} = 49/36\ R_{\rm in}$.  The
assumption of optical thickness implies that each annulus emits
radiation as a black body with temperature: $T(R) =
[F(R)/\sigma]^{1/4}$.  The maximum temperature is therefore
(Eq. \ref{eq-ad-maxflux}):
\begin{equation}
  \label{eq-ad-scaling-temp}
  \frac{T_{\rm max}}{\rm [K]} = 
        3.46 \times 10^4
        \cpar{\frac{\eta}{0.1}}^{3/4}
        \cpar{\frac{M}{10^9 M_{\sun}}}^{-1/2}
        \cpar{\frac{\dot{M}}{M_{\sun}\ {\rm yr}^{-1}}}^{1/4}\\
\end{equation}

The emitted spectrum is a superposition of black body spectra:
\begin{equation}
  \begin{aligned}
    \label{eq-ad-lumdens}
    L_{\nu} = 2 \times
    \int_{R_{\rm in}}^{R_{\rm out}} 2 \pi\ R\ {\rm d} R\ \pi B[\nu, T(R)]\\
     = 4 \pi^2 R_{\rm in}^2 {\cal P}^{3/4} 
     \int_{1}^{x_{\rm out}} x\ {\rm d}x\ B \spar{\frac{\nu}{{\cal P}^{1/4}},
       \cpar{\frac{\tilde{F}(x)}{\sigma}}^{1/4}}
  \end{aligned}
\end{equation}
where $B[\nu, T(R)]$ is the Planck function.  The spectrum profile is
completely determined by the dimensionless integral, the only
dependences on physical parameters (${\cal P}$) being the
characteristic frequency ($\propto {\cal P}^{1/4}$) and the overall
normalization ($\propto R_{\rm in}^2 {\cal P}^{3/4}$).  The disk
spectra are therefore self--similar, and the peak frequency and
luminosity scale as:
\begin{equation}
  \begin{aligned}
    \label{eq-ad-scaling}
    \frac{\nu_{\rm p}}{\rm [Hz]} =
         {\cal A}
         \cpar{\frac{\eta}{0.1}}^{3/4}
         \cpar{\frac{M}{10^9 M_{\sun}}}^{-1/2}
         \cpar{\frac{\dot{M}}{M_{\sun}\ {\rm yr}^{-1}}}^{1/4}\\
    \frac{\nu_{\rm p} L_{\nu_{\rm p}}}{\rm [erg\ s^{-1}]} =
         {\cal B}
         \cpar{\frac{\eta}{0.1}}
         \cpar{\frac{\dot{M}}{M_{\sun}\ {\rm yr}^{-1}}}\\
  \end{aligned}
\end{equation}
where $\nu_{\rm p}$ is the frequency of the peak in the $\nu L_{\nu}$
representation, $\log {\cal A} = 15.25$ and $\log {\cal B} = 45.36$.
By introducing the Eddington ratio $\ell = L_{\rm d} / L_{\rm Edd}$
(with $L_{\rm Edd} = 1.3 \times 10^{47} (M / 10^9 M_{\sun})$ erg
s$^{-1}$), the previous equations can be rewritten as:
\begin{equation}
  \begin{aligned}
  \label{eq-ad-scaling-edd}
  \frac{\nu_{\rm p}}{\rm [Hz]} = 
       {\cal A}
       \cpar{\frac{\eta}{0.1}}^{1/2}
       \cpar{\frac{M}{10^9 M_{\sun}}}^{-1/4}
       \cpar{\frac{\ell}{0.04}}^{1/4}\\
  \frac{\nu_{\rm p} L_{\nu_{\rm p}}}{\rm [erg\ s^{-1}]} =
       {\cal B}
       \cpar{\frac{M}{10^9 M_{\sun}}}
       \cpar{\frac{\ell}{0.04}}\\
  \end{aligned}
\end{equation}
Notice that, for a given value of $\eta$, an estimate of the
luminosity and peak frequency allows to determine the physical
parameters $M$ and $\dot{M}$.

The spectrum of an AD is shown in Fig. \ref{fig-template-ssad} (black
solid line): the superposition of black body spectra, weighted by the
surface of emitting annuli produces a ``flat'' spectrum with
slope\footnote{The slope $\alpha_\nu = 1/3$ is achieved only for
  $R_{\rm out} \rightarrow \infty$.  For a finite value of the outer
  radius of the disk, such as the one used in this work $R_{\rm out} =
  2 \times 10^3 R_{\rm g}$, a more realistic value for the slope is
  $\alpha_\nu \sim 1/4$.} $\alpha_\nu \sim 1/3$; at highest
frequencies, the Wien spectrum from the inner ring dominates, and the
overall spectrum decays exponentially.

The self--similarity of AD spectra implies the existence of relations
among quantities at the peak frequencies in the $L_{\nu}$ and the $\nu
L_{\nu}$ representations ($\nu_{\rm p^\prime}$ and $\nu_{\rm p}$
respectively), and the disk luminosity $L_{\rm d}$:
\begin{equation}
  \label{eq-ad-scaling-q}
  \frac{\nu_{\rm p}}{\nu_{\rm p^\prime}} = 3.1  
  \hspace{1cm}
  \frac{L_{\nu_{\rm p}}}{L_{\nu_{\rm p^\prime}}} = 0.66
  \hspace{1cm}
  \frac{\nu_{\rm p} L_{\nu_{\rm p}}}{L_{\rm d}} = 0.5
\end{equation}
Also, note that the peak luminosity $\nu_{\rm p} L_{\nu_{\rm p}}$ is
independent from the actual value of $R_{\rm out}$, as long as $R_{\rm
  out} \gtrsim 10\, R_{\rm in}$.  The relation between the maximum
temperature in the disk and the color temperature of the AD spectrum
(i.e. the black body temperature associated to the peak frequency
$\nu_{\rm p^\prime}$) is $T_{\rm max} = 3.5\ T_{\rm col}$.

\subsection{Peak shift}
\label{sec-ad-peakshift}
\begin{figure}
  \includegraphics[width=.5\textwidth]{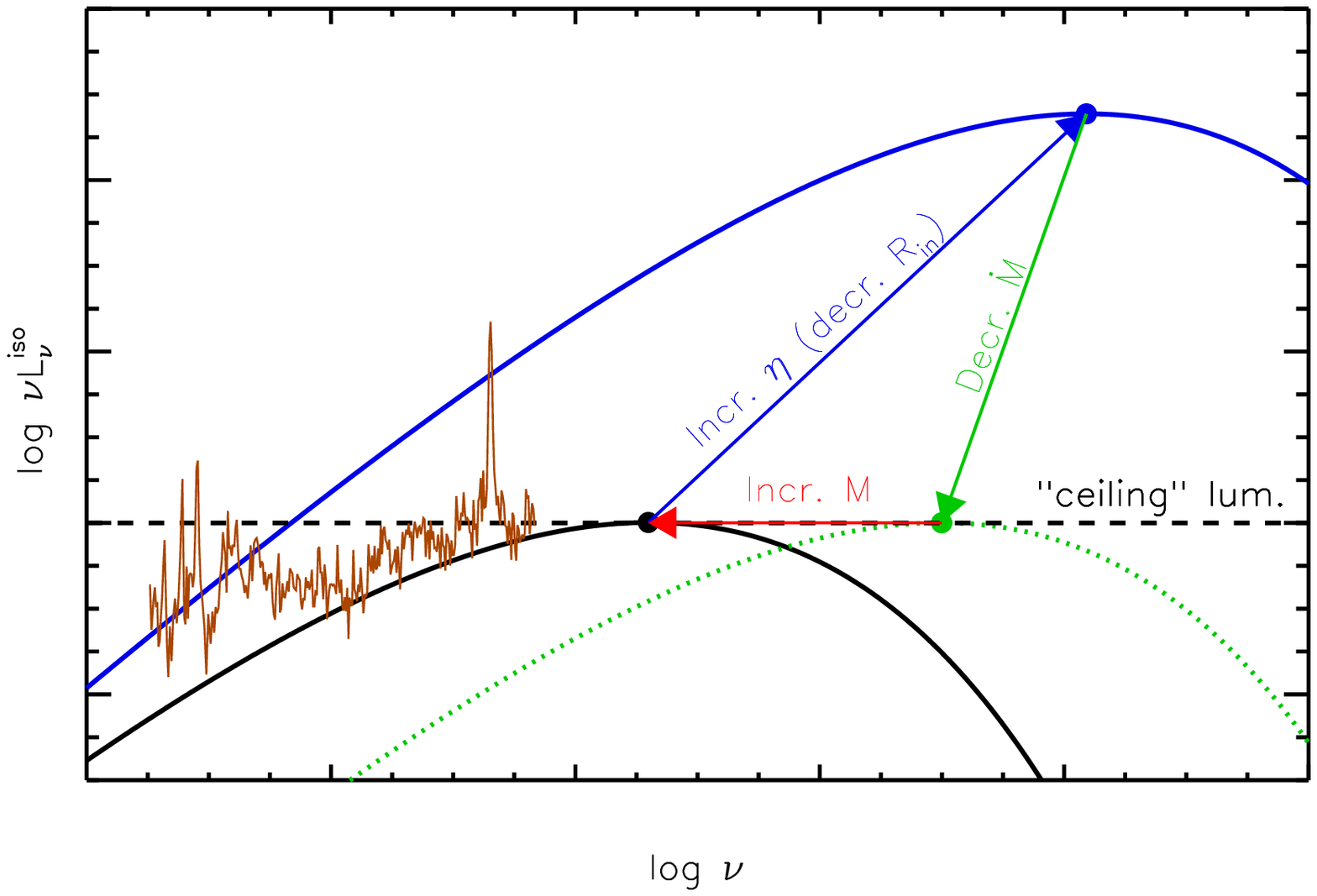}
  \caption{The black hole mass estimate provided by the AD modeling
    procedures (\S \ref{sec-method}) show a linear dependence on the
    assumed value of the radiative efficiency: $M \propto \eta$ (case
    (v) in \S \ref{sec-ad-peakshift}).  The physical interpretation of
    this dependence is shown in the figure: the AD model (black line)
    identified with either the \llll{} or \best{} procedure is in
    agreement with observed data (brown line).  In particular the peak
    lies at the ``ceiling'' luminosity level determined by broad line
    luminosities (Eq. \ref{eq-line2disclum-rlnls1}) and the AD
    spectrum is in agreement with the AGN continuum.  By increasing
    the $\eta$ parameter (i.e. decreasing the inner radius of the disk
    $R_{\rm in}$) the peak shifts to higher frequencies and
    luminosities as radiation comes from the inner, hotter radii (blue
    line).  This new model would no longer be in agreement with the
    ``ceiling'' luminosity argument, therefore we must decrease the
    accretion rate (green line), leaving $M$ and $\eta$ unchanged.
    Still, the obtained spectrum is not in agreement with the observed
    continuum, therefore we must decrease the ``temperature'' of the
    spectrum (\S \ref{sec-caldofreddo}), by increasing $M$ (red line).
    The final AD model is again in agreement with observed data, but
    has higher values of $\eta$ and $M$ (and a lower value of
    $\dot{M}$).}
  \label{fig-diskscale}
\end{figure}

The physical parameters $\eta$, $M$ and $\dot{M}$ uniquely identify
the frequency and luminosity of the spectral peak
(Eq. \ref{eq-ad-scaling} or \ref{eq-ad-scaling-edd}).  Variations of
one or more of these parameters will shift the peak along specific
directions, whose slope in a $\log \nu L_{\nu}$ vs. $\log \nu$ plot is
given by:
\begin{equation}
  \label{eq-ad-slope-peakshift}
  \alpha = \frac{{\rm d} \log \nu_{\rm p} L_{\nu_{\rm p}}}
         {{\rm d} \log  \nu_{\rm p}}
\end{equation}
Here is a list of peak shift relations used in this work:
\begin{enumerate}
\item vertical shift ($\alpha = \infty$): variations in $M$, $\dot{M}$
  with constant $\dot{M} / M^2$ ratio (fixed $\eta$);
\item horizontal shift ($\alpha = 0$): variations in $M$ with constant
  $\dot{M}$ (fixed $\eta$);
\item $\alpha=4$: variations in $\dot{M}$, with constant $M$ (fixed
  $\eta$);
\item $\alpha=-4$: variations in $M$ and $\dot{M}$, with constant
  $\ell \propto \dot{M}/M$ (fixed $\eta$);
\item no shift: variations in all parameters, with constant $\eta
  \dot{M}$ and $\eta / M$.
\end{enumerate}
In particular, case (v) is used in \S \ref{sec-mass-discrepancy} to
show that if the actual radiative efficiency $\eta$ is greater than
hypothesized in \S \ref{sec-method-hyp}, then our black hole mass
estimate is a lower limit.  The physical interpretation of case (v) is
depicted in Fig. \ref{fig-diskscale}.
% -------------------------------------------------------
% -------------------------------------------------------
By following the black hole mass estimation method outlined \S
\ref{sec-method-auto} we identify an AD model (black line) in
agreement with observed data (brown line): the peak lies at the
``ceiling'' luminosity level determined by broad line luminosities
(Eq. \ref{eq-line2disclum-rlnls1}) and the AD spectrum is in agreement
with the observed continuum.  This model relies on the assumption of
radiative efficiency $\eta \sim 0.1$.  However, the actual value of
efficiency may be different.  By increasing the $\eta$ parameter
(i.e. decreasing the inner radius of the disk $R_{\rm in}$) the peak
shifts to higher frequencies and luminosities as radiation comes from
the inner, hotter radii (blue line).  This new model would no longer
be in agreement with the ``ceiling'' luminosity argument, therefore we
must decrease the accretion rate (green line), leaving $M$ and $\eta$
unchanged.  Still, the obtained spectrum is not in agreement with the
observed continuum, therefore we must decrease the ``temperature'' of
the spectrum (\S \ref{sec-caldofreddo}), by increasing $M$ (red line).
The final AD model is again in agreement with observed data, but has
higher values of $\eta$ and $M$ (and a lower value of $\dot{M}$).

\subsection{Observational properties}
\label{sec-obsprop}
The emission from the whole (geometrically thin) disk is anisotropic
since the observed flux is proportional to the projected area seen by
the observer, i.e. $F_{\nu} \propto \cos \theta$, where $\theta$ is
the viewing angle. By requiring $\oint_{\rm Sph.} F_{\nu} D_{\rm L}^2
d \Omega = L_{\nu}$ we obtain:
\begin{equation}
  \label{eq-ad-lumfluxrel}
  L_{\nu} = \frac{2 \pi D_{\rm L}^2 F_{\nu_o}}{(1+z) \cos \theta}
\end{equation}
where $D_{\rm L}$ is the luminosity distance, $\nu_o = \nu / (1+z)$ is
the observed frequency and $F_{\nu_o}$ is the observed flux density.
Note that the luminosity--flux relation for a thin disk is different
from the isotropic case, in particular the relation between the
``isotropic equivalent'' luminosity and the real luminosity is:
\begin{equation}
  \label{eq-ad-iso-corr}
  L_{\nu}^{\rm iso} = 2 \cos \theta L_\nu
\end{equation}
The observed flux is therefore (from Eq. \ref{eq-ad-lumdens},
\ref{eq-ad-lumfluxrel}):
\begin{equation}
  \label{eq-ad-fluxdens}
  F_{\nu_o} = \frac{4 \pi h \nu_o^3}{c^2 D_{\rm L}^2} (1+z) \cos \theta
  \int_{R_{\rm in}}^{R_{\rm out}} \frac{R\ {\rm d} R}{\exp{(h \nu / k T) - 1}}
\end{equation}
The model for the observed spectrum has four parameters: $M$,
$\dot{M}$, $R_{\rm in}$ and $\cos \theta$ (the value of $R_{\rm out}$
is not important here) which are related to quantities in
Eq. \ref{eq-ad-fluxdens} through the temperature distribution given in
Eq. \ref{eq-ad-flux}.  Not all parameter can be constrained
observationally, since the viewing angle is degenerate with both
$\dot{M}$ and $M$.  Hence we are forced to make a simplifying
assumption about the inclination angle: since we are interested in
Type 1 AGN, we assume that the viewing angle is in the range 0--45
deg, i.e. the aperture of the obscuring torus. If observed at a
greater angle, the source would likely be classified as a Type 2
AGN. The average value of $\cos \theta$ (where $\theta$ is measured
from the disk normal) is:
  \begin{equation}
    \label{eq-ad-mean-cost}
    \langle \cos \theta \rangle = 
    \frac{1 + \cos \theta_{\rm max}}{2}
  \end{equation}
By setting $\theta_{\rm max}$=45 deg we obtain $\langle \cos \theta
\rangle$=0.854, corresponding to an average viewing angle of $\sim$30
deg.  With this assumption Eq. \ref{eq-ad-iso-corr} reads:
\begin{equation}
  \label{eq-ad-iso-corr2}
  L_{\nu}^{\rm iso} \sim 1.7\ L_\nu
\end{equation}

\subsection{General relativistic corrections}
\label{sec-GRcorrections}
The general relativistic model for the accretion disk is described in
\citet{1973-novikov-ssad} and \citet{1974-Page-ad-fluxdist}.  The
differences with respect to the AD model influencing the observational
appearance of the spectrum are:
\begin{enumerate}
\item the innermost stable circular orbit (isco) depends on the spin
  parameter $a = J c / G M^2$.  For a non--rotating black hole ($a=0$)
  $R_{\rm isco} = 6 R_{\rm g}$.  The maximum spin of an accreting
  black hole is $a=0.998$ \citep{1974-thorne-BHspin0.998}, with
  $R_{\rm isco} = 1.24 R_{\rm g}$.  The binding energy of a particle
  at $R_{\rm isco}$ in units of the particle rest--mass is
  \citep[e.g.][]{1975-Cunningham-RelativEffectsOnADSpectrum}:
  \begin{equation}
    \label{eq-eta-gr}
    \eta_{\rm gr} = 1 - \sqrt{1 - \frac{2}{3} \frac{R_{\rm g}}{R_{\rm isco}}}
  \end{equation}
  i.e. $\eta_{\rm gr}(a=0) \sim 0.06$ and $\eta_{\rm gr}(a=0.998) \sim
  0.32$.  This is expected to be the maximum possible value for
  the radiative efficiency (compare Eq. \ref{eq-ad-flux-norm}).

\item the different radial distribution of energy flux
  \citep{1974-Page-ad-fluxdist,1997-Zhang-reviewKerrEq} with respect
  to Eq. \ref{eq-ad-flux}.  The resulting spectrum is still the
  superposition of black body spectra;

\item the spectrum received by distant observers is influenced by
  gravitational redshift, Doppler boost and gravitational bending of
  light \citep{1975-Cunningham-RelativEffectsOnADSpectrum}.

\end{enumerate}
\citet{2005-li-kerrbb} have developed a package to synthesize the
observed spectrum for an optically thick, geometrically thin accretion
disk around a Kerr black hole, taking into account all these effects.
The code is available as the model {\tt KERRBB} within the X-ray data
reduction package XSPEC \citep{1996-Arnaud-xspec}.  In the following
we will compare the spectral profiles of both the ``classical'' and
``relativistic'' models, and show that the differences are negligible
for the purpose of our work.

We compute the accretion disk flux, as received by an observer at a
given distance, using both the the SS73 and {\tt KERRBB} models.  The
black hole mass, accretion rate and distance of the observer will be
kept fixed for all the considered models.

\begin{figure*}
  \includegraphics[width=\textwidth]{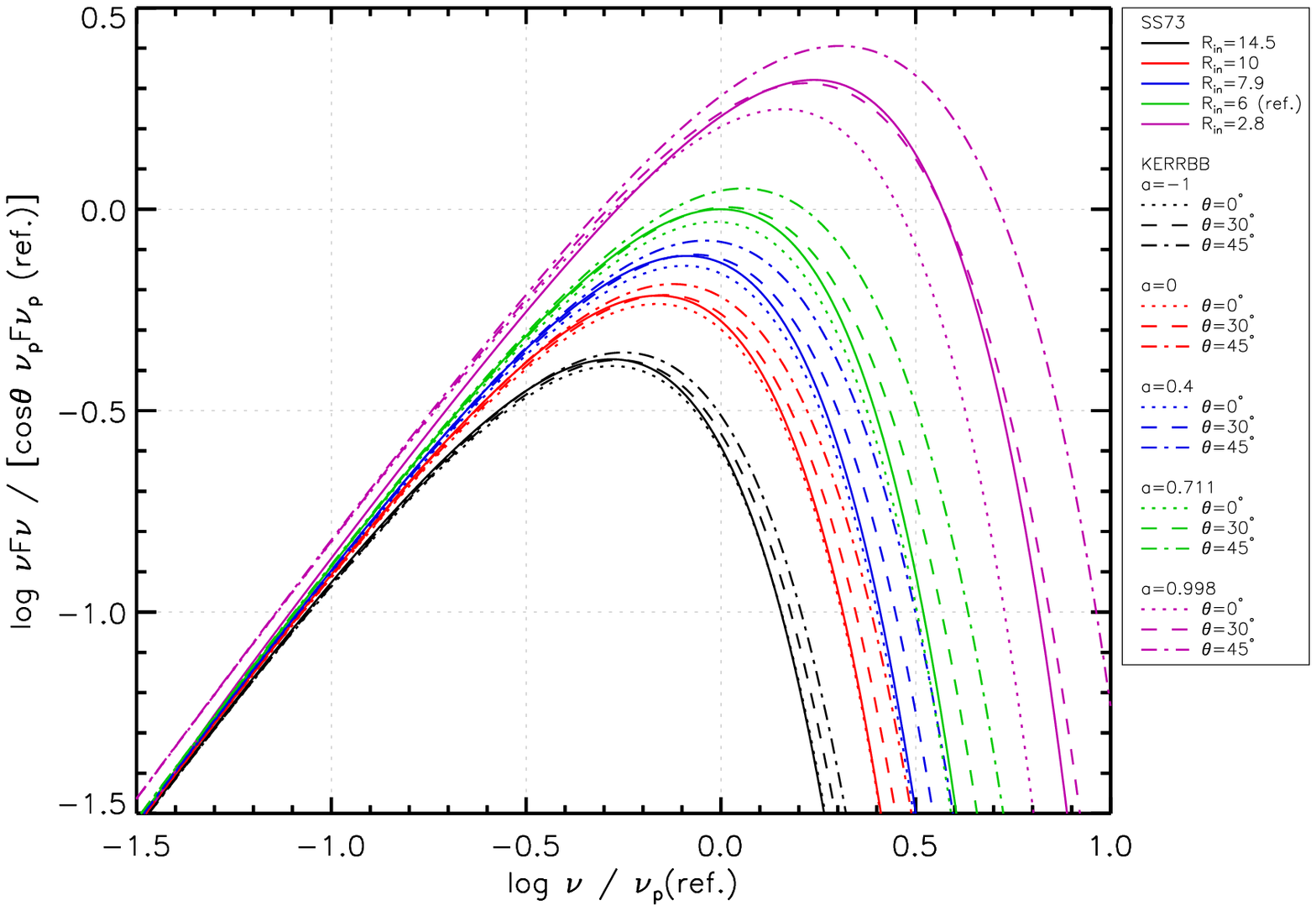}
  \caption{Comparison between the SS73 (\citealt{1973-ssad}, solid
    lines) and the {\tt KERRBB} (\citealt{2005-li-kerrbb}, dotted,
    dashed and dot--dashed lines) accretion disk spectra.  The SS73
    spectrum with $R_{\rm in} = 6\, R_{\rm g}$ is used as a reference
    spectrum: all other spectra are normalized by its peak luminosity.
    We consider three viewing angles $\theta=0^\circ, 30^\circ$ and
    $45^\circ$, and normalize all spectra by $\cos \theta$.  This
    completely removes the dependence of the SS73 on the viewing
    angle.  The {\tt KERRBB} show a residual dependence on the viewing
    angle (dotted, dashed and dot--dashed lines).  The colors identify
    a value for the inner radius of the SS73 model (respectively
    $R_{\rm in} / R_{\rm g} = 14.5, 10, 7.9, 6$ and 2.8) and for the
    black hole spin of the {\tt KERRBB} model (respectively $a=-1, 0,
    0.4, 0.7, 0.998$).  The inner radii for the SS73 models have been
    chosen in order to allow the SS73 spectra to resemble as close as
    possible the {\tt KERRBB} spectra, at given values of the spin.  The
    resulting empirical relation between $R_{\rm in}$ and the
    radiative efficiency of the {\tt KERRBB} model
    (Eq. \ref{eq-eta-gr}) is given in Eq. \ref{eq-ad-rinVsSpin}.}
  \label{fig-ad-cmp}
\end{figure*}

We consider five SS73 AD models, by varying the inner radius of the
disk $R_{\rm in}$.  The values of $R_{\rm in}$ has been chosen in
order to reduce the discrepancies between SS73 and {\tt KERRBB} models
(see below).  We take the model with $R_{\rm in} = 6 R_{\rm g}$ as a
reference spectrum, and normalize all other SS73 spectra by the
luminosity of its peak ($\nu_{\rm p} F_{\nu_{\rm p,ref}}$).  These
spectra are shown with solid lines in Fig. \ref{fig-ad-cmp}.
Note that the only dependence on the viewing angle $\theta$ for the
SS73 model is due to the projected area seen by the observer, i.e. to
a factor $\cos \theta$.  By plotting spectra normalized by $\cos
\theta$ we completely remove this dependence.

Then we consider five groups of {\tt KERRBB} models, by varying the
spin of the black hole: $a=-1, 0, 0.4, 0.7$ and 0.998.  These values
span the entire range of allowed values for the spin of an accreting
black hole \citep{1974-thorne-BHspin0.998}.  For each value of the
spin, we consider three different viewing angles: $\theta=0^\circ,
30^\circ$ and $45^\circ$.  All spectra are normalized by the
luminosity $\nu_{\rm p} F_{\nu_{\rm p,ref}}$ of the reference SS73
model discussed above, and by $\cos \theta$.  These spectra are shown
with dotted, dashed and dot--dashed lines in Fig. \ref{fig-ad-cmp}.
Note that the {\tt KERRBB} models show a residual dependence on the
viewing angle, due to light bending and Doppler boosting.

The values of $R_{\rm in}$ has been chosen in order to allow the SS73
spectra to resemble as close as possible the {\tt KERRBB} spectra, at
given values of spin.  The profile of the normalized spectra are
indeed very similar (spectra of the same color), the differences being
at most $\pm$0.1 dex for the highest value of spin ($a=0.998$).  The
(empirical) relation between the $R_{\rm in}$ in the SS73 model and
the radiative efficiency of the corresponding {\tt KERRBB} model is:
% -------------------------------------------------------
\begin{equation}
  \label{eq-ad-rinVsSpin}
  \frac{R_{\rm in}}{R_{\rm g}} = \frac{1}{2 \eta_{\rm gr}} + 1.25
\end{equation}
% -------------------------------------------------------
where $\eta_{\rm gr}$ is given by Eq. \ref{eq-eta-gr}.

From the observational point of view, the SS73 AD model with $R_{\rm
  in} = 6 R_{\rm g}$ (used throughout this work) mimics the {\tt
  KERRBB} model with spin $a \sim 0.7$ ($\eta_{\rm gr} \sim 0.1$), as
long as frequencies below the peak are concerned.  Therefore, the
results of our work are not influenced by having neglected the general
relativistic corrections in modeling the accretion disk spectrum.

\newpage
\section{Figures: spectral fitting}
\label{app-fig-spec}
This appendix is a collection of the figures related to the spectral
fitting procedure discussed in \S \ref{sec-gasf}.  On the left panels
we show the whole rest frame wavelength range, while on the right
panels we show a detailed view on the H$\beta$, \oiii{} and \mgii{}
regions.  The SDSS data and associated uncertainties are shown with
black squares and grey lines respectively.  Also shown are the fitting
models (red lines), as well as the individual components: the AGN
continuum (black), the galaxy template (cyan), the jet component (as
extrapolated from WISE photometry, purple), the iron templates
(orange), the broad (blue) and narrow (green) emission lines, and the
additional emission lines (grey).  In lower part of left panels we
show the residuals in units of data uncertainties.  The red lines show
the cumulative $\chi^2_{\rm red}$ (values on right axis).

\begin{figure*}%
  \centering
  \subfloat[][]{
    \includegraphics[width=.58\textwidth]{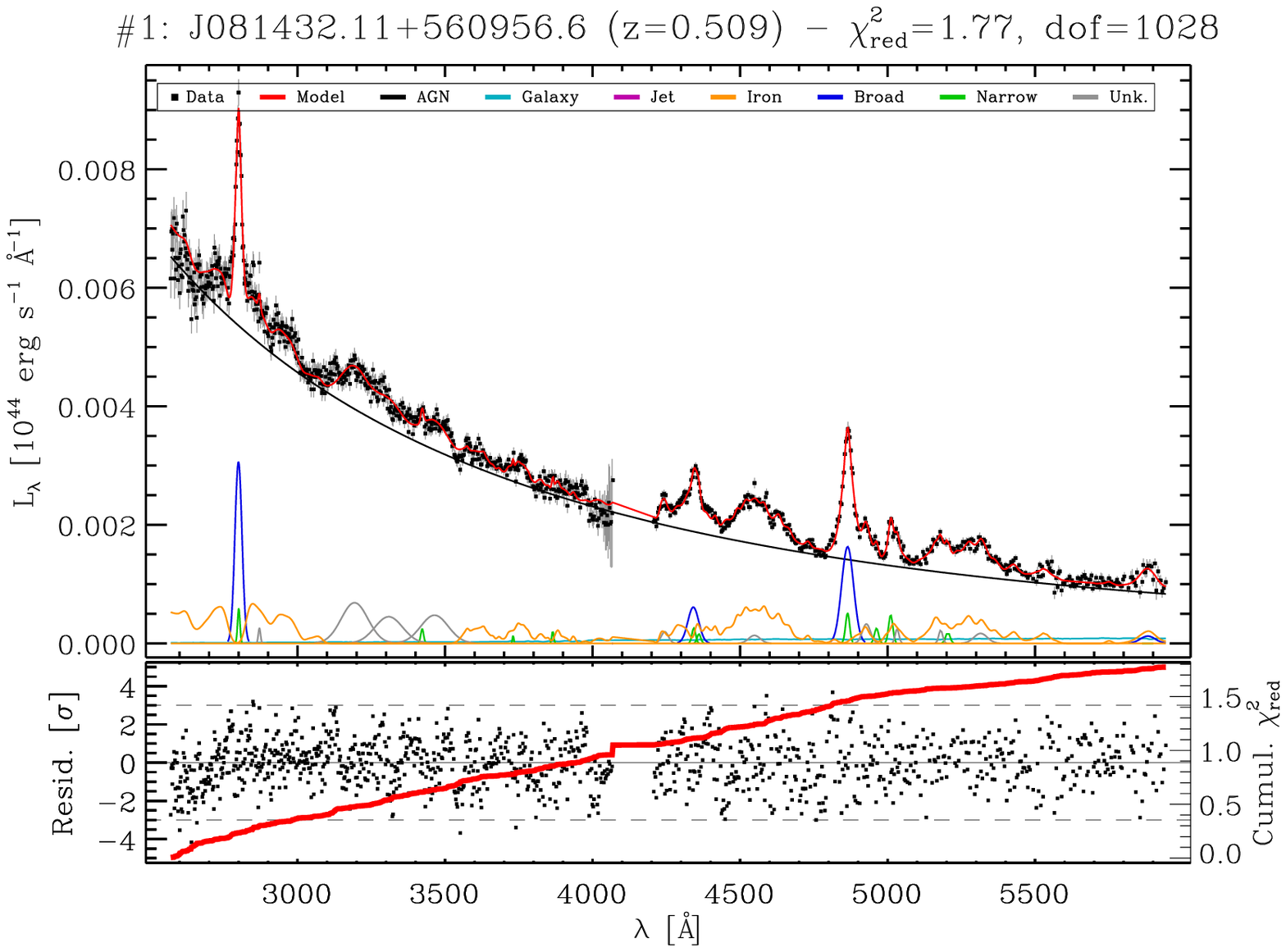}
    \includegraphics[width=.28\textwidth]{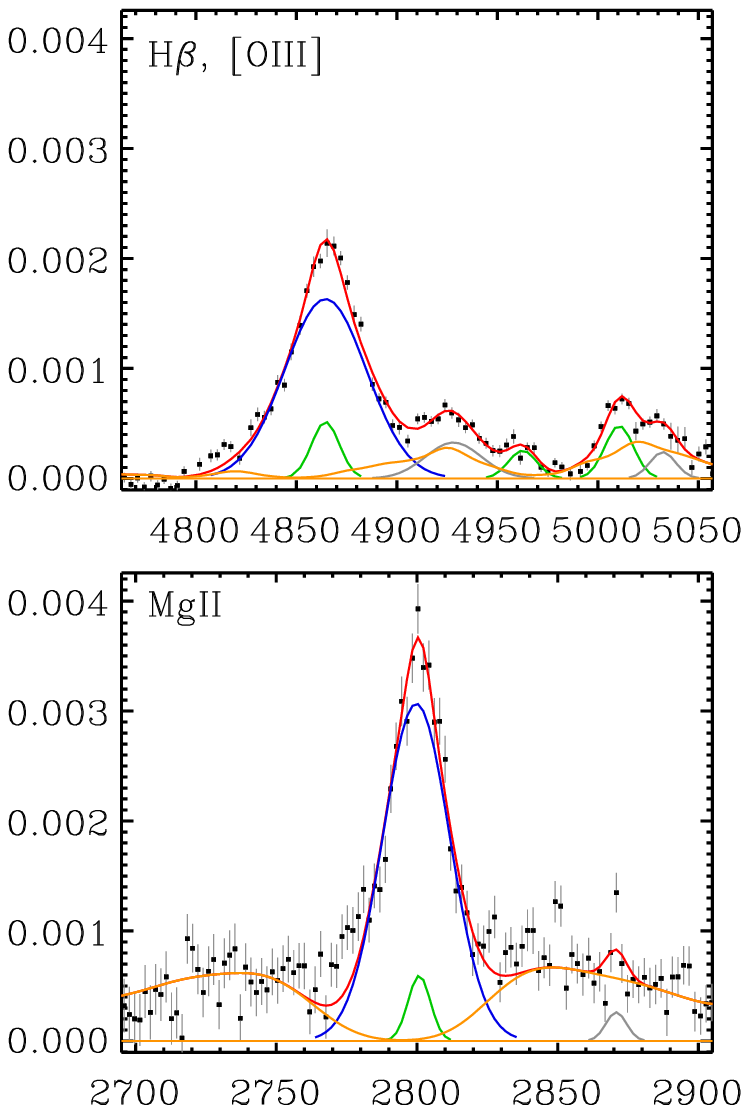}}\\
  \vspace{-0.7cm} \subfloat[][]{
    \includegraphics[width=.58\textwidth]{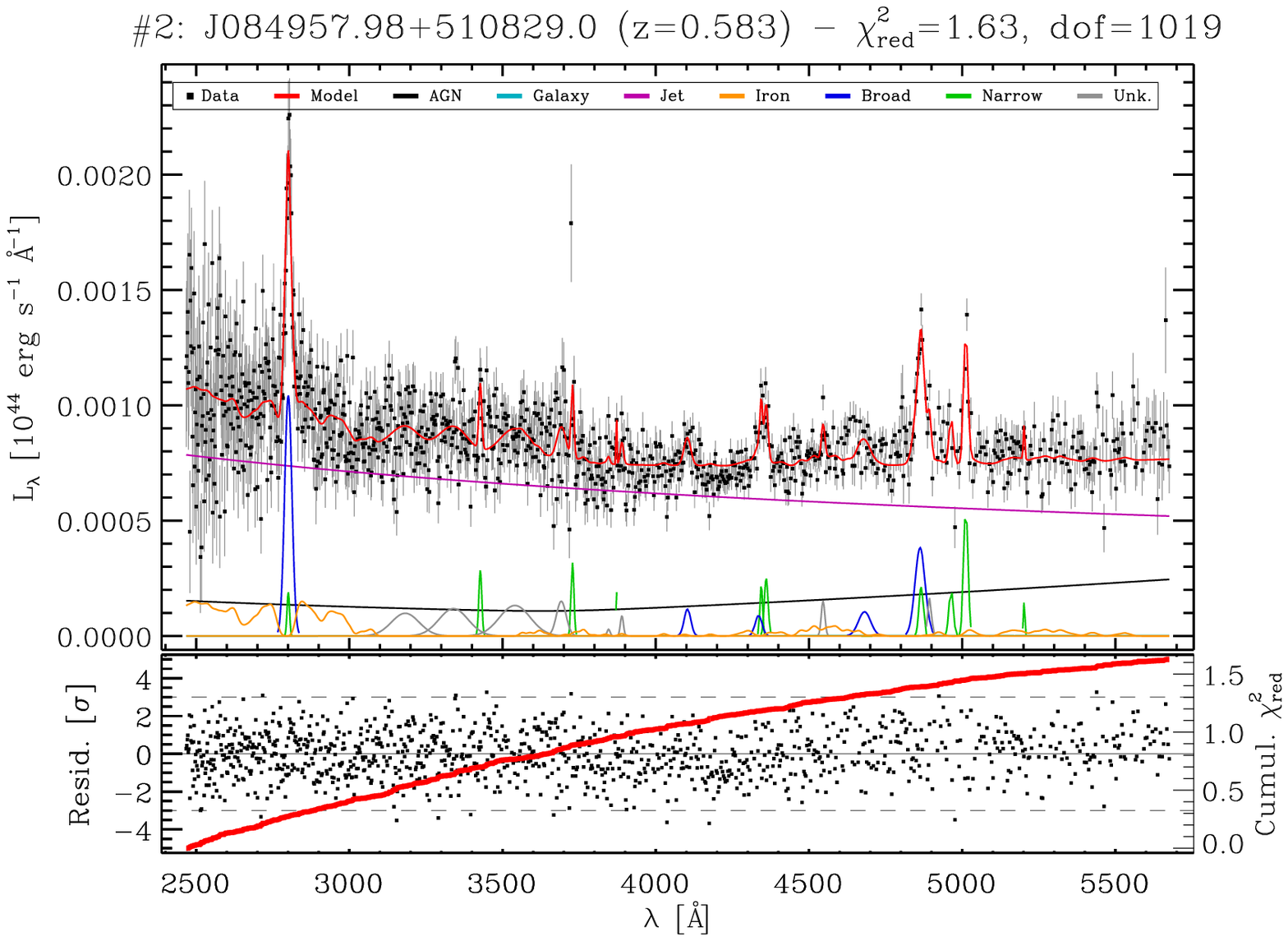}
    \includegraphics[width=.28\textwidth]{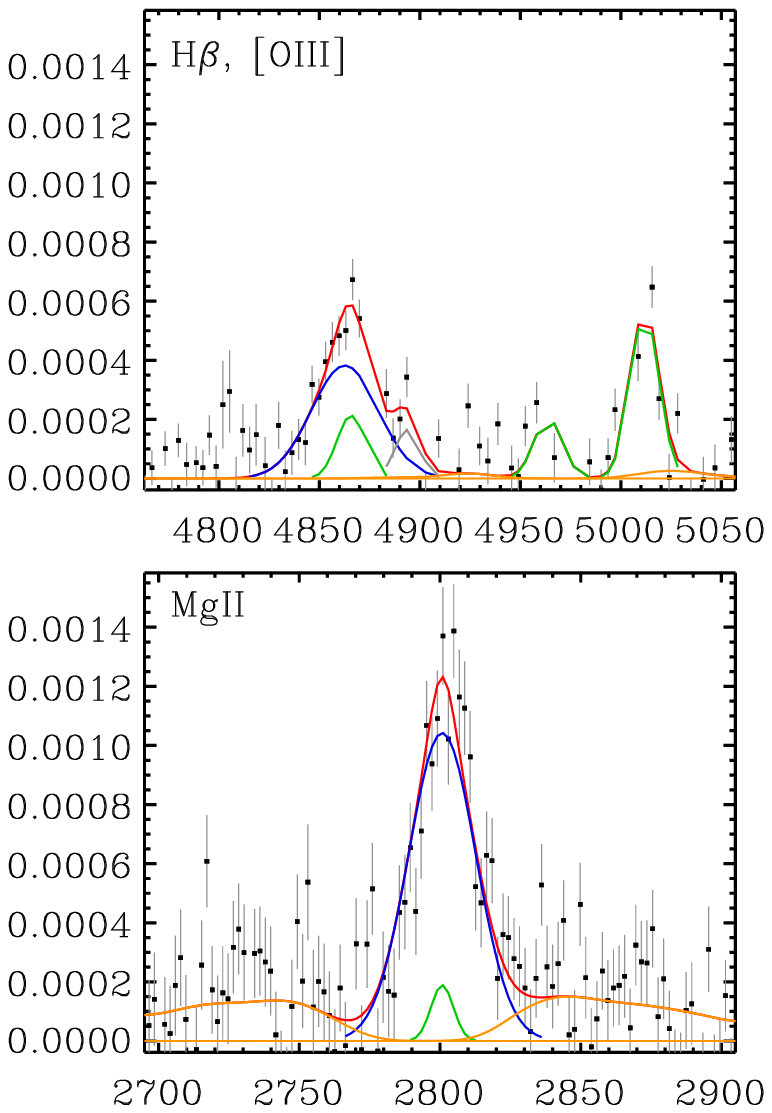}}\\
  \vspace{-0.7cm} \subfloat[][]{
    \includegraphics[width=.58\textwidth]{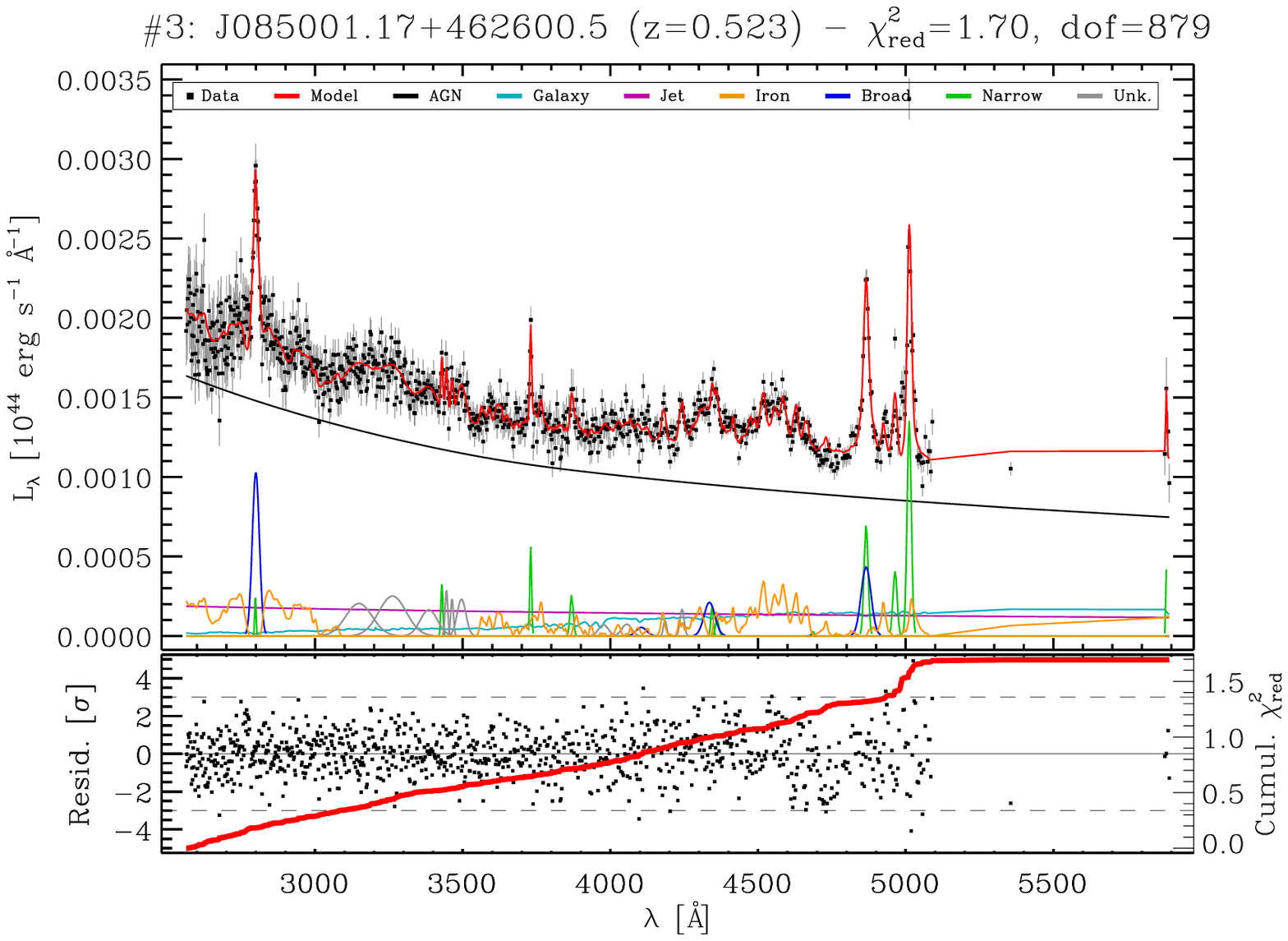}
    \includegraphics[width=.28\textwidth]{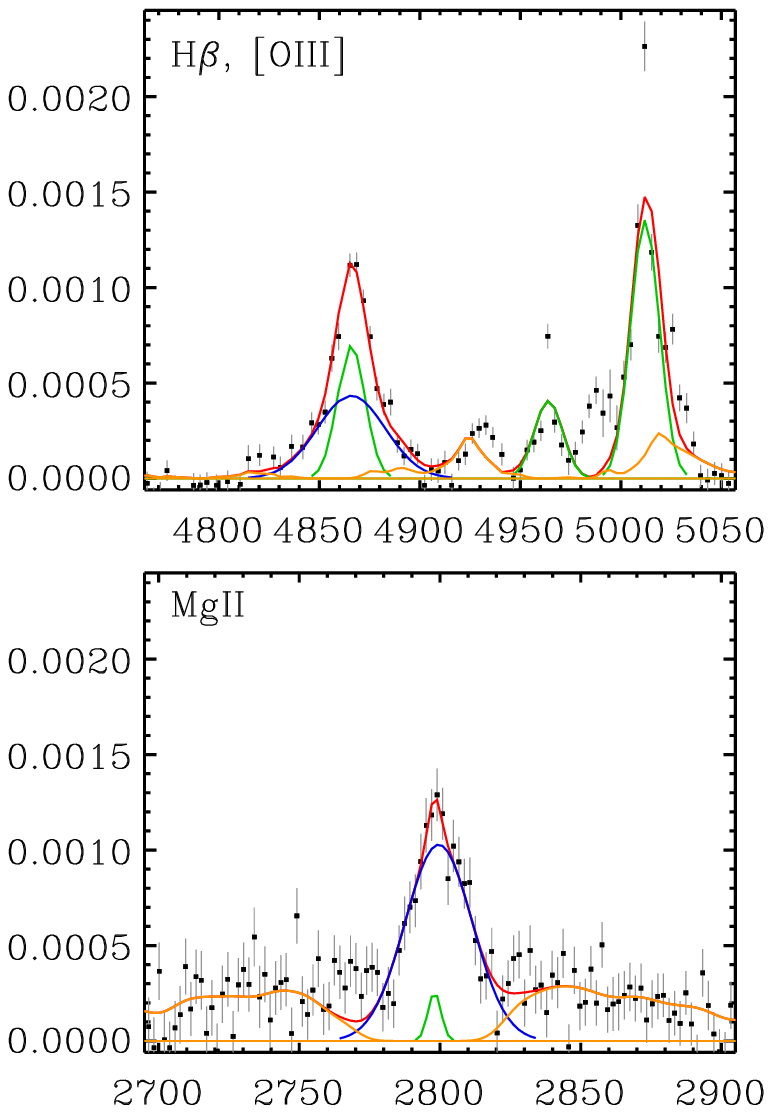}}\\
  \caption{Results of the spectral fitting procedure (\S
  \ref{sec-gasf}, App. \ref{app-fig-spec}).}%
  \label{fig-spec}%
\end{figure*}

\begin{figure*}%
  \ContinuedFloat
  \centering
  \subfloat[][]{
    \includegraphics[width=.58\textwidth]{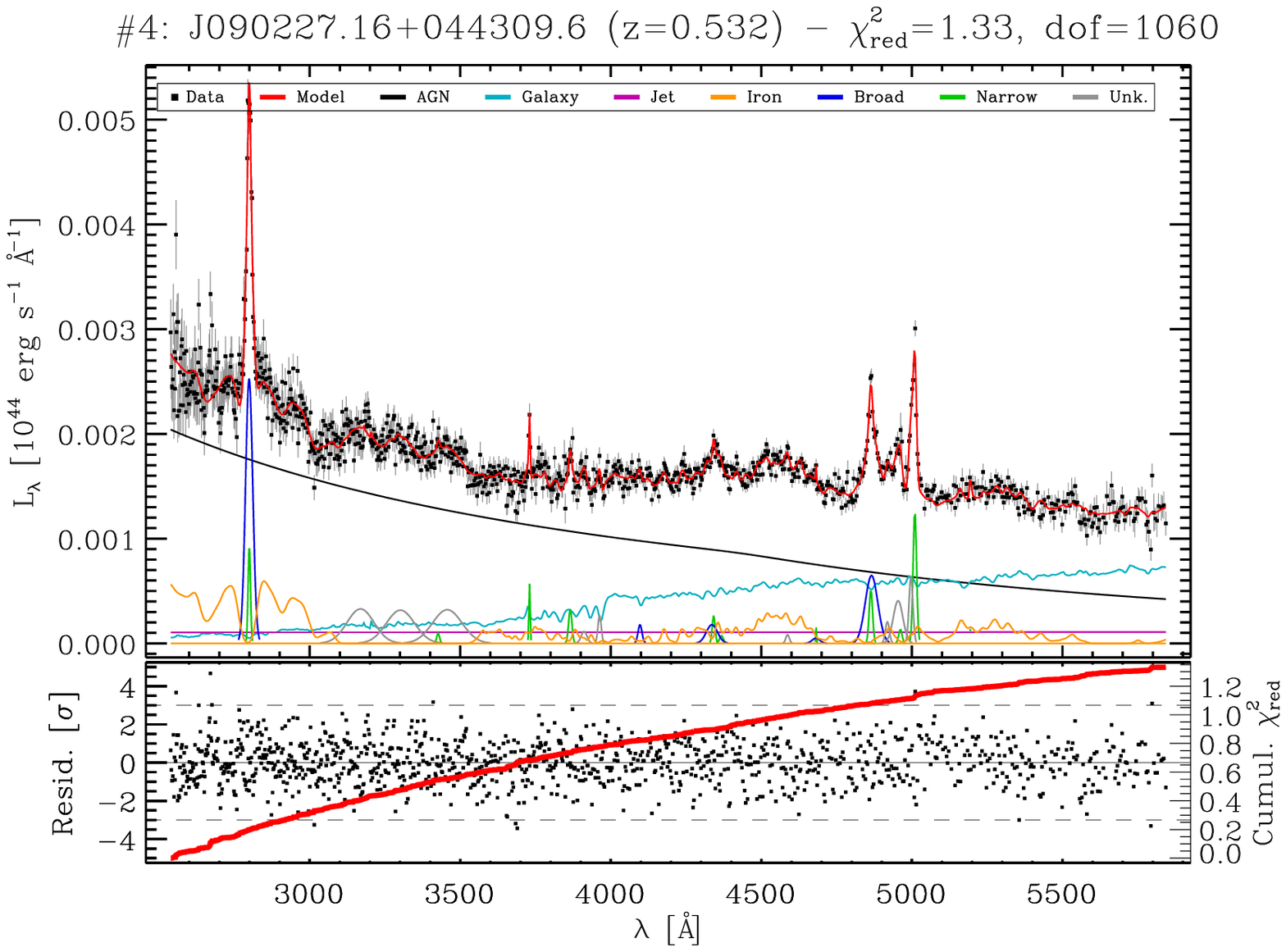}
    \includegraphics[width=.28\textwidth]{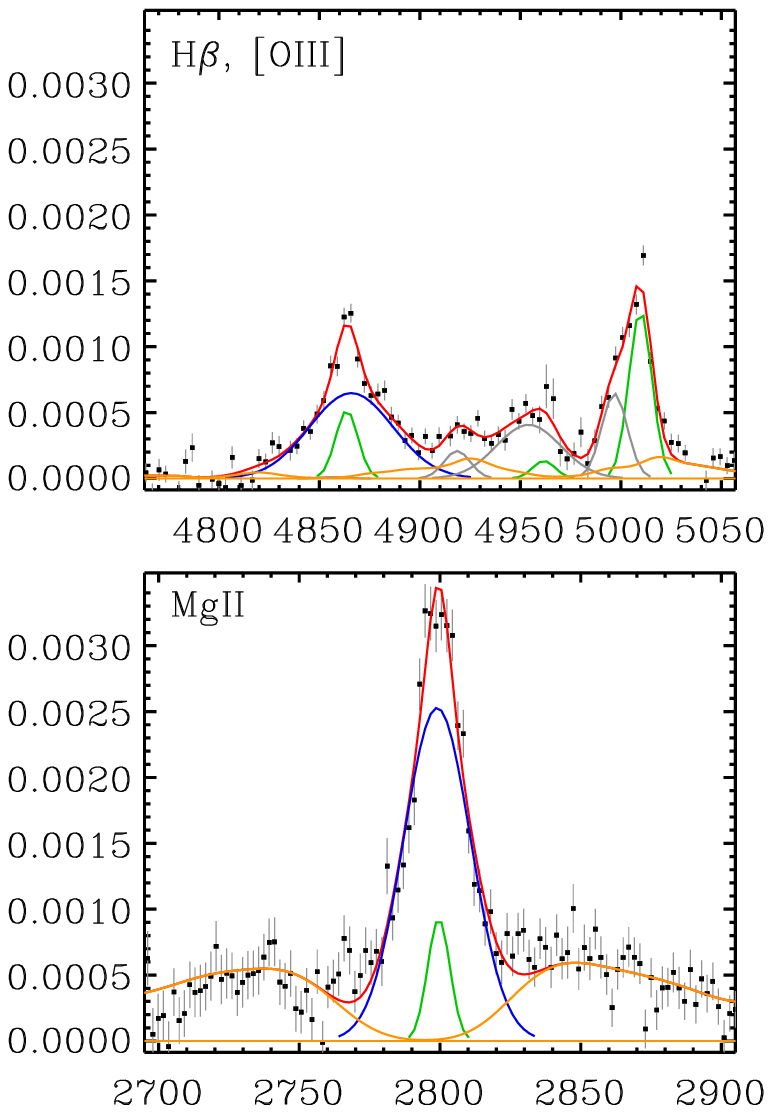}}\\
  \vspace{-0.7cm} \subfloat[][]{
    \includegraphics[width=.58\textwidth]{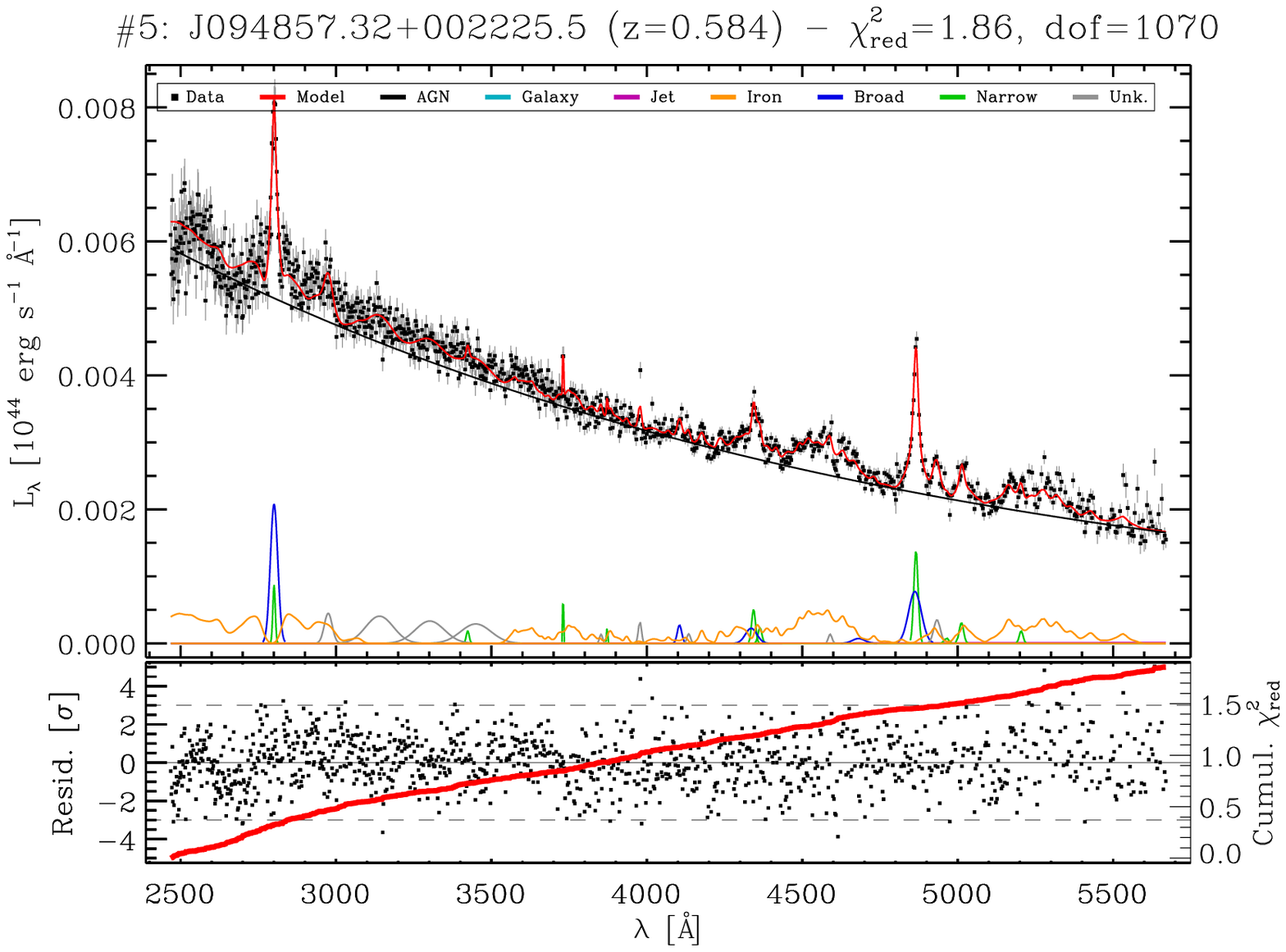}
    \includegraphics[width=.28\textwidth]{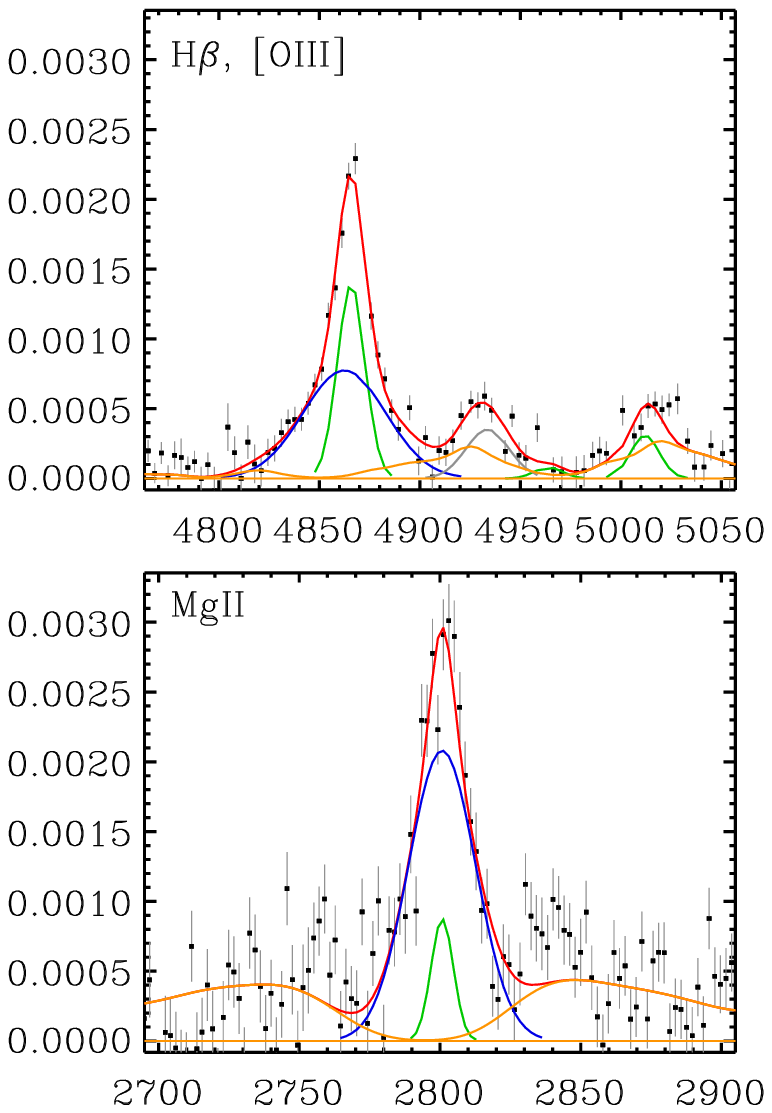}}\\
  \vspace{-0.7cm} \subfloat[][]{
    \includegraphics[width=.58\textwidth]{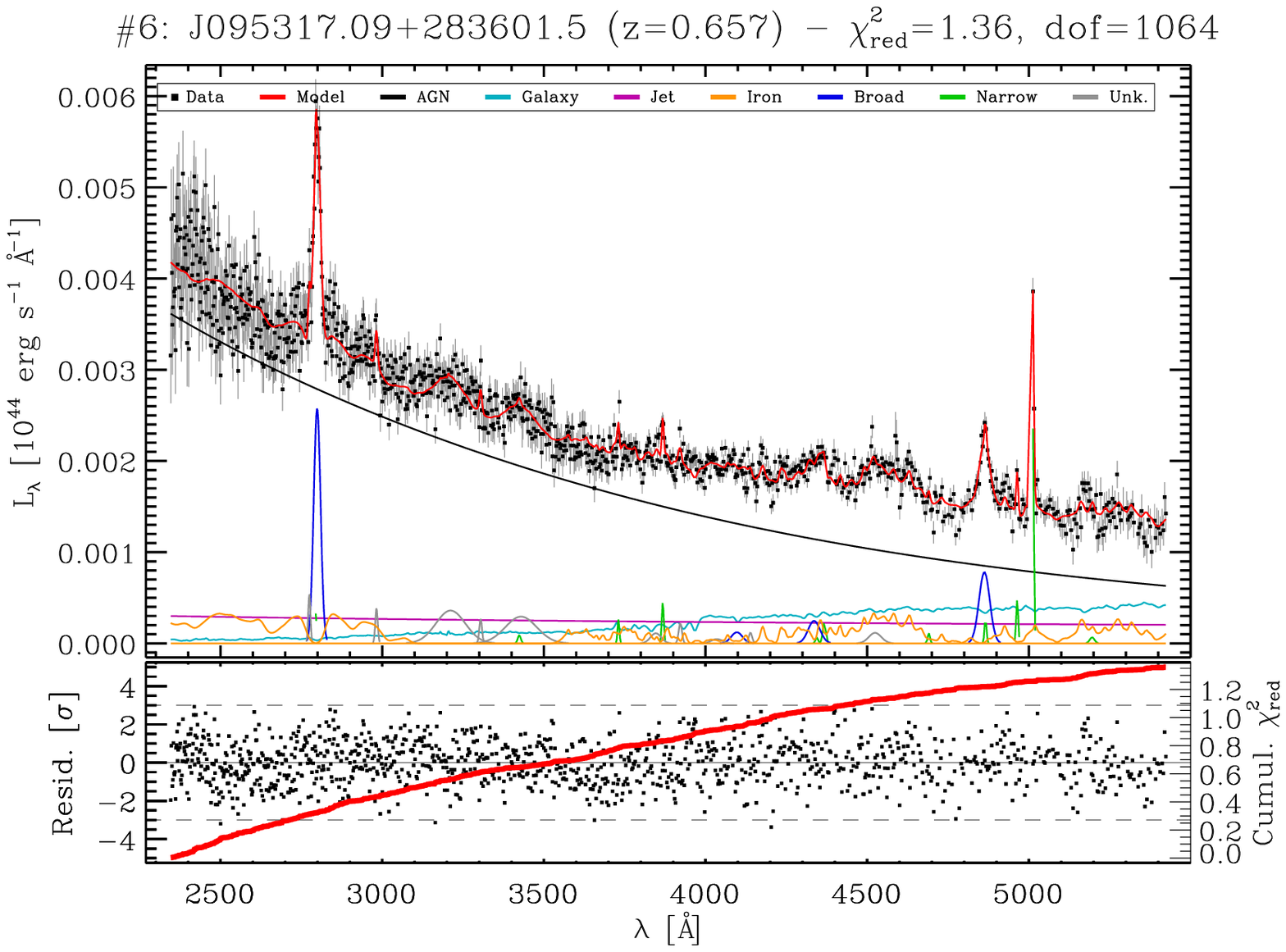}
    \includegraphics[width=.28\textwidth]{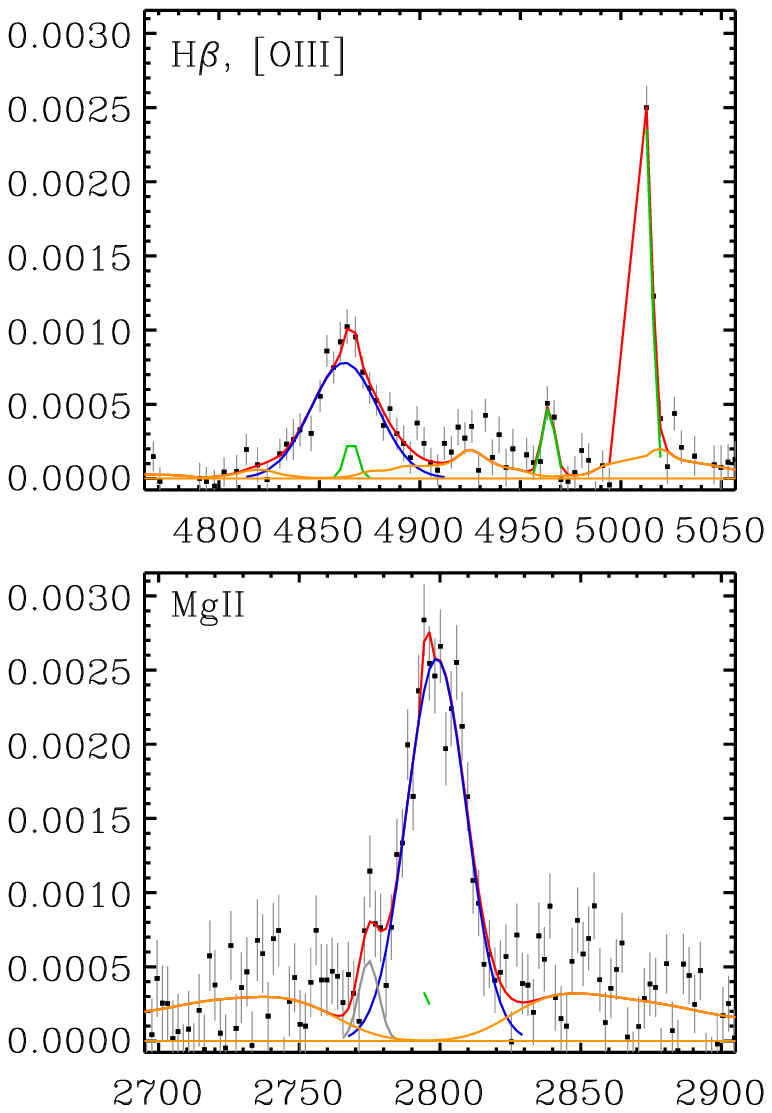}}\\
  \caption{(continued)}%
\end{figure*}

\begin{figure*}%
  \ContinuedFloat
  \centering
  \subfloat[][]{
    \includegraphics[width=.58\textwidth]{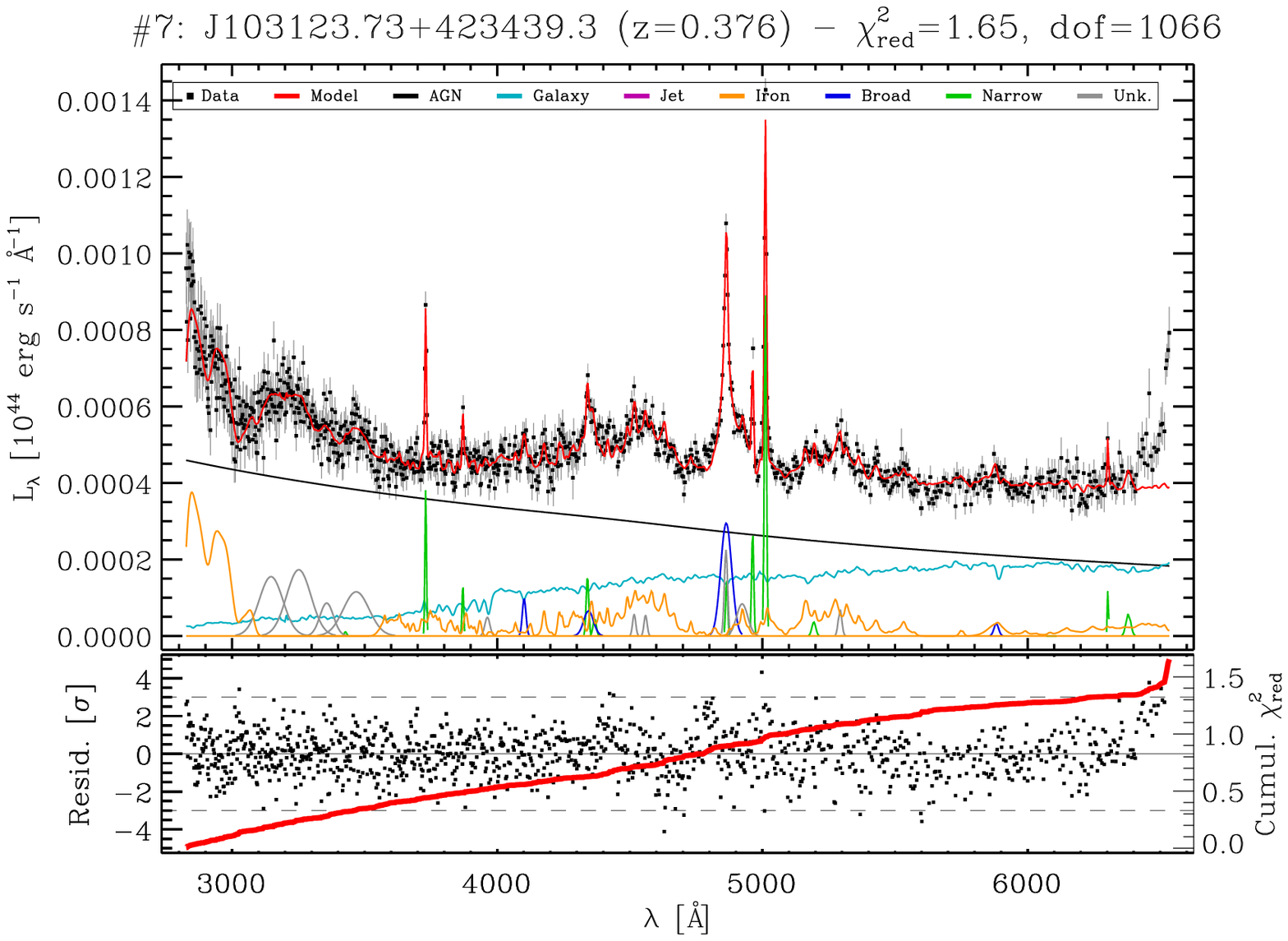}
    \includegraphics[width=.28\textwidth]{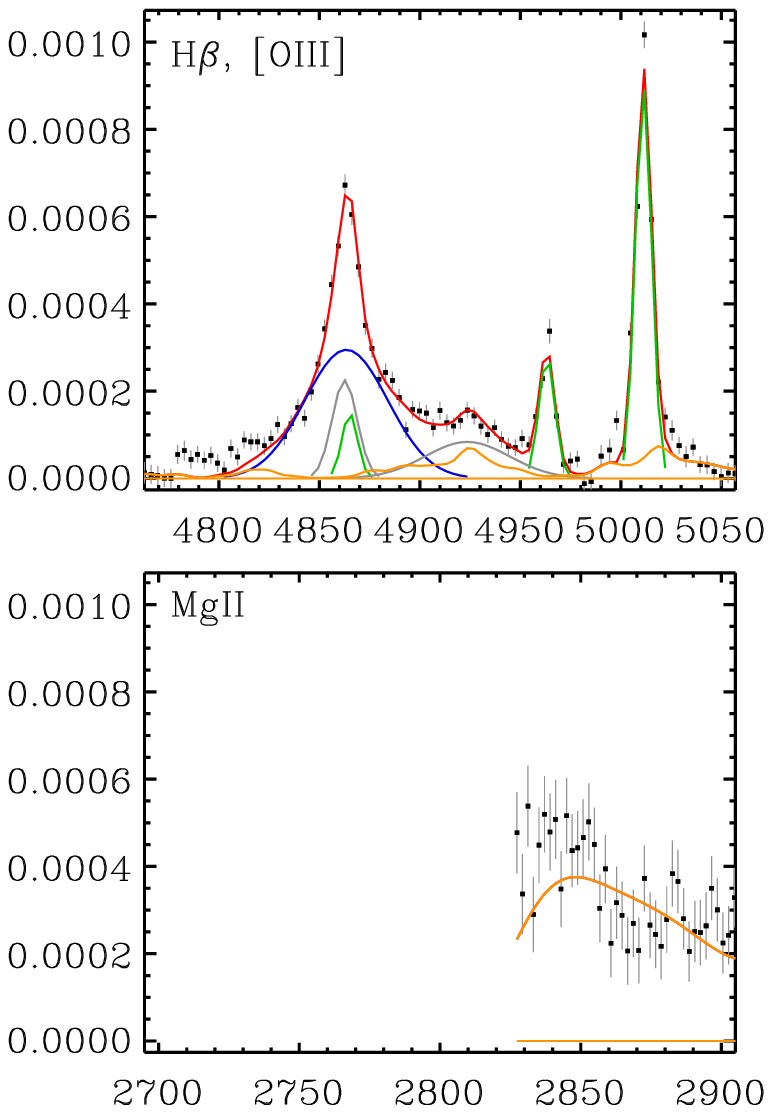}}\\
  \vspace{-0.7cm} \subfloat[][]{
    \includegraphics[width=.58\textwidth]{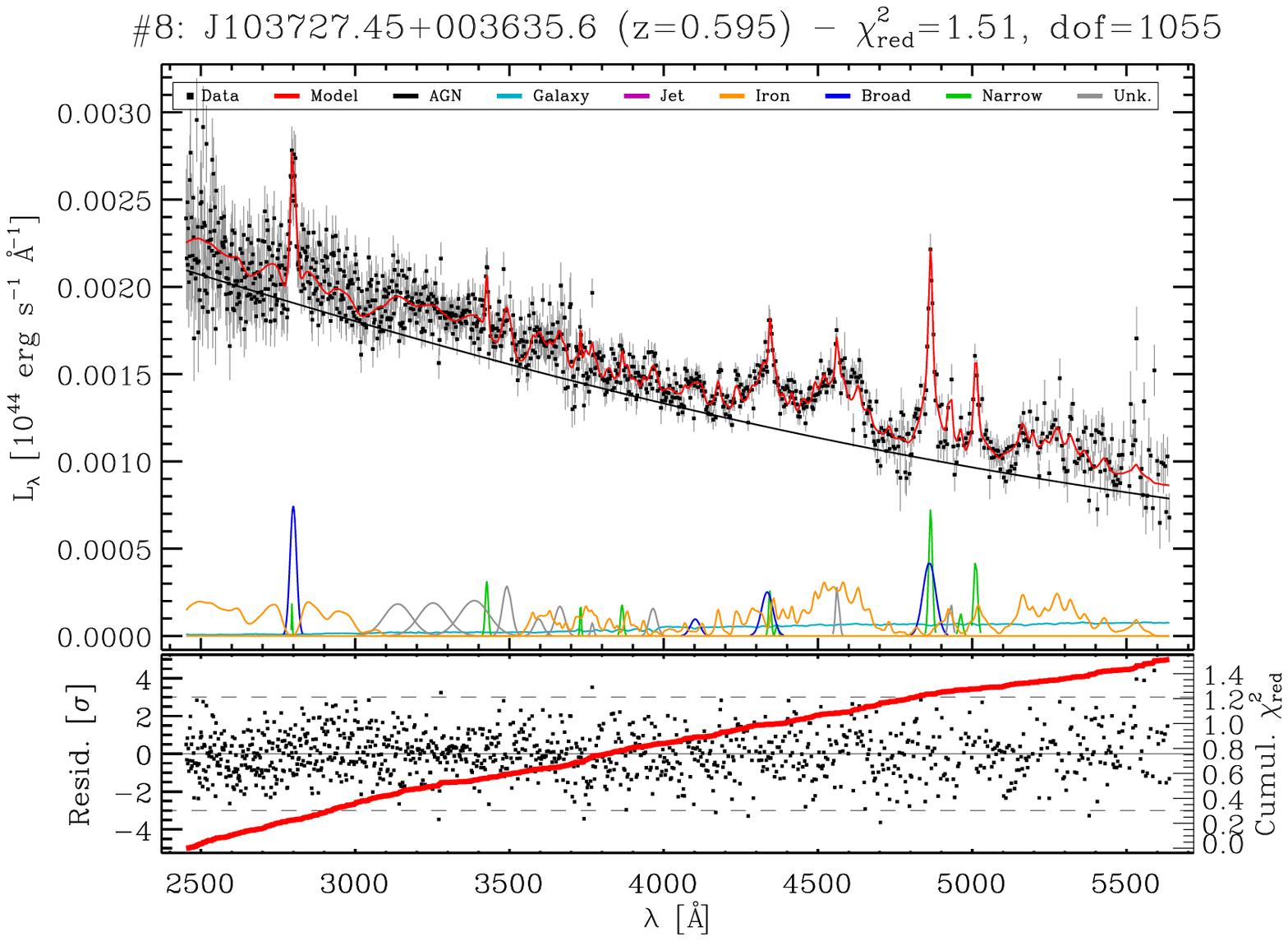}
    \includegraphics[width=.28\textwidth]{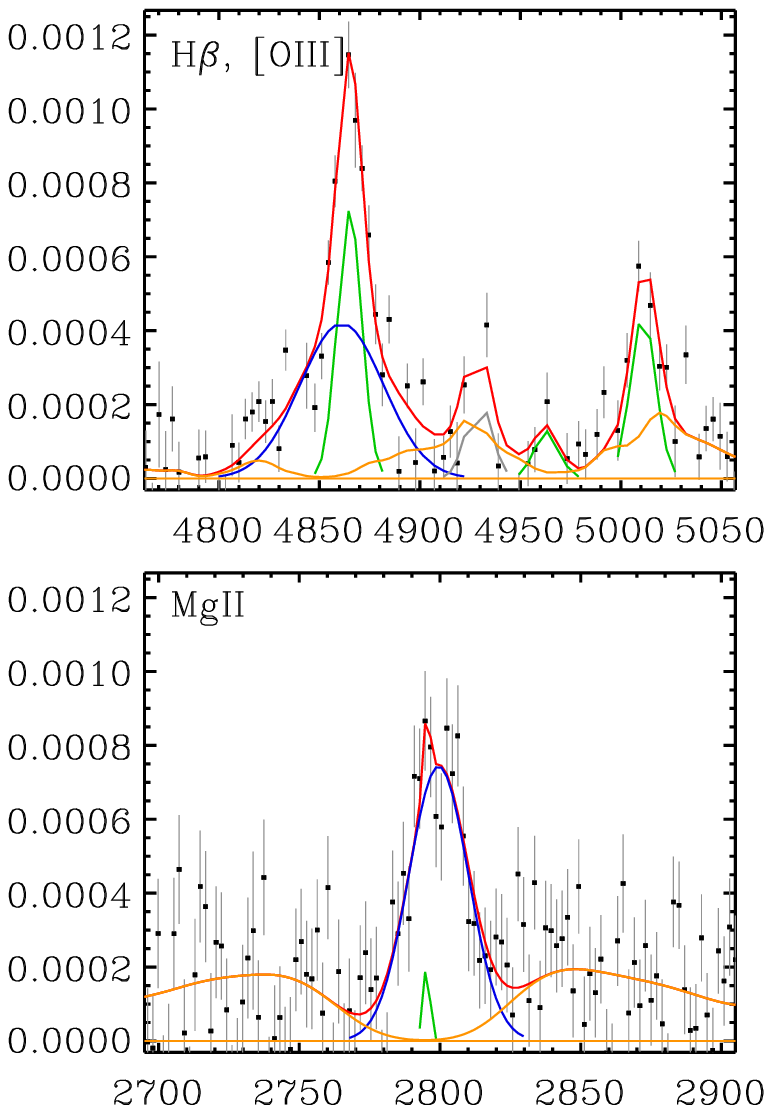}}\\
  \vspace{-0.7cm} \subfloat[][]{
    \includegraphics[width=.58\textwidth]{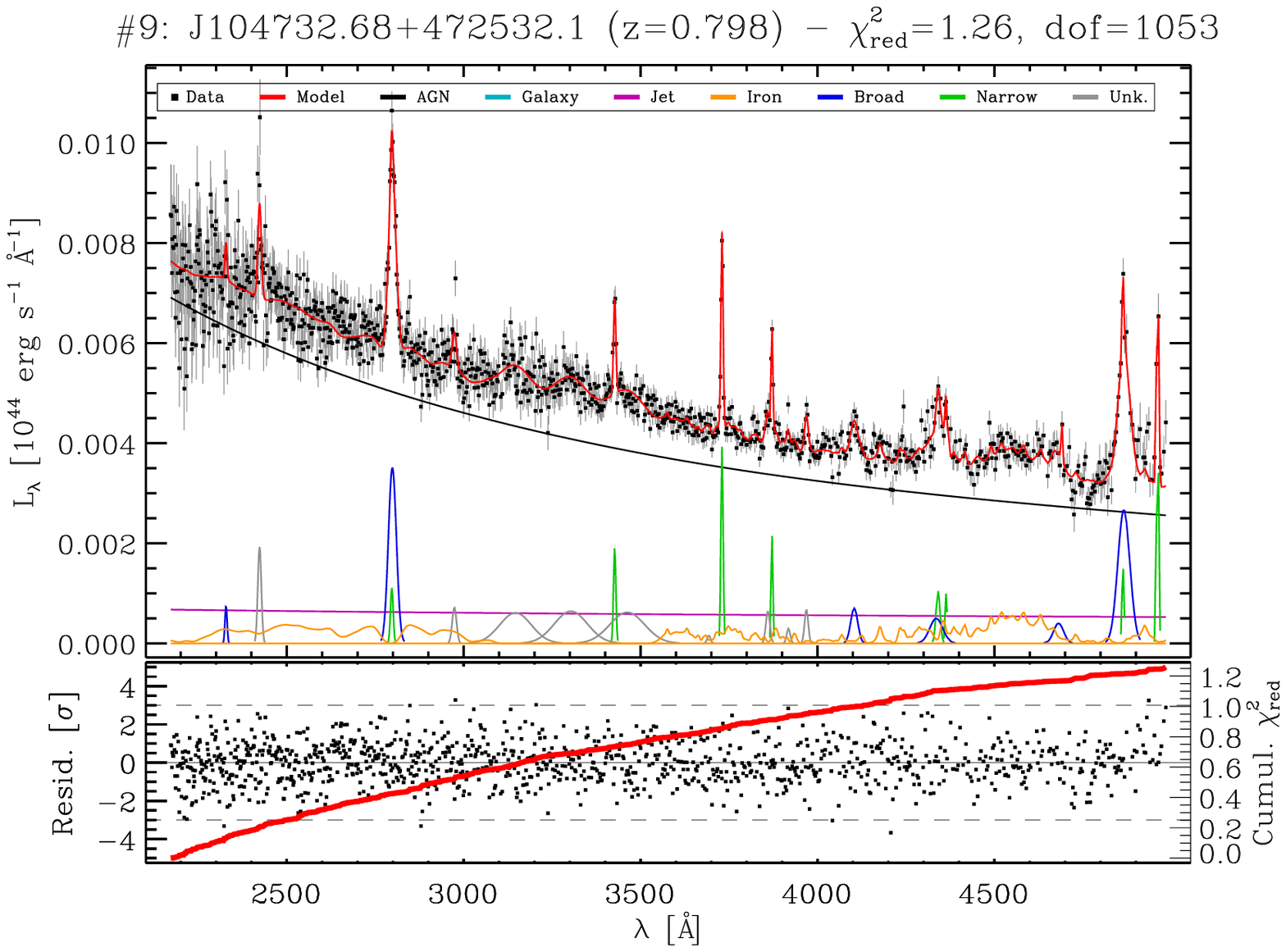}
    \includegraphics[width=.28\textwidth]{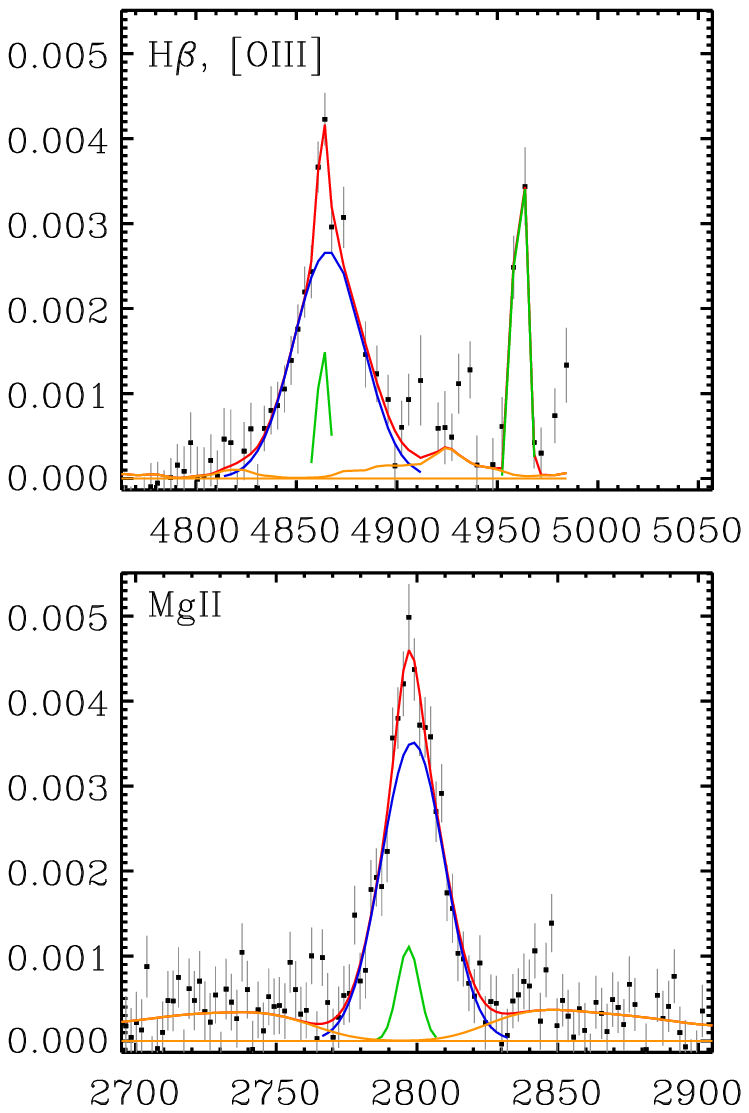}}\\
  \caption{(continued)}%
\end{figure*}

\begin{figure*}%
  \ContinuedFloat
  \centering
  \subfloat[][]{
    \includegraphics[width=.58\textwidth]{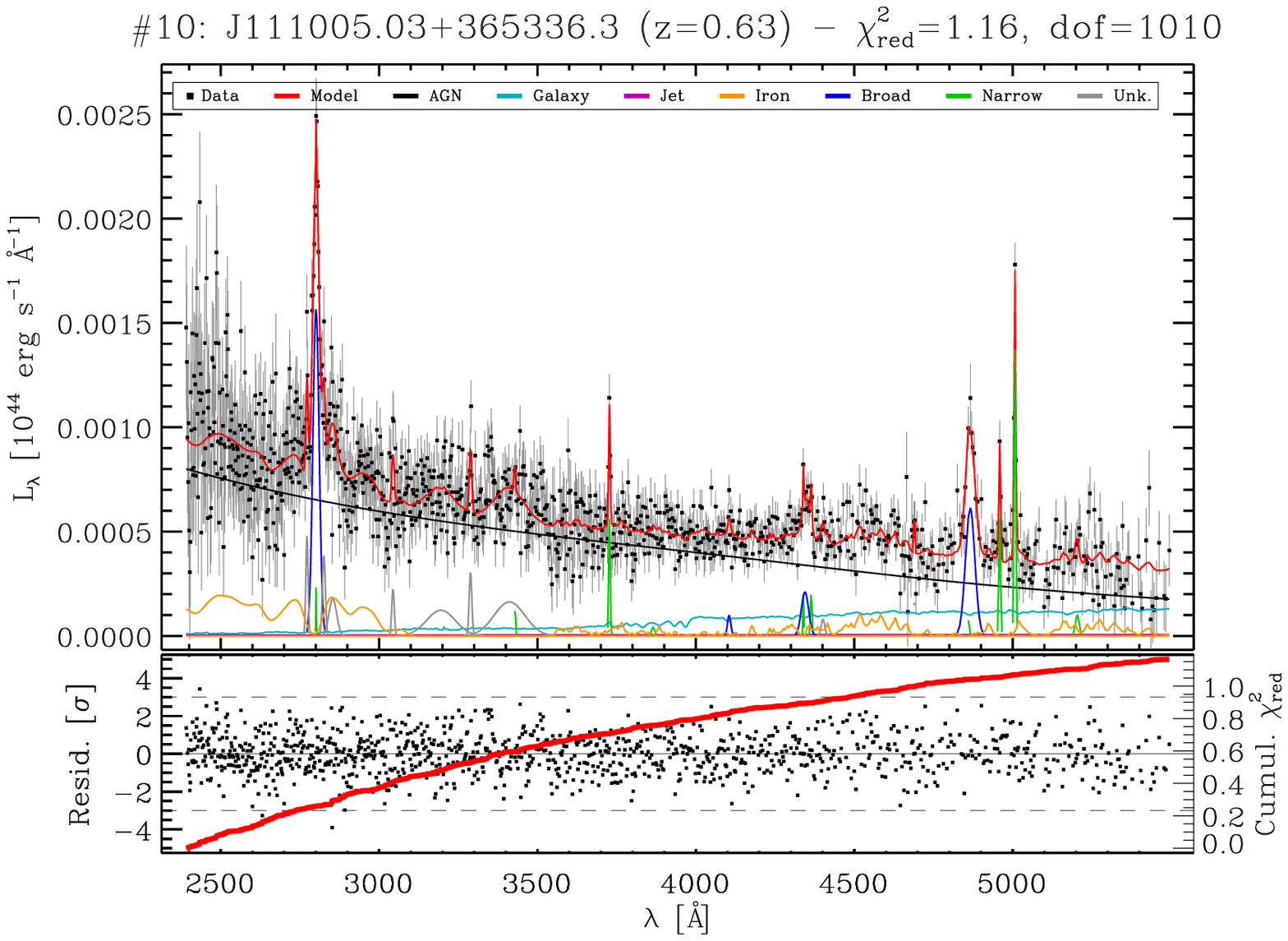}
    \includegraphics[width=.28\textwidth]{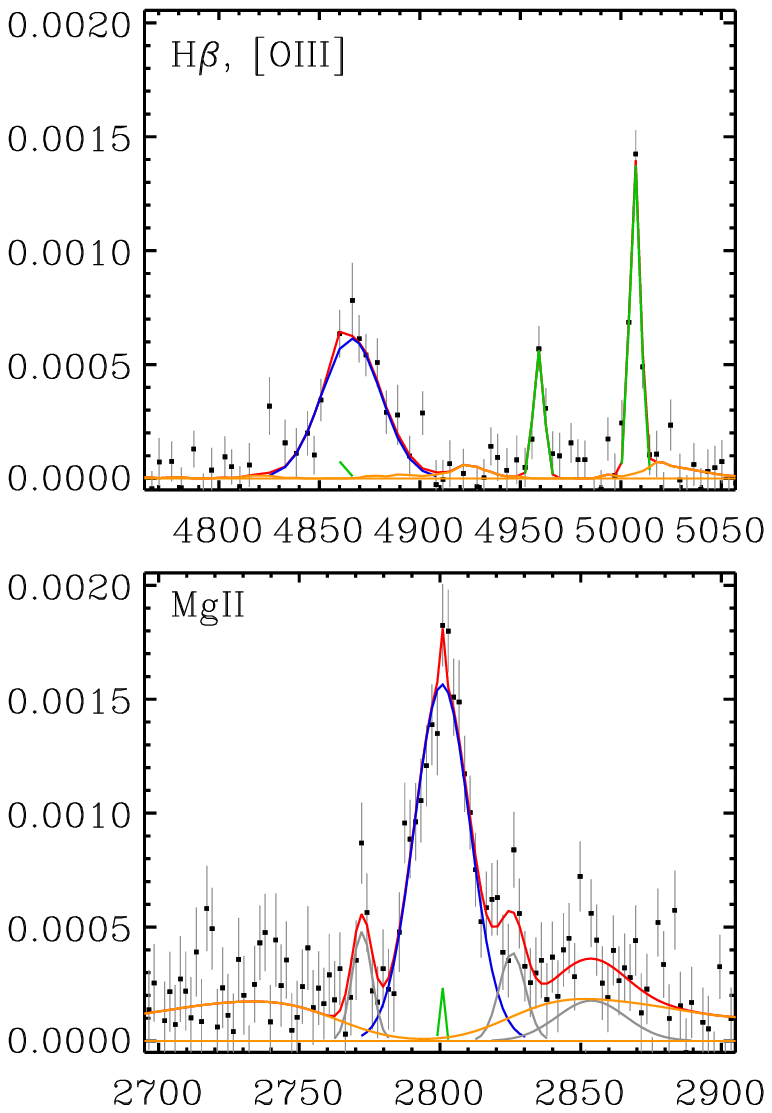}}\\
  \vspace{-0.7cm} \subfloat[][]{
    \includegraphics[width=.58\textwidth]{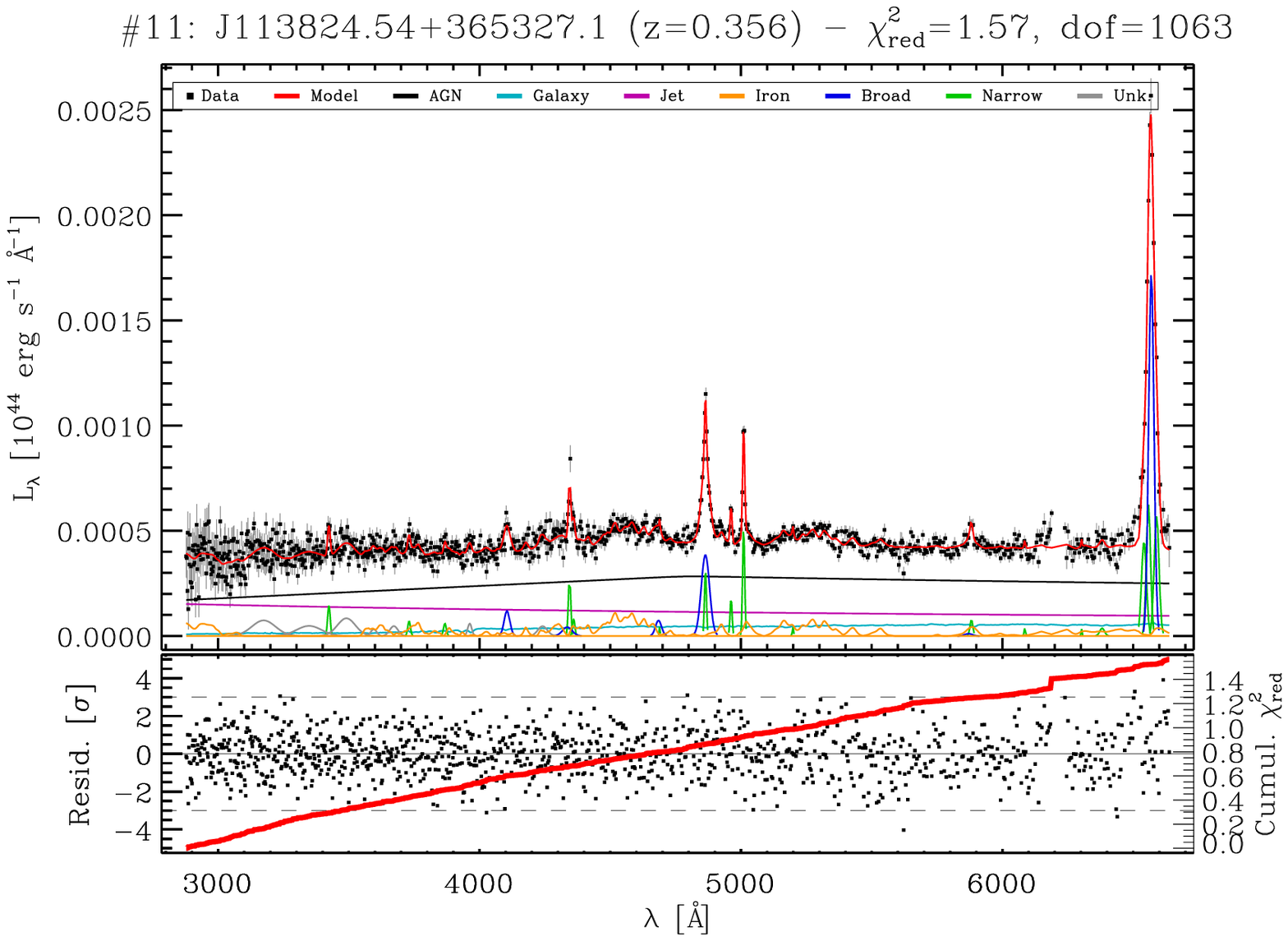}
    \includegraphics[width=.28\textwidth]{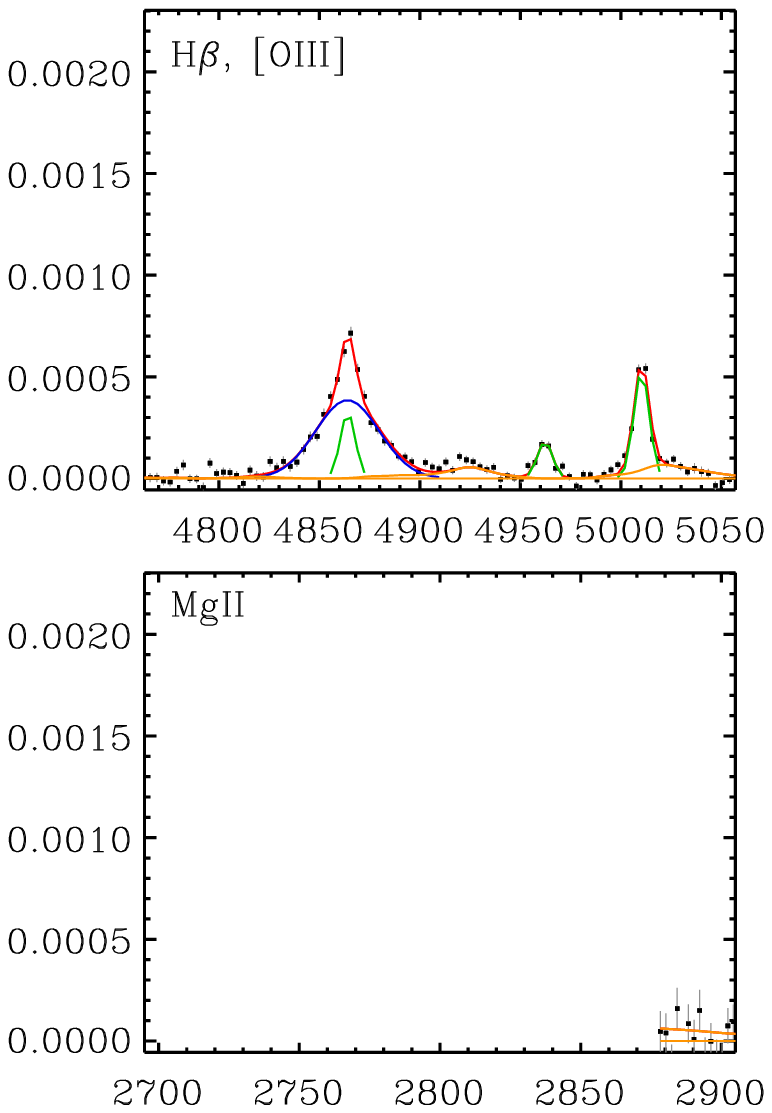}}\\
  \vspace{-0.7cm} \subfloat[][]{
    \includegraphics[width=.58\textwidth]{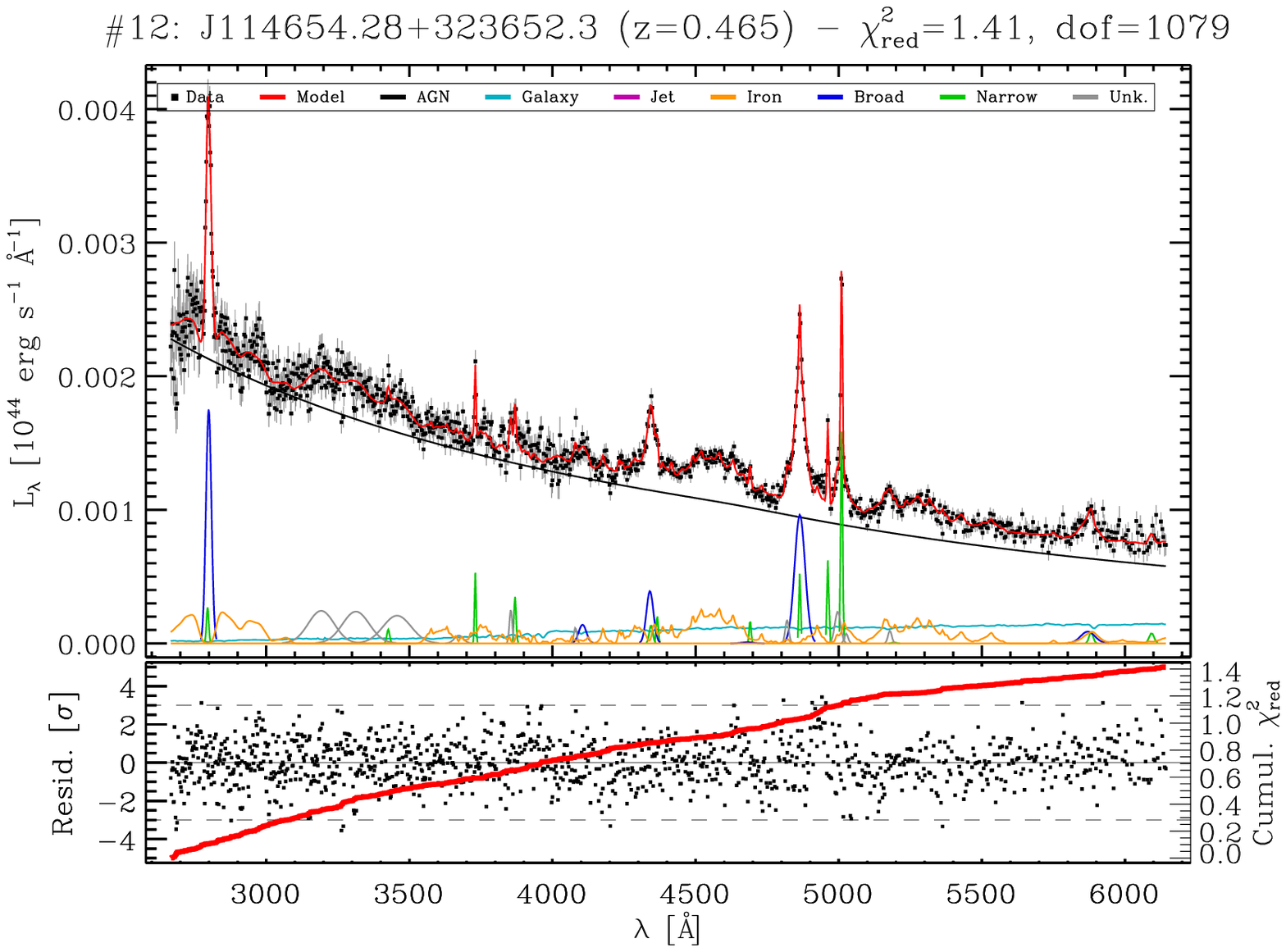}
    \includegraphics[width=.28\textwidth]{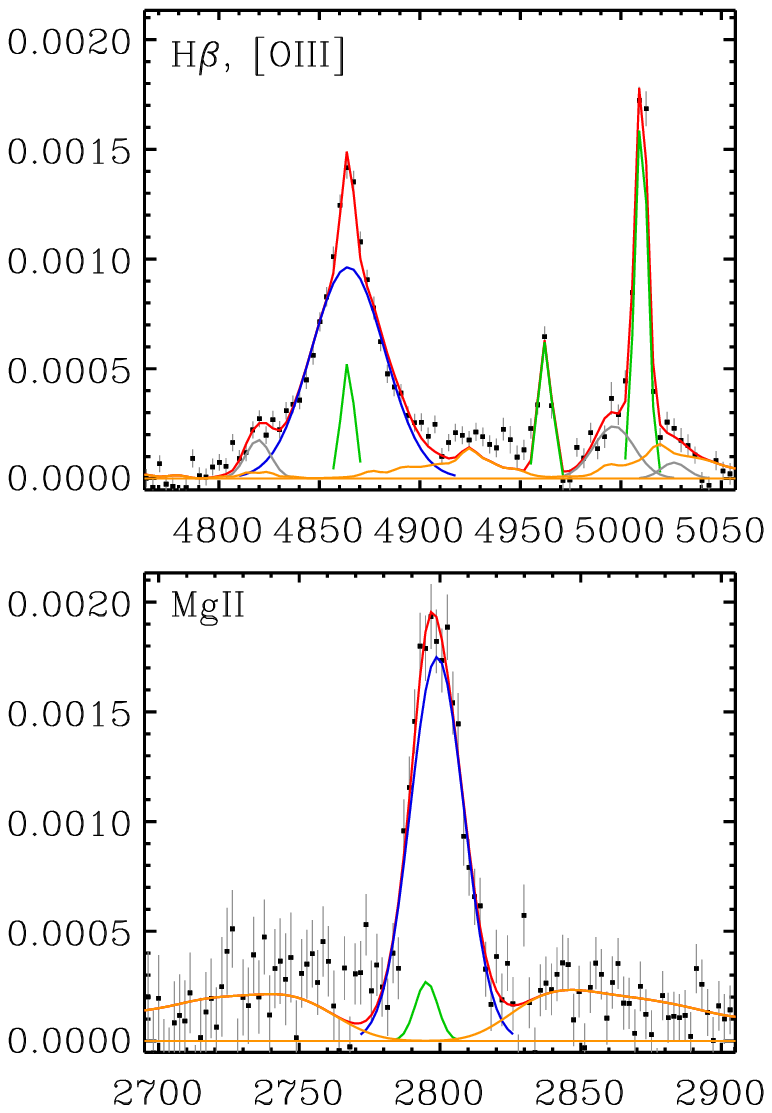}}\\
  \caption{(continued)}%
\end{figure*}

\begin{figure*}%
  \ContinuedFloat
  \centering
  \subfloat[][]{
    \includegraphics[width=.58\textwidth]{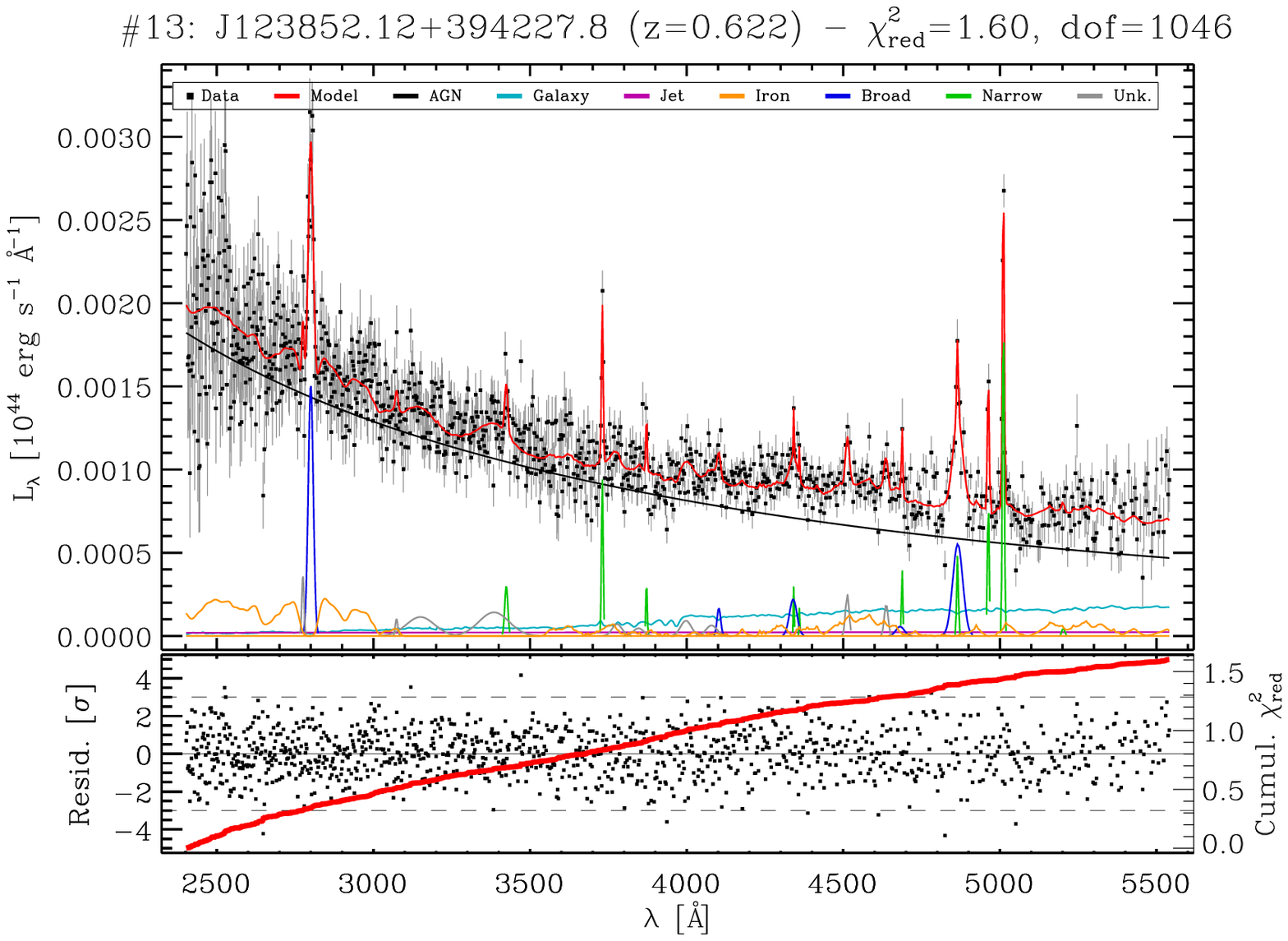}
    \includegraphics[width=.28\textwidth]{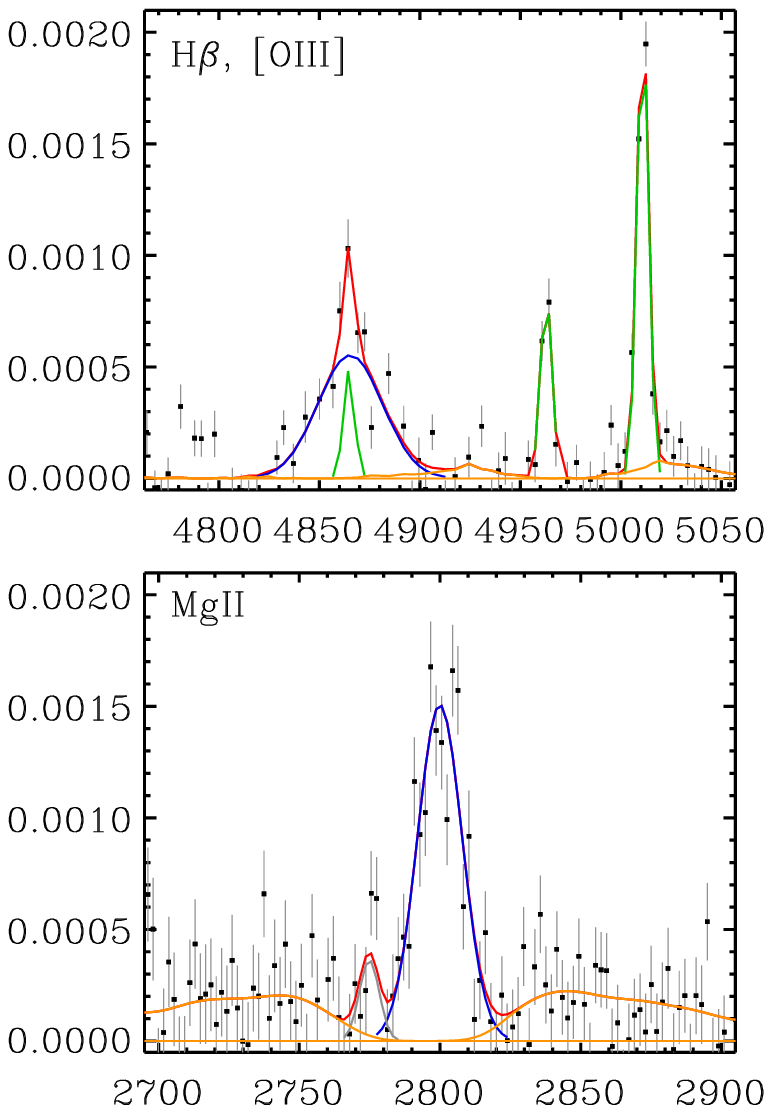}}\\
  \vspace{-0.7cm} \subfloat[][]{
    \includegraphics[width=.58\textwidth]{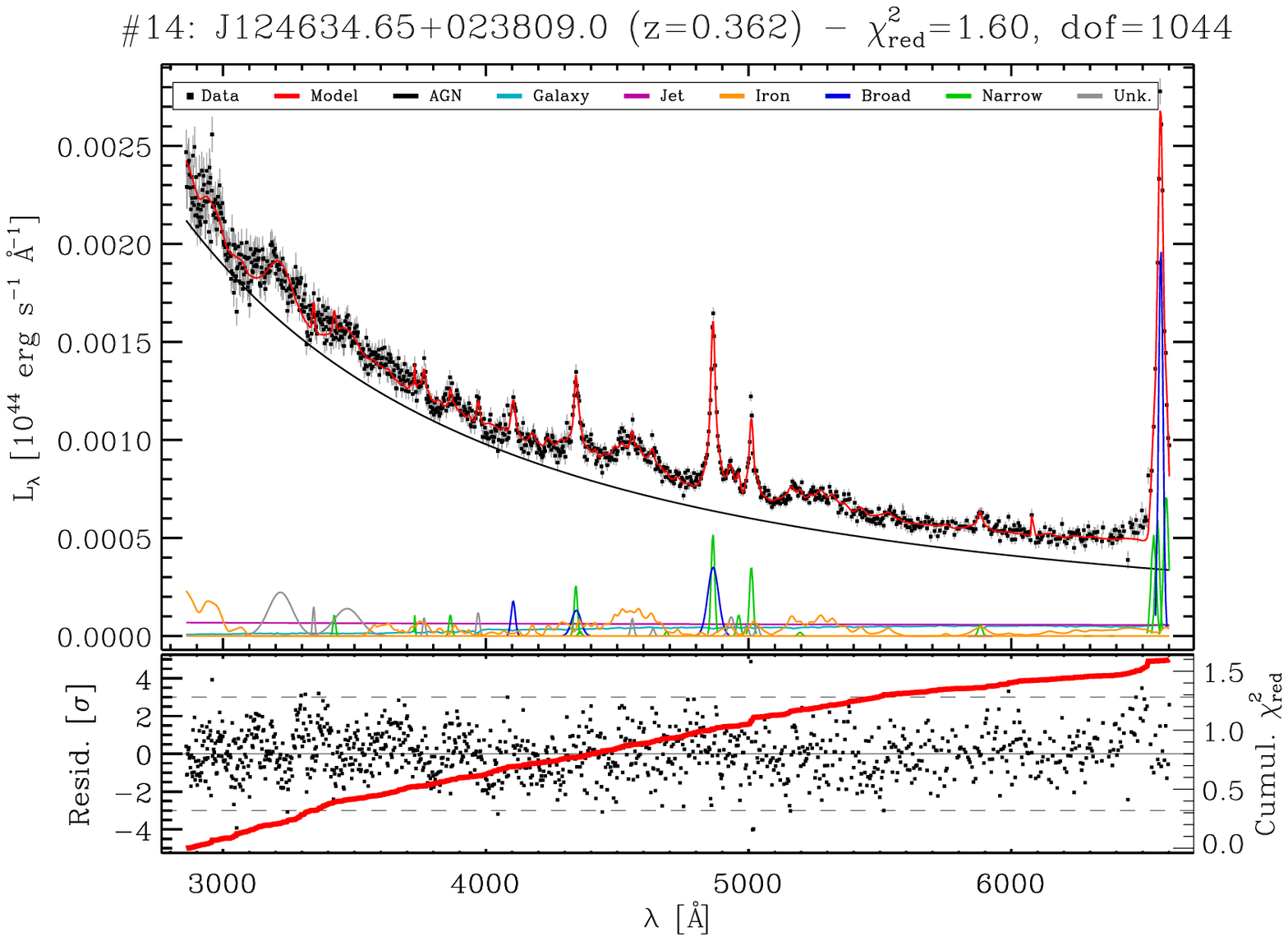}
    \includegraphics[width=.28\textwidth]{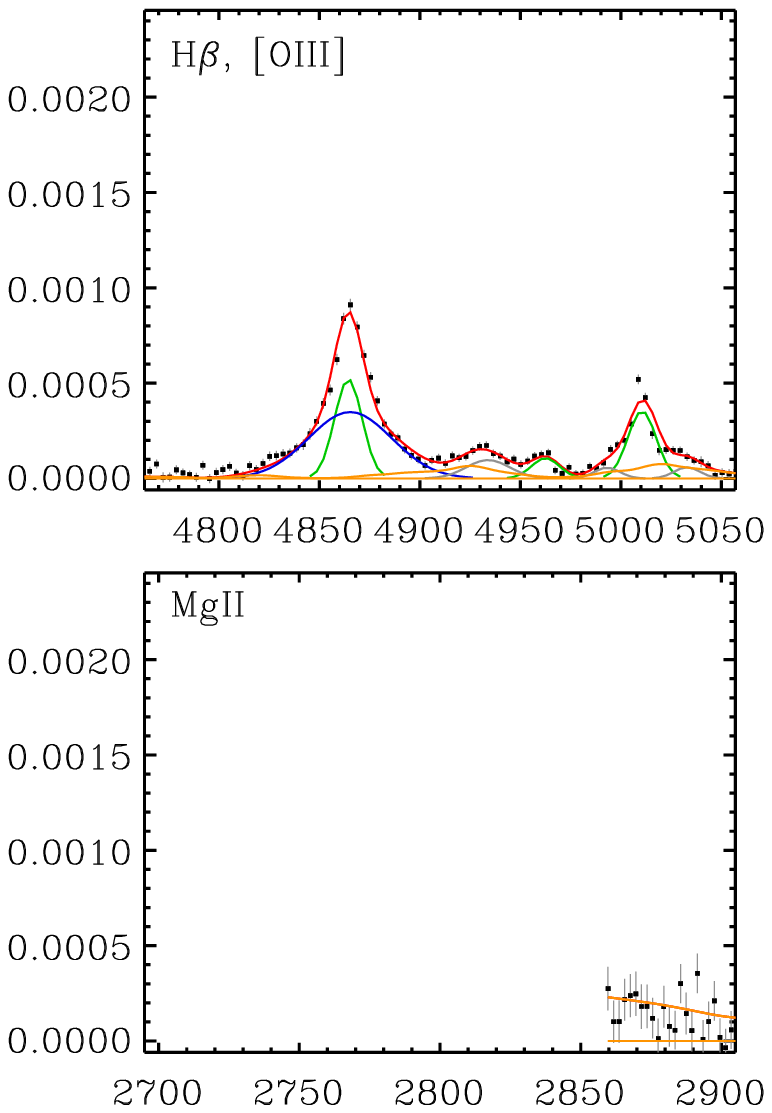}}\\
  \vspace{-0.7cm} \subfloat[][]{
    \includegraphics[width=.58\textwidth]{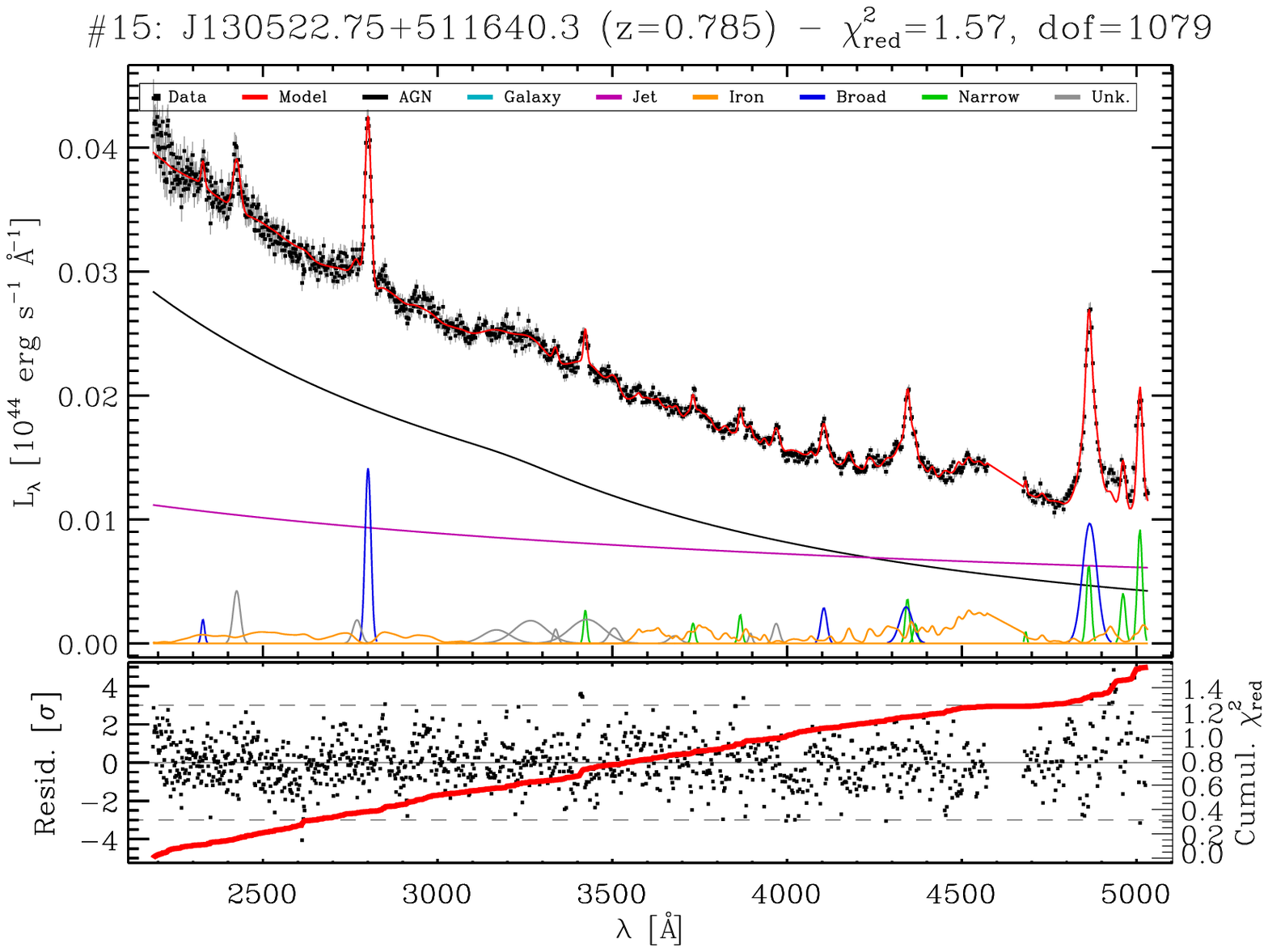}
    \includegraphics[width=.28\textwidth]{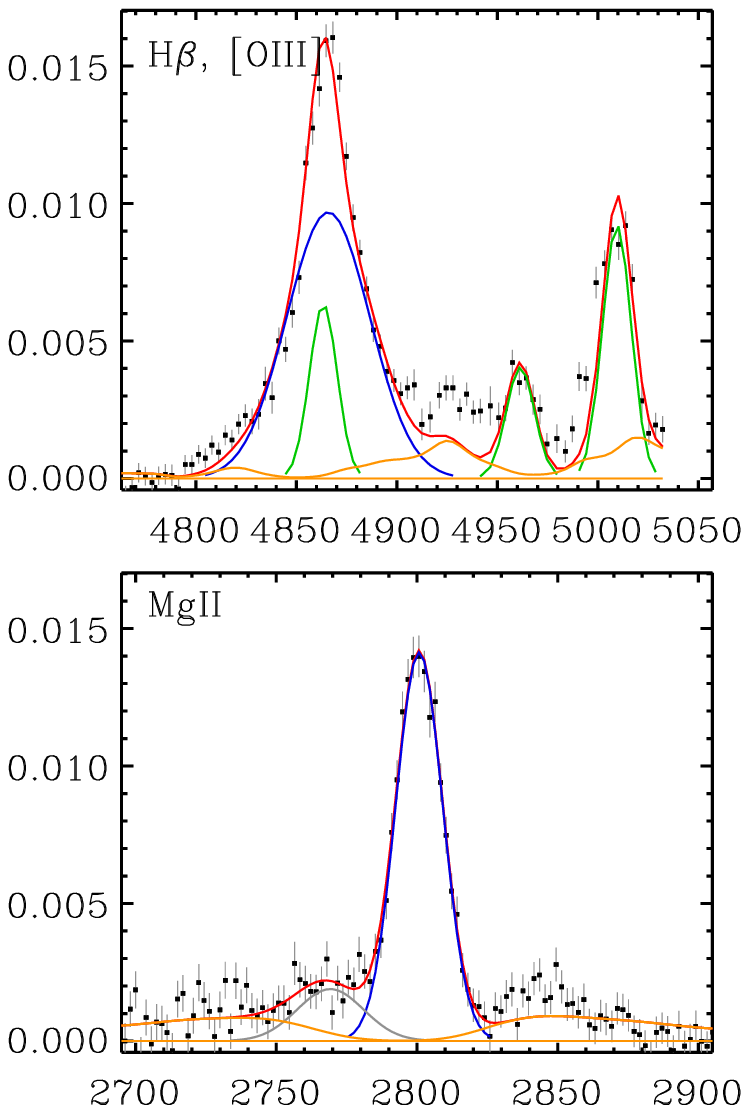}}\\
  \caption{(continued)}%
\end{figure*}

\begin{figure*}%
  \ContinuedFloat
  \centering
  \subfloat[][]{
    \includegraphics[width=.58\textwidth]{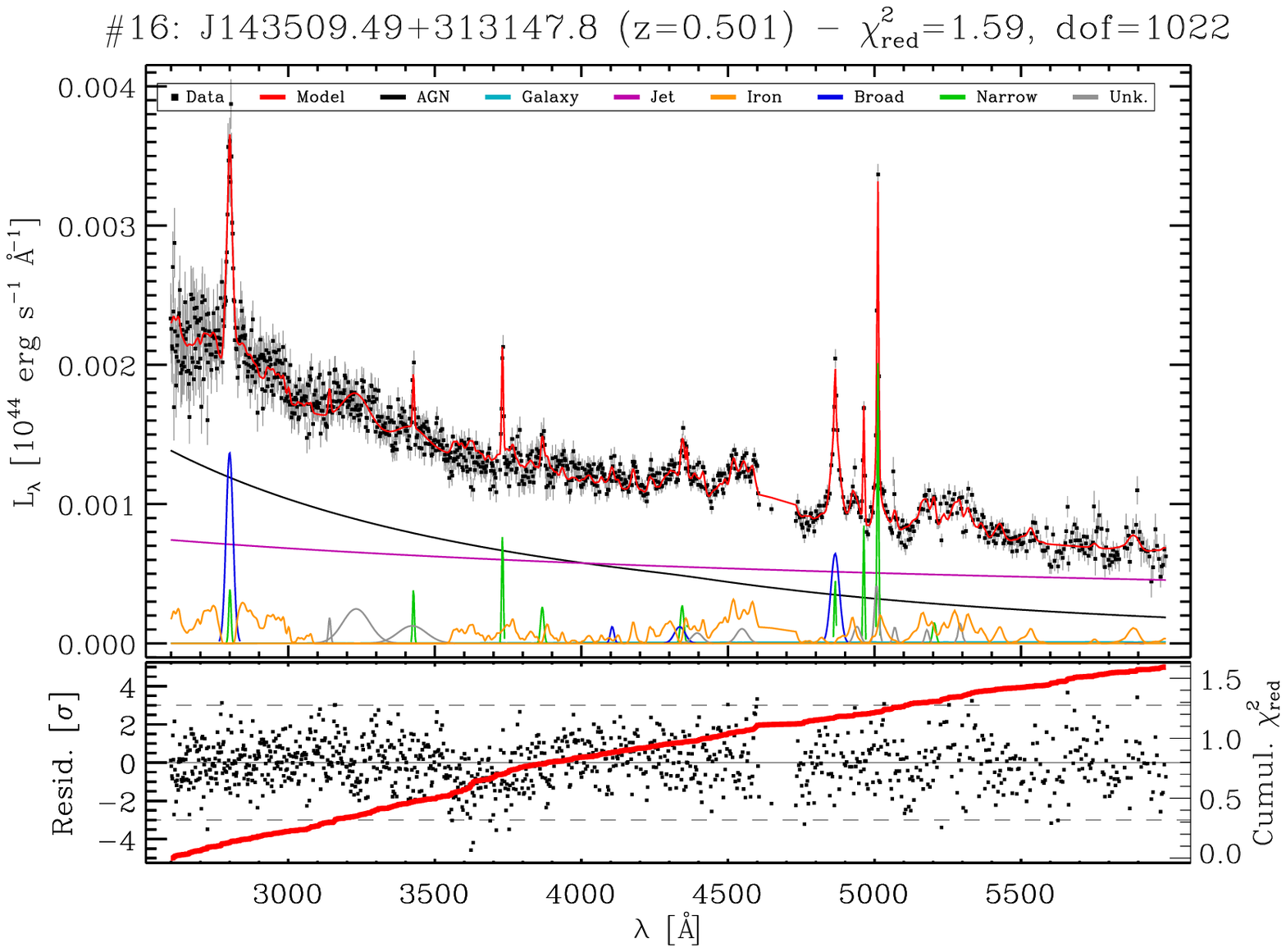}
    \includegraphics[width=.28\textwidth]{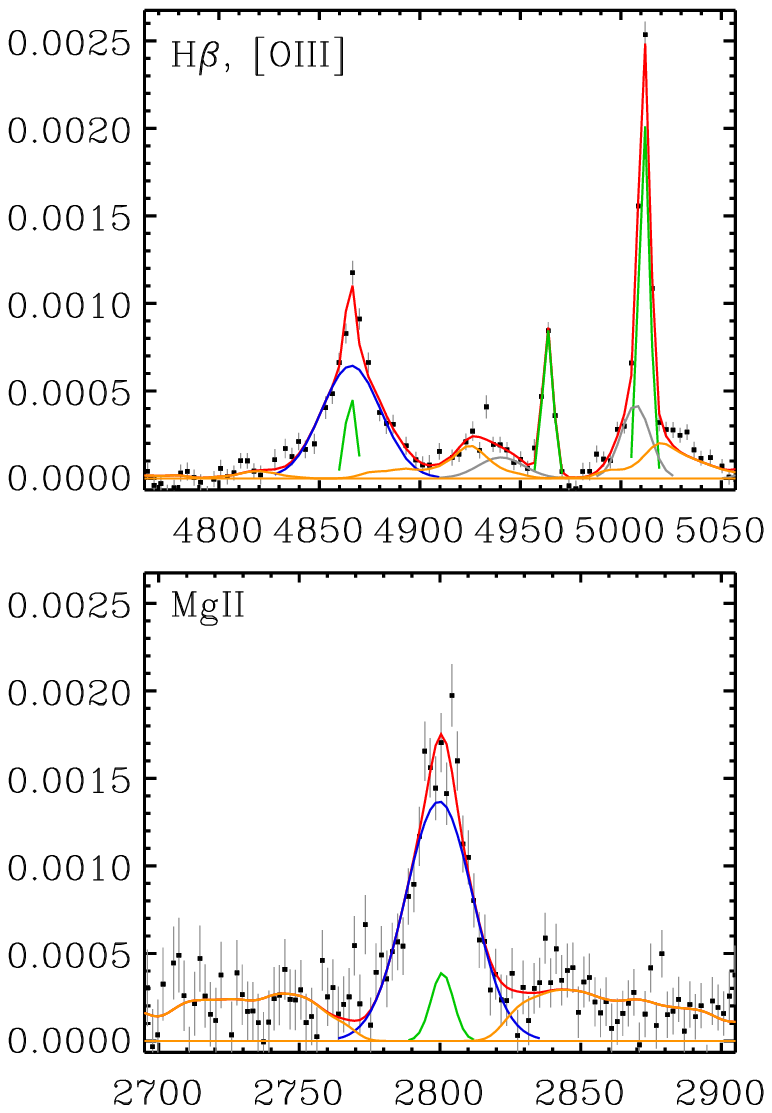}}\\
  \vspace{-0.7cm} \subfloat[][]{
    \includegraphics[width=.58\textwidth]{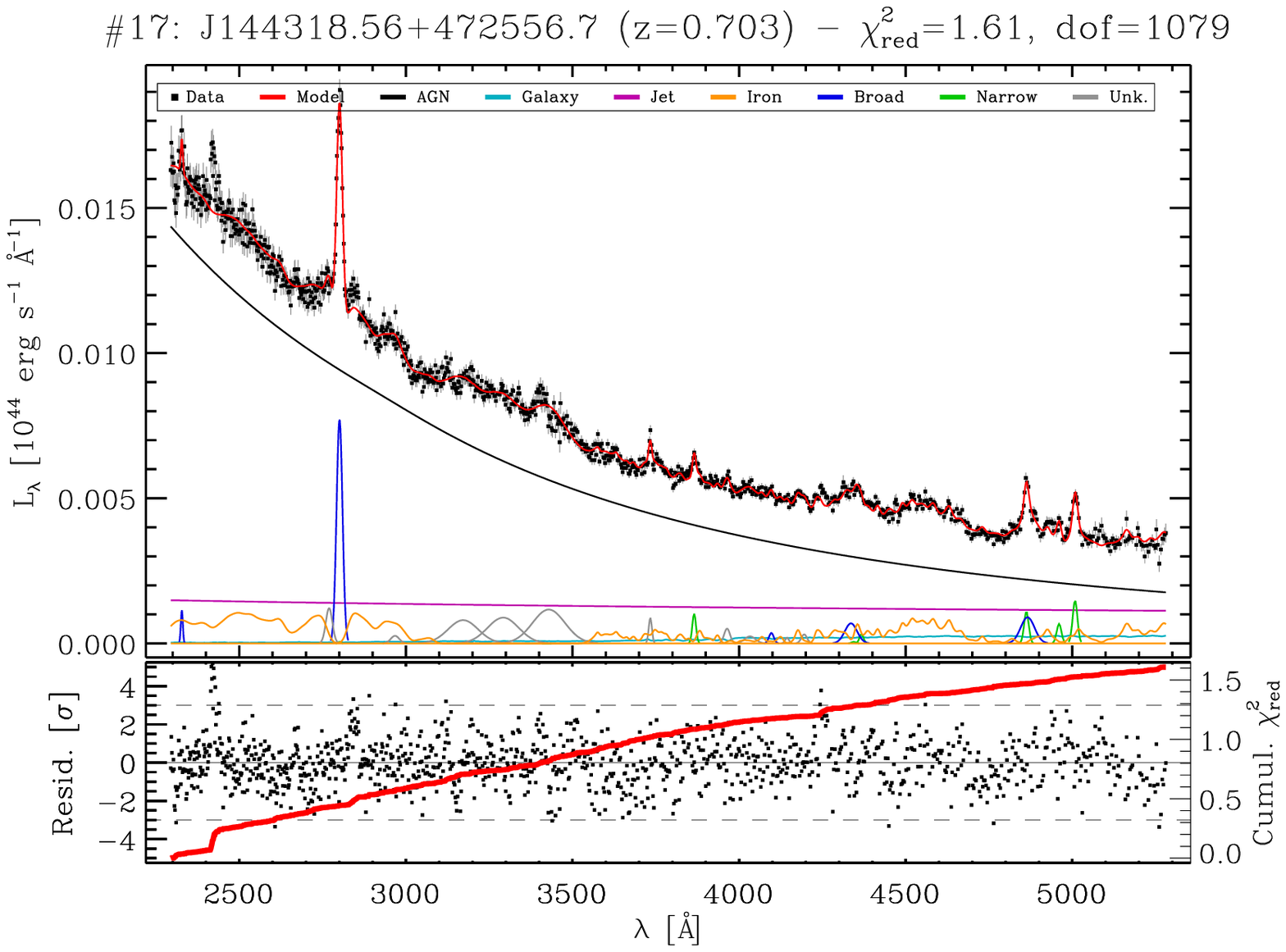}
    \includegraphics[width=.28\textwidth]{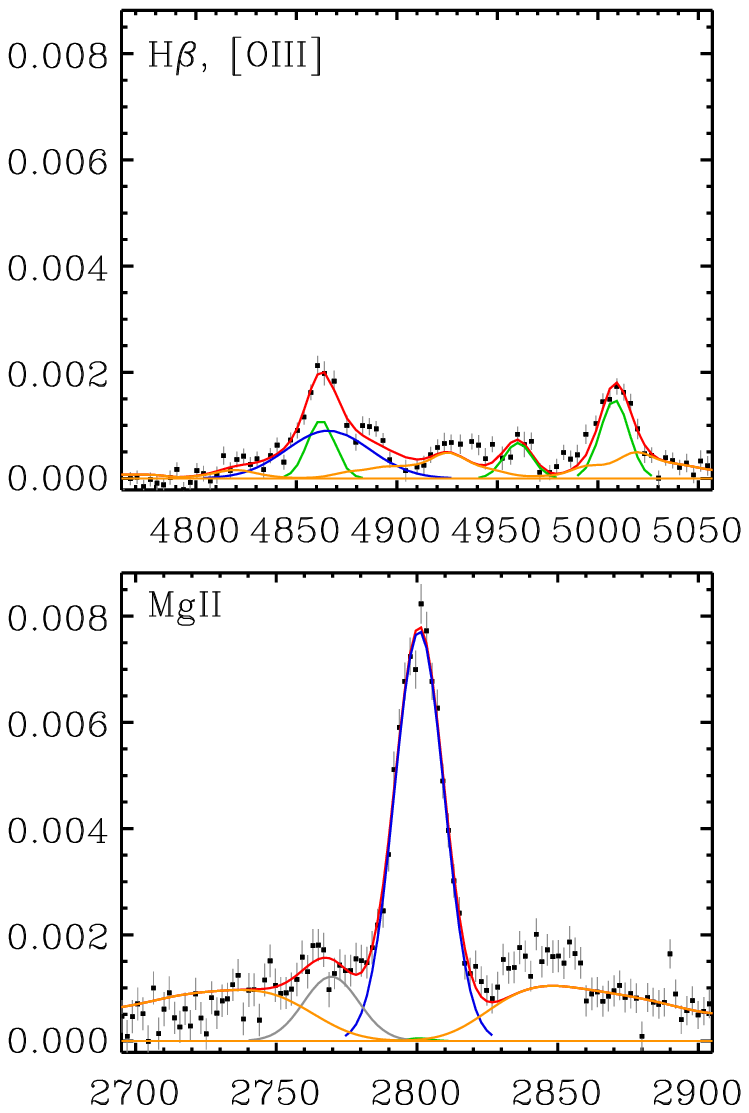}}\\
  \vspace{-0.7cm} \subfloat[][]{
    \includegraphics[width=.58\textwidth]{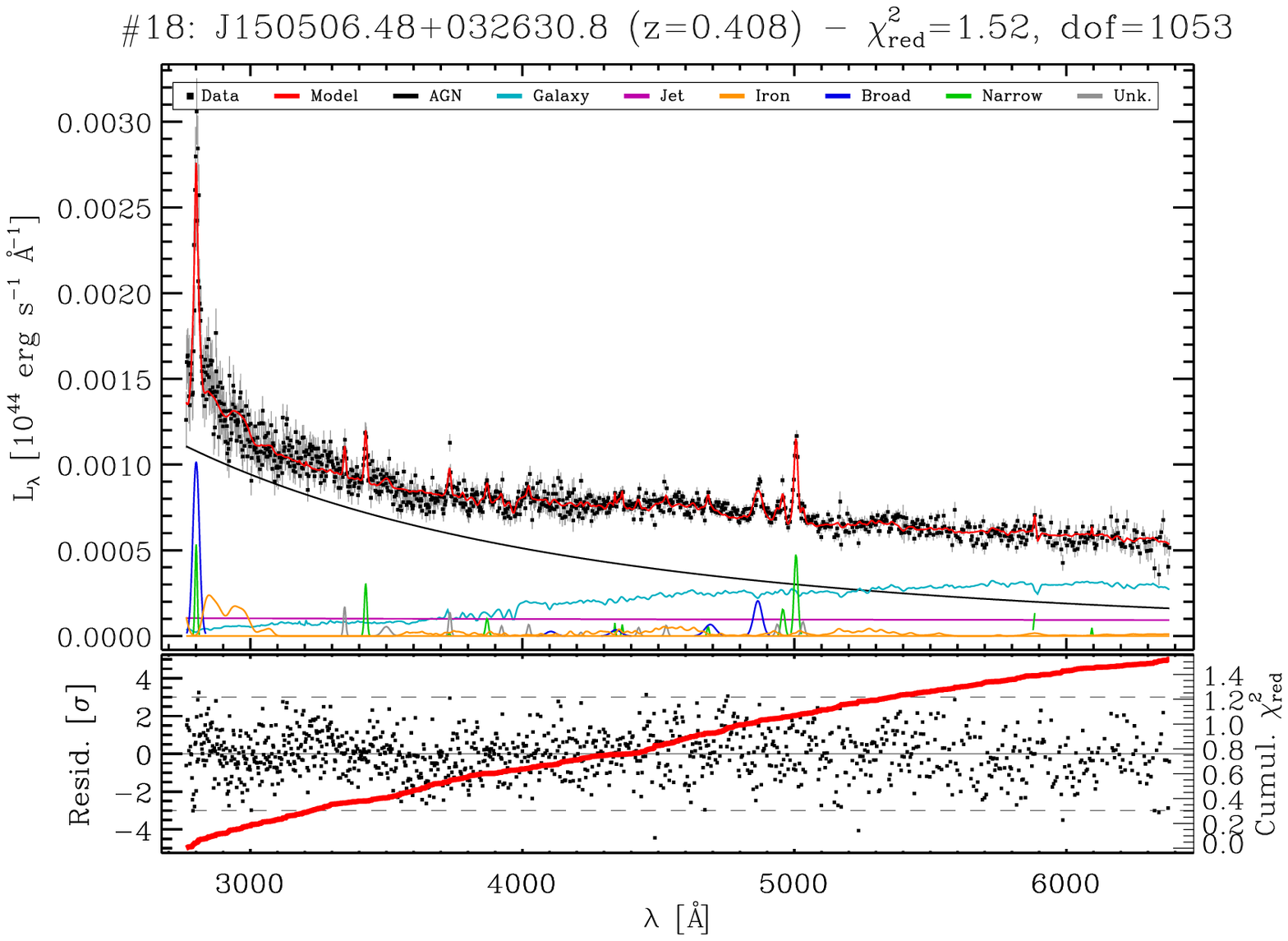}
    \includegraphics[width=.28\textwidth]{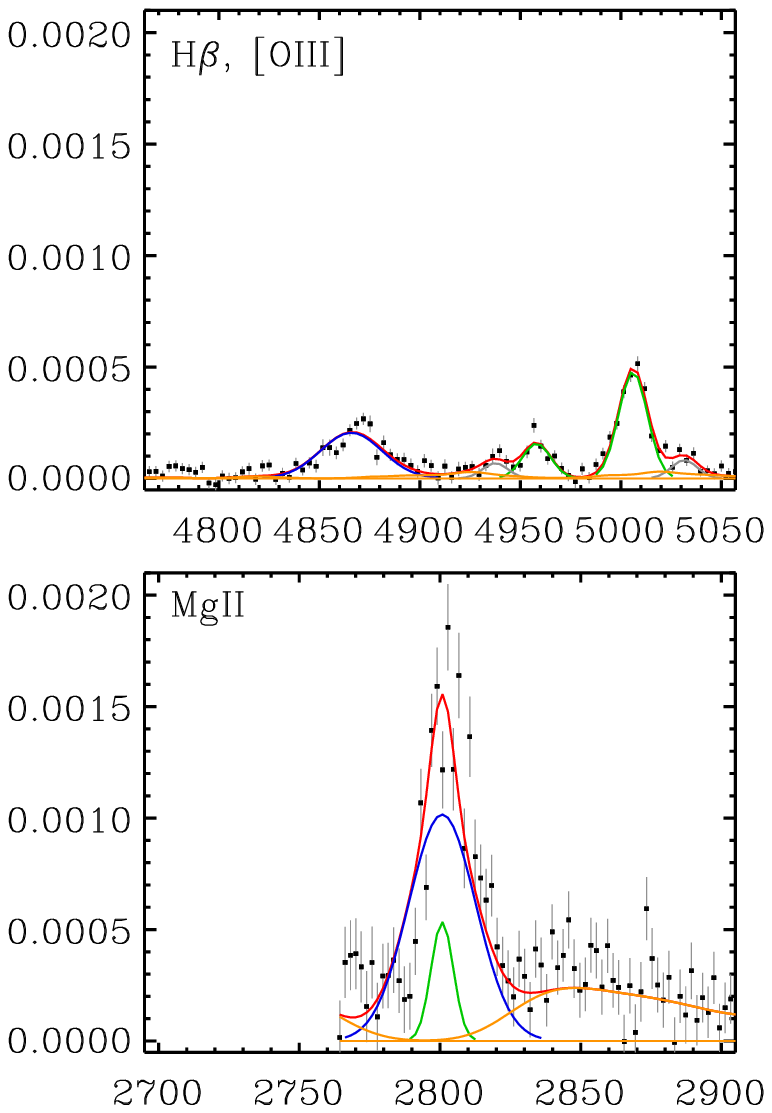}}\\
  \caption{(continued)}%
\end{figure*}

\begin{figure*}%
  \ContinuedFloat
  \centering
  \subfloat[][]{
    \includegraphics[width=.58\textwidth]{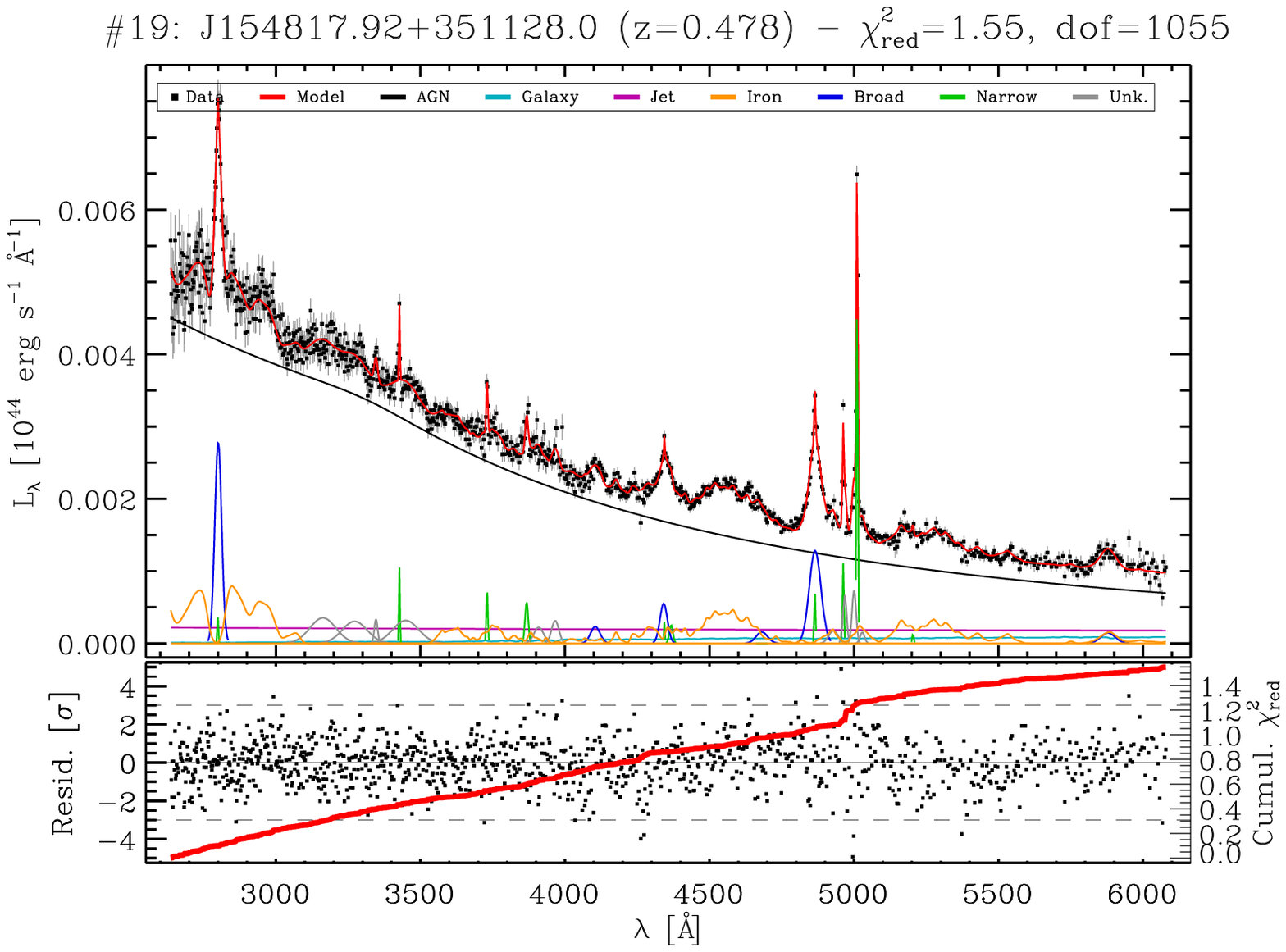}
    \includegraphics[width=.28\textwidth]{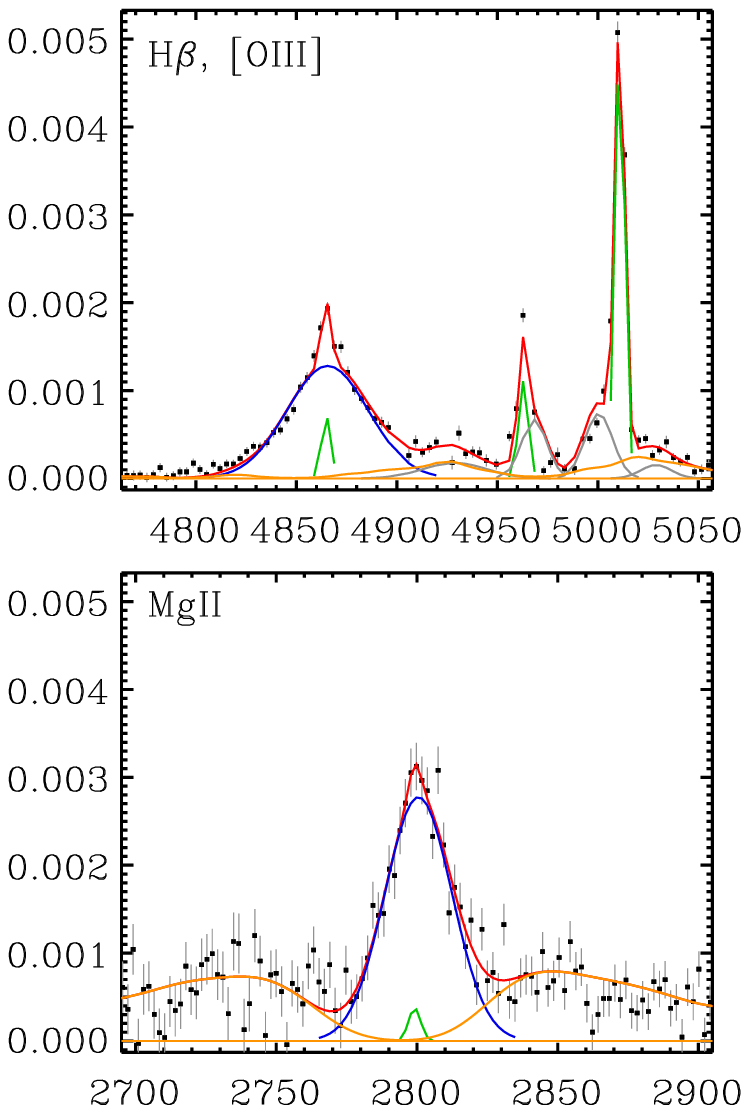}}\\
  \vspace{-0.7cm} \subfloat[][]{
    \includegraphics[width=.58\textwidth]{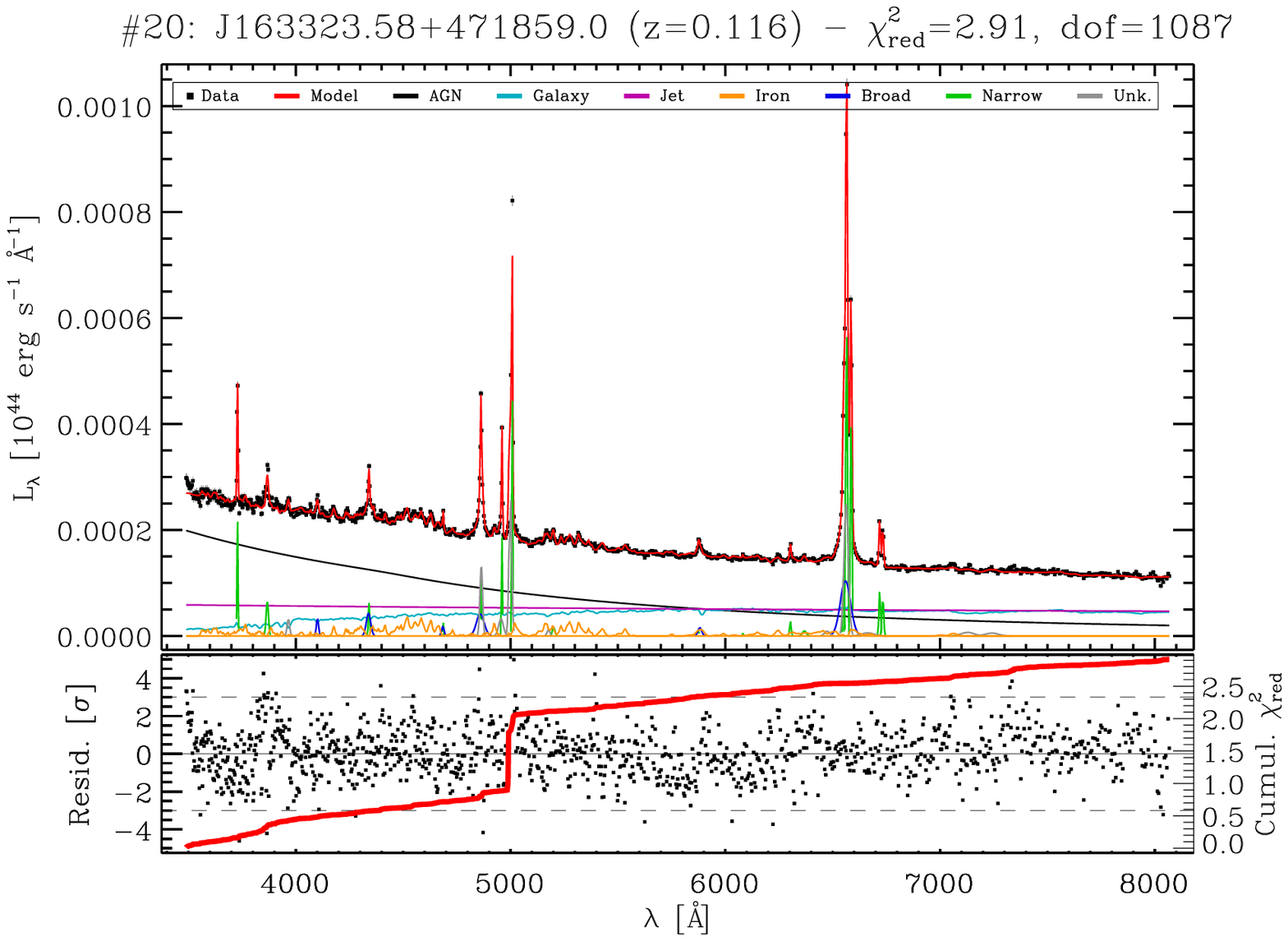}
    \includegraphics[width=.28\textwidth]{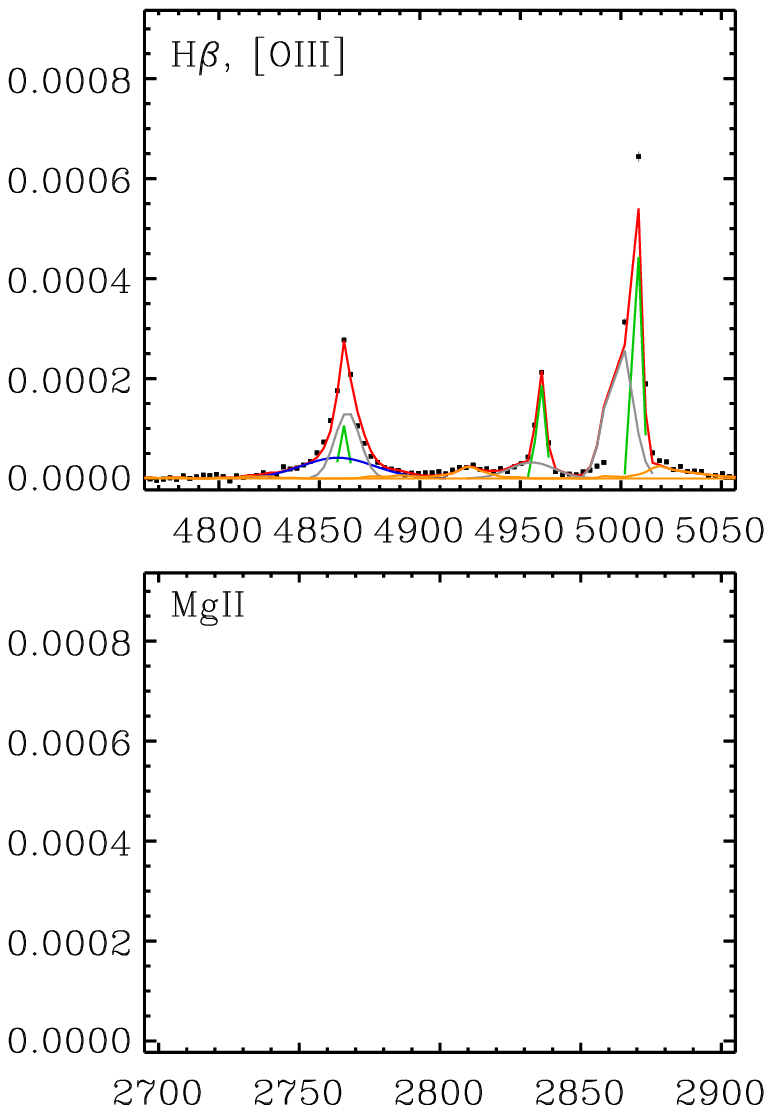}}\\
  \vspace{-0.7cm} \subfloat[][]{
    \includegraphics[width=.58\textwidth]{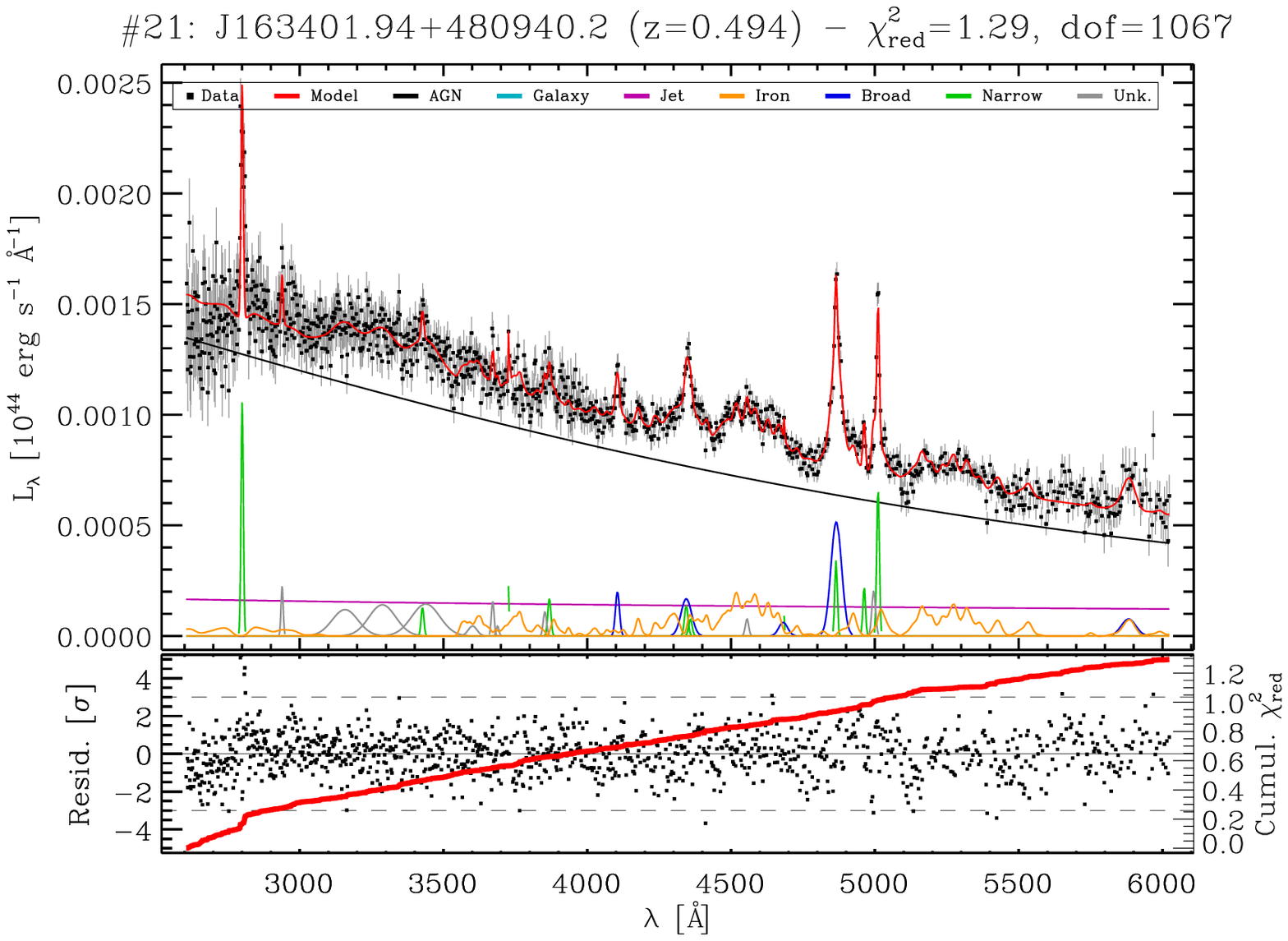}
    \includegraphics[width=.28\textwidth]{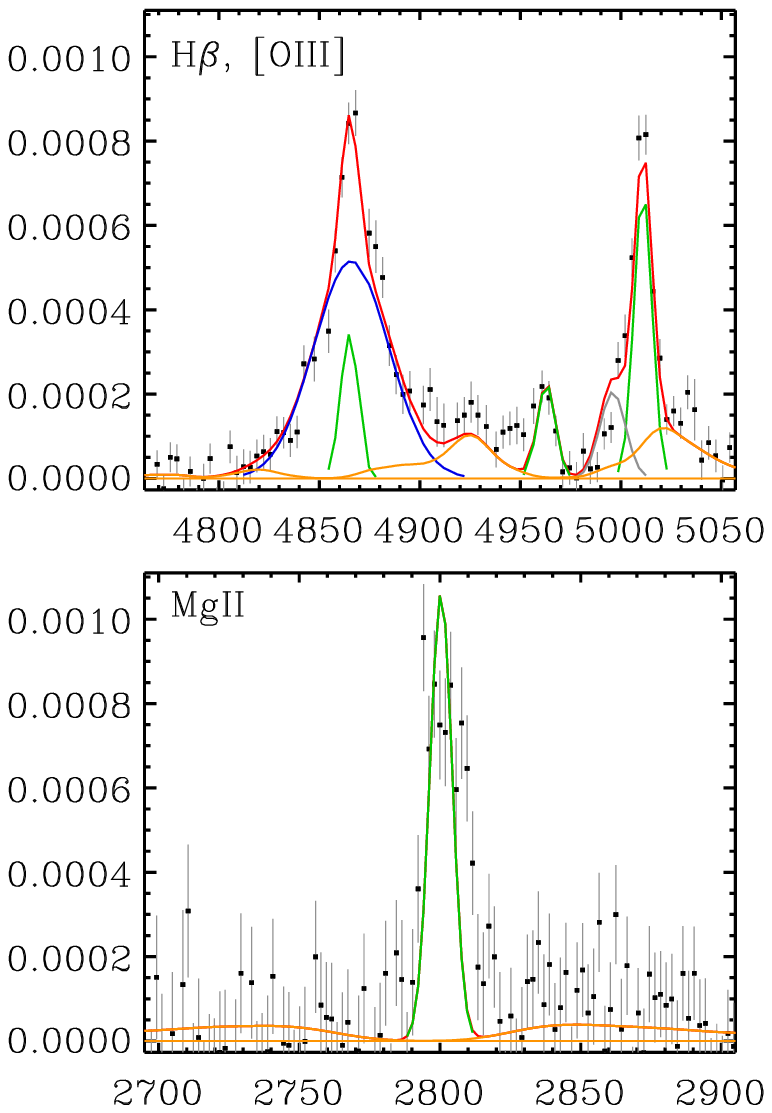}}\\
  \caption{(continued)}%
\end{figure*}

\begin{figure*}%
  \ContinuedFloat
  \centering
  \subfloat[][]{
    \includegraphics[width=.58\textwidth]{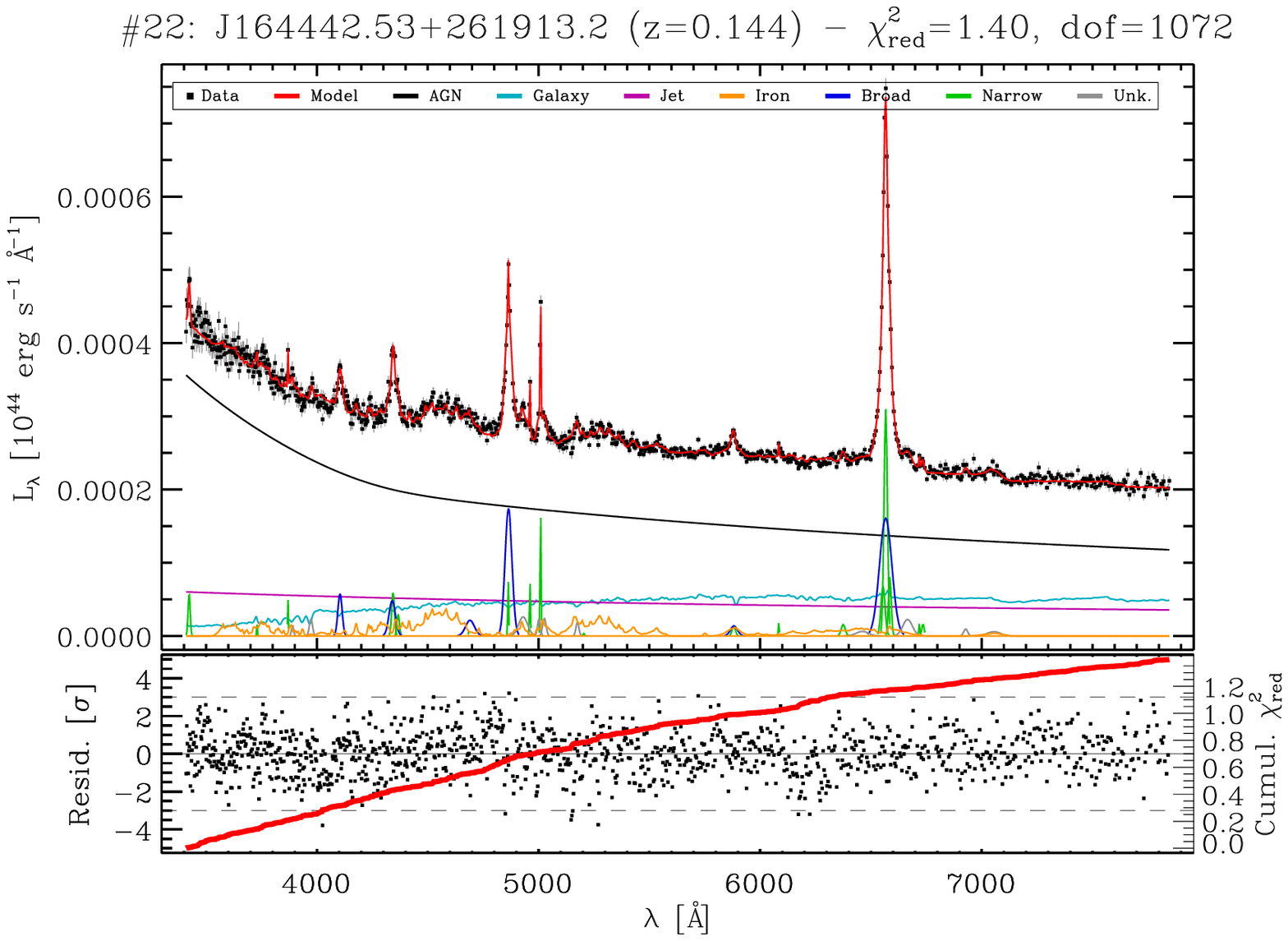}
    \includegraphics[width=.28\textwidth]{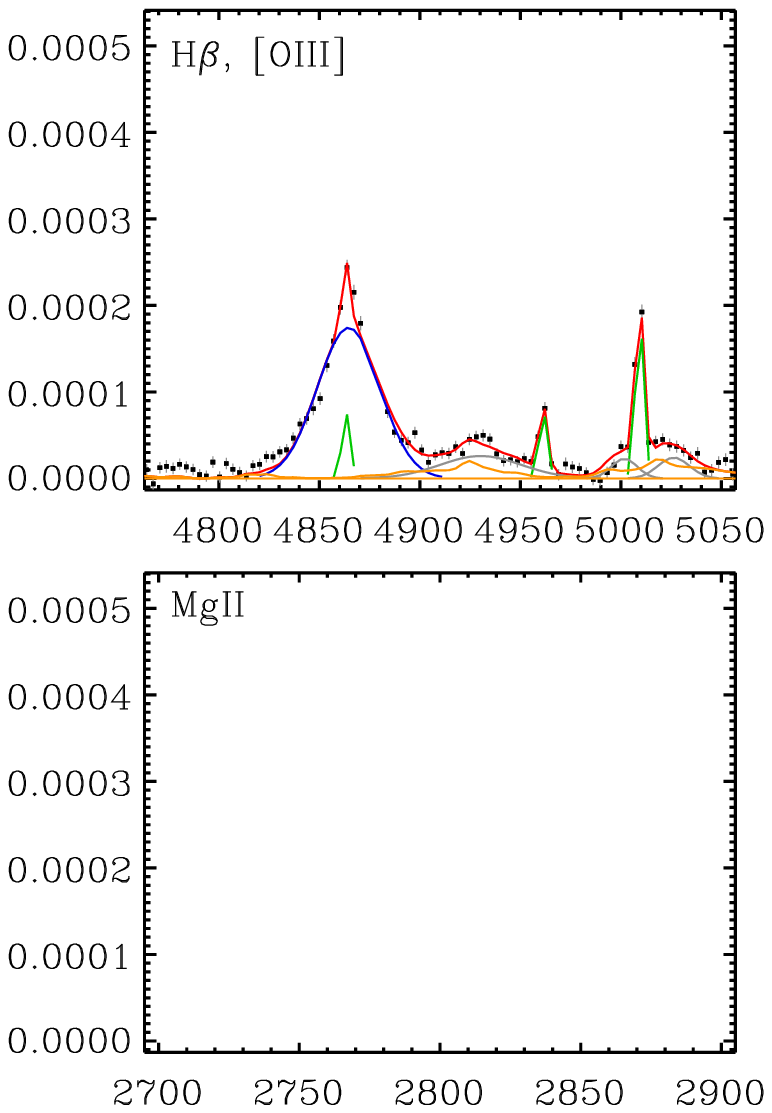}}\\
  \vspace{-0.7cm} \subfloat[][]{
    \includegraphics[width=.58\textwidth]{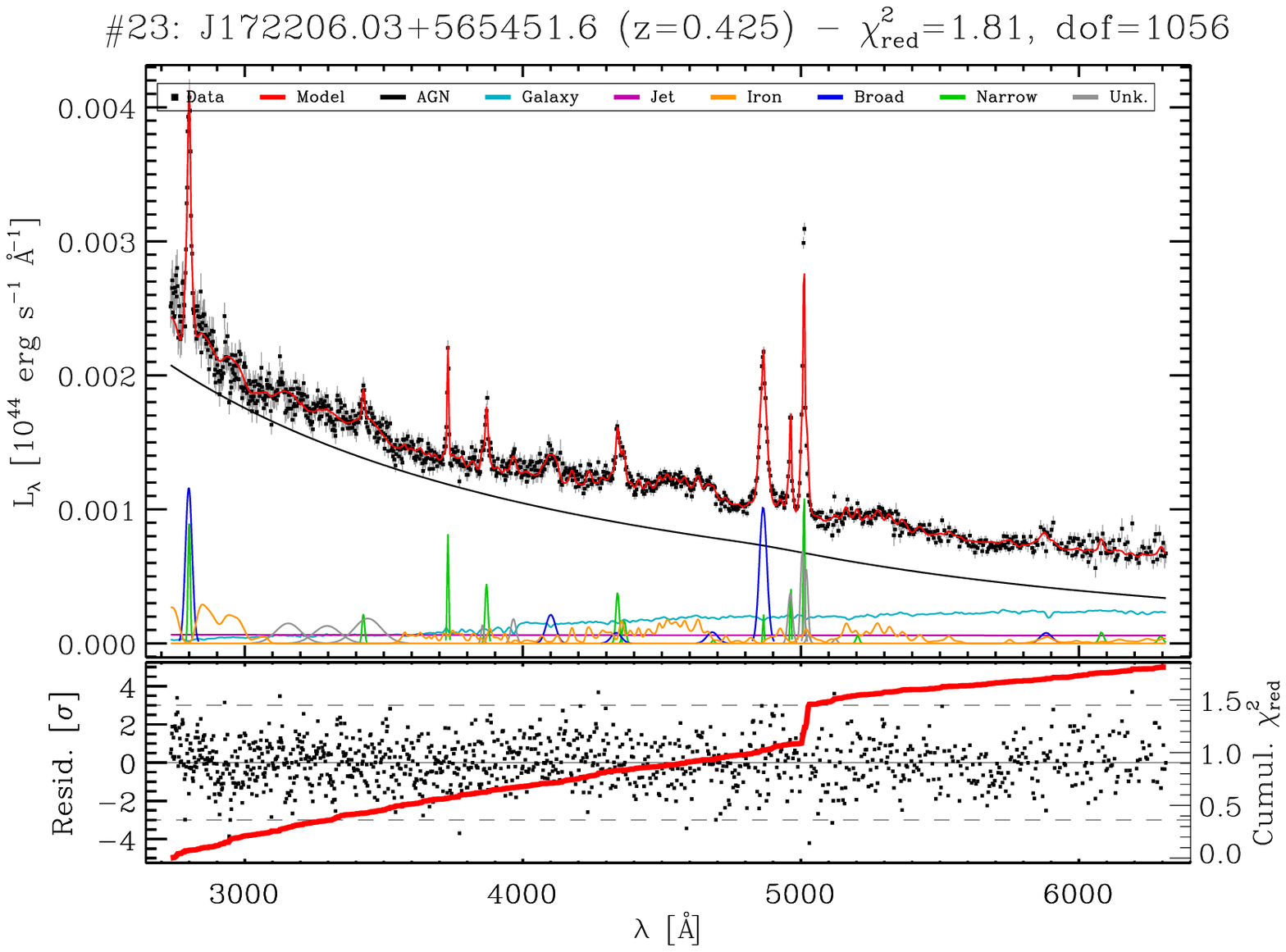}
    \includegraphics[width=.28\textwidth]{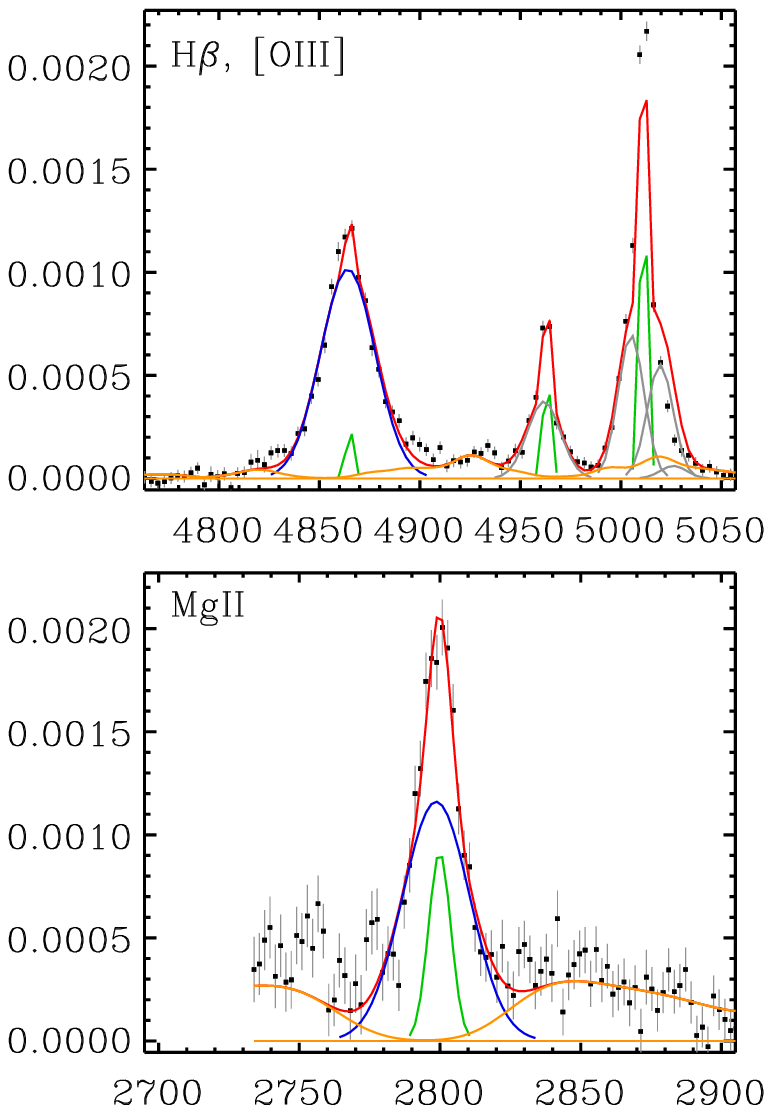}}\\
  \caption{(continued)}%
\end{figure*}

\newpage
\section{Figures: black hole mass estimation}
\label{app-fig-mass}
This appendix is a collection of the figures related to the black hole
mass estimation procedures described in \S \ref{sec-method}, adopting
the same notation as in Fig. \ref{fig-example}.

\begin{figure*}%
  \centering
  \subfloat[][]{
    \includegraphics[width=.5\textwidth]{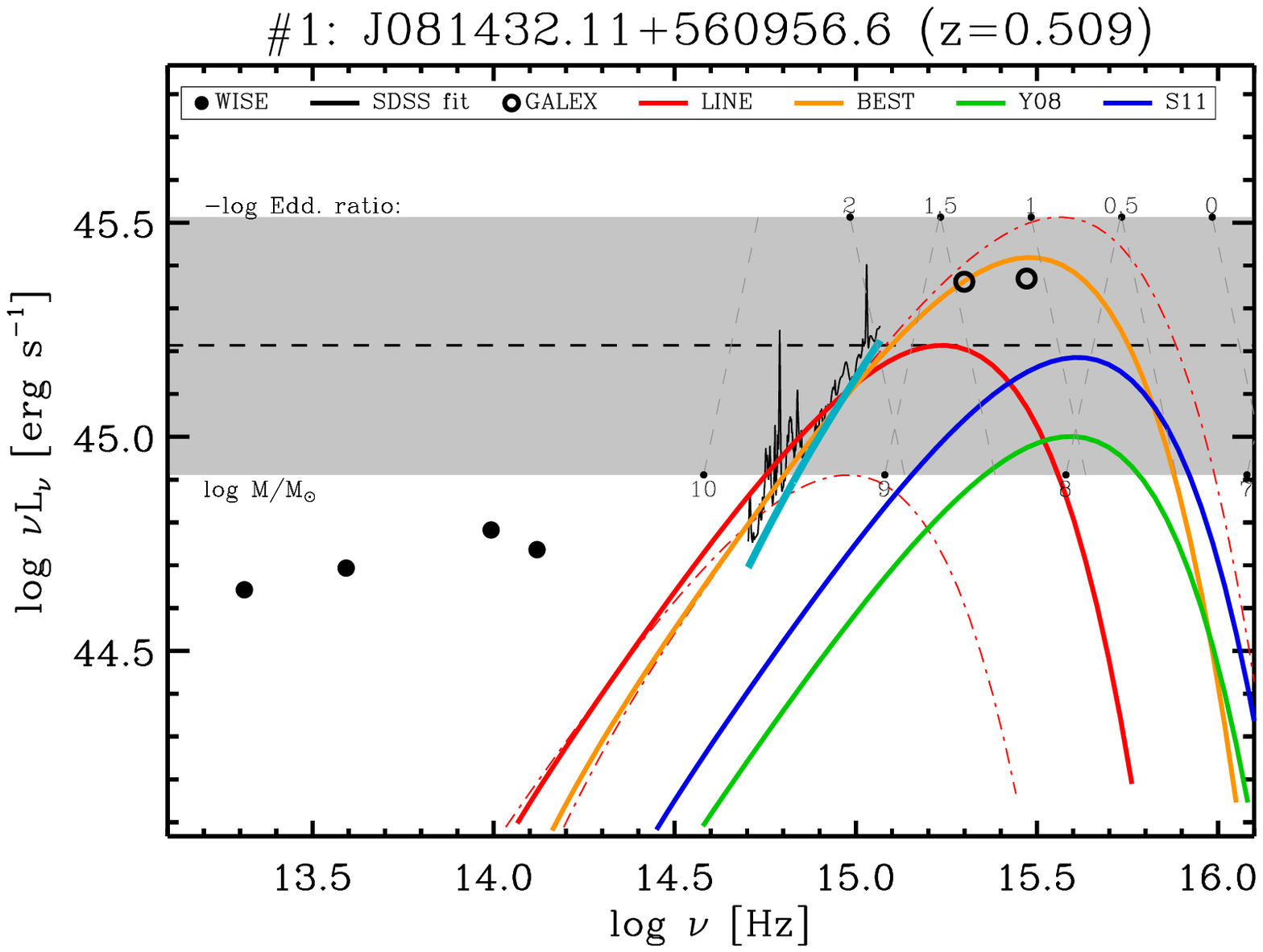}
    \includegraphics[width=.5\textwidth]{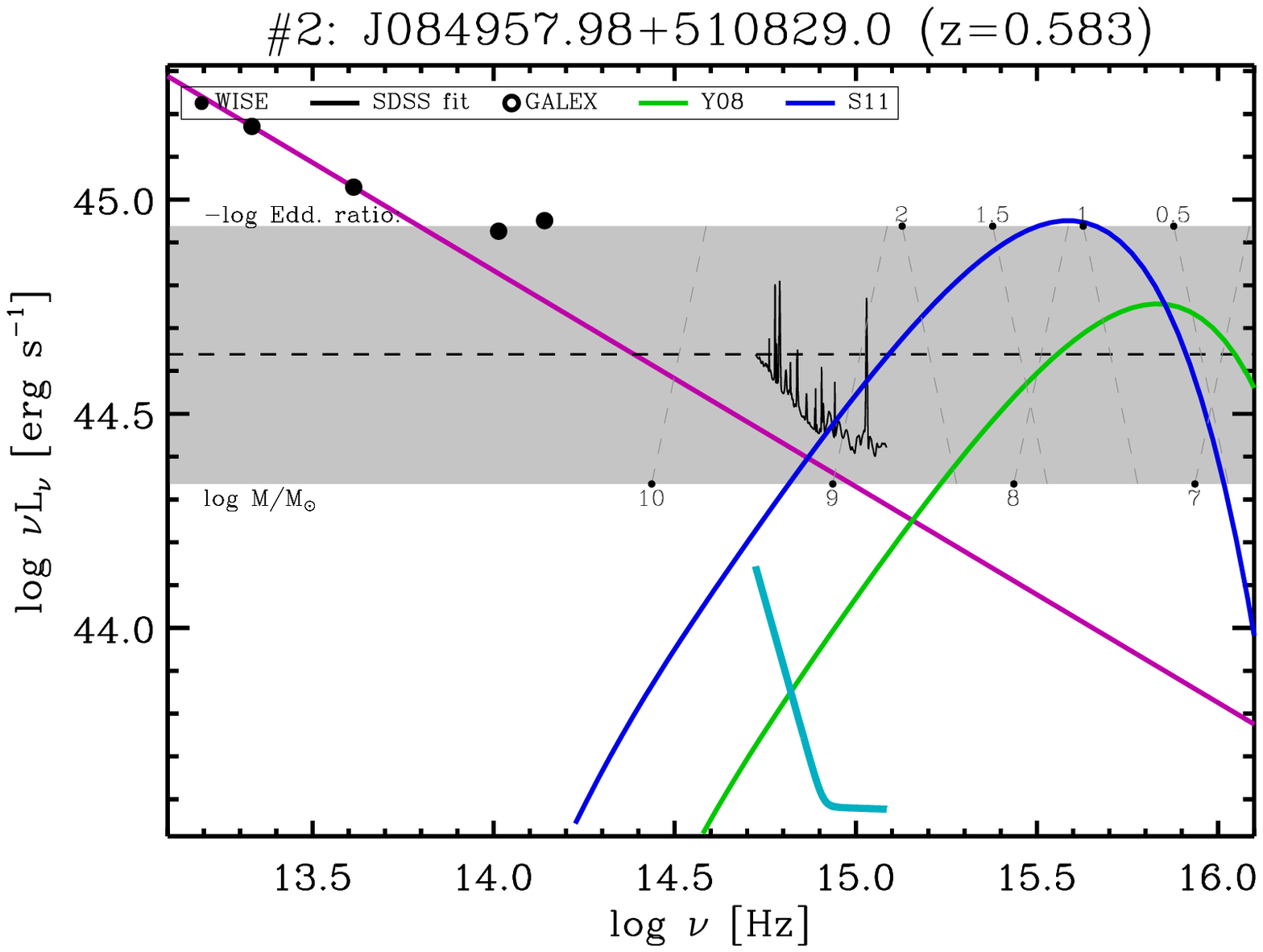}}\\
  \vspace{-0.7cm} \subfloat[][]{
    \includegraphics[width=.5\textwidth]{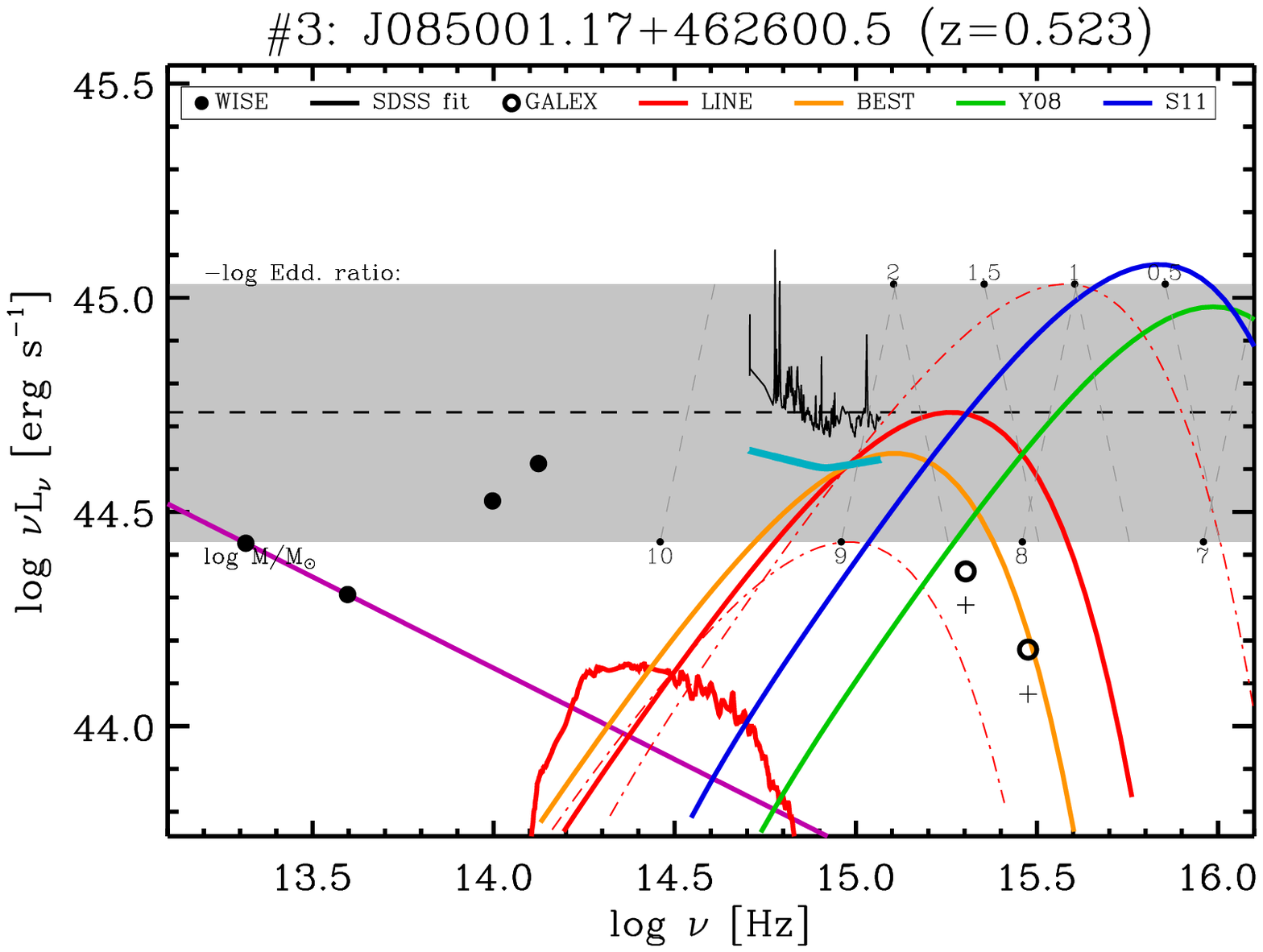}
    \includegraphics[width=.5\textwidth]{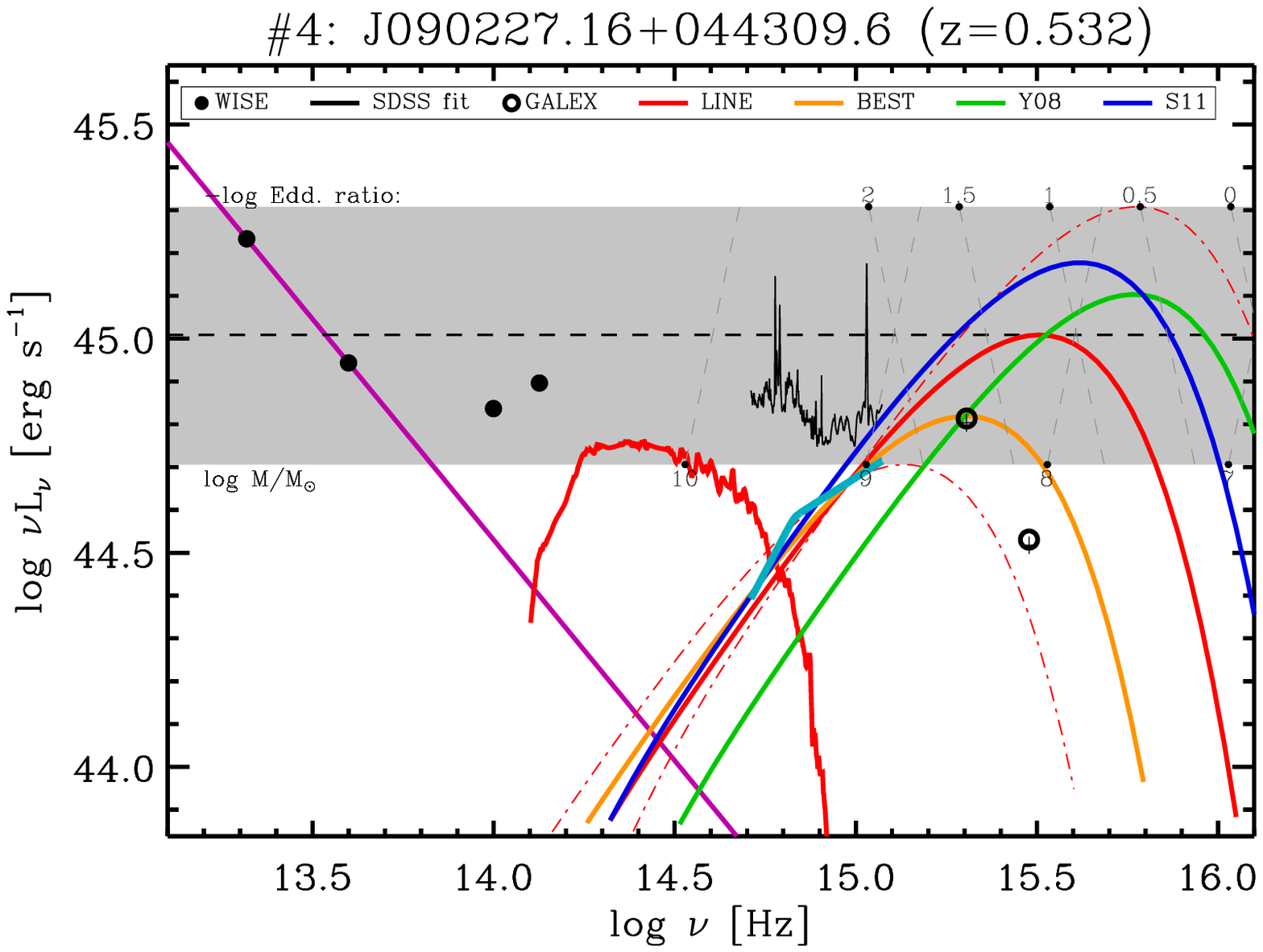}}\\
  \vspace{-0.7cm} \subfloat[][]{
    \includegraphics[width=.5\textwidth]{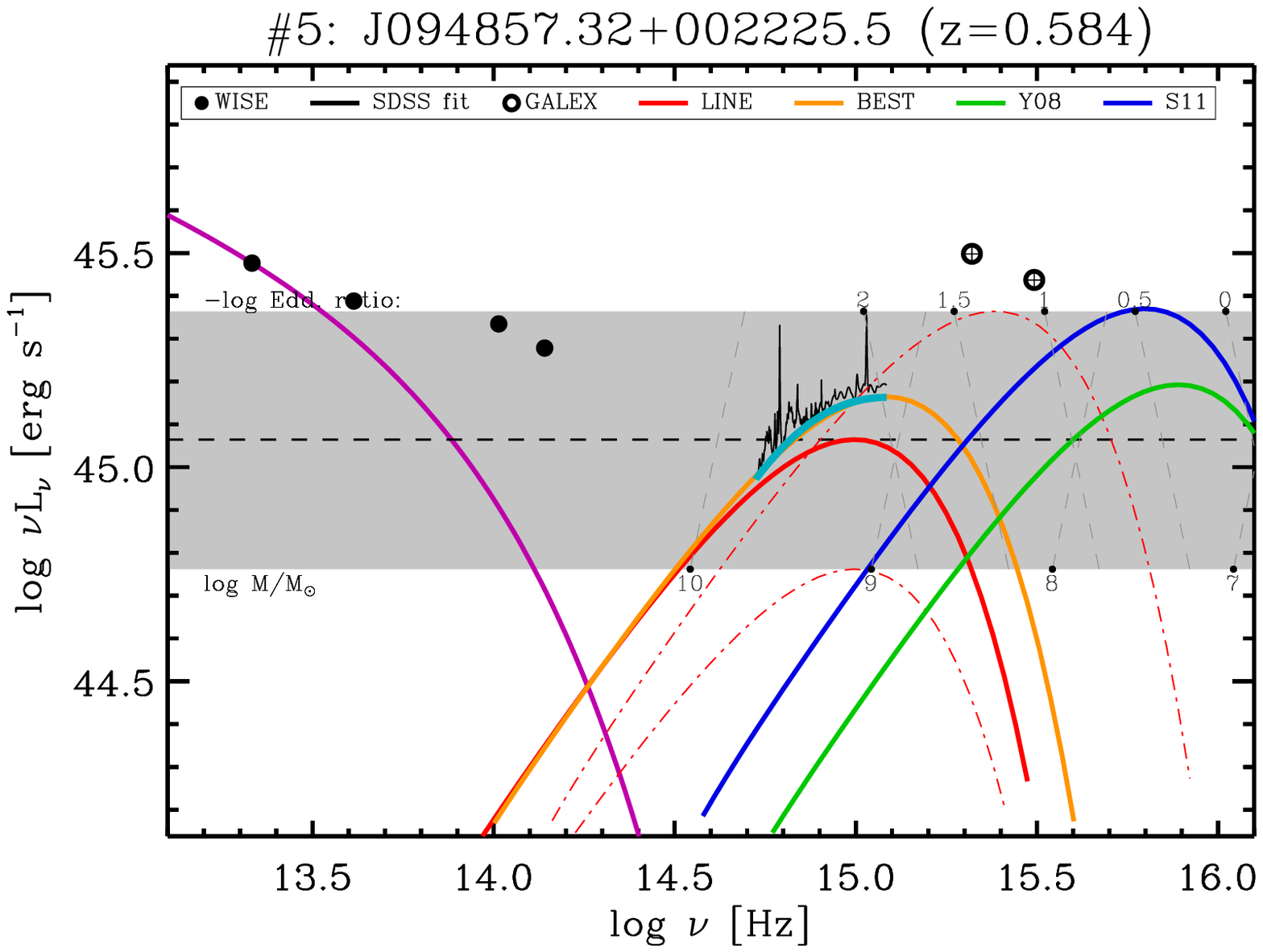}
    \includegraphics[width=.5\textwidth]{gasf_show_id_5_ssad}}\\
  \caption{Results of the black hole mass estimation procedures (\S
  \ref{sec-method}, App. \ref{app-fig-mass}).  Notation is the same as in Fig. \ref{fig-example}.}%
  \label{fig-mass}%
\end{figure*}

\begin{figure*}%
  \ContinuedFloat
  \centering
  \subfloat[][]{
    \includegraphics[width=.5\textwidth]{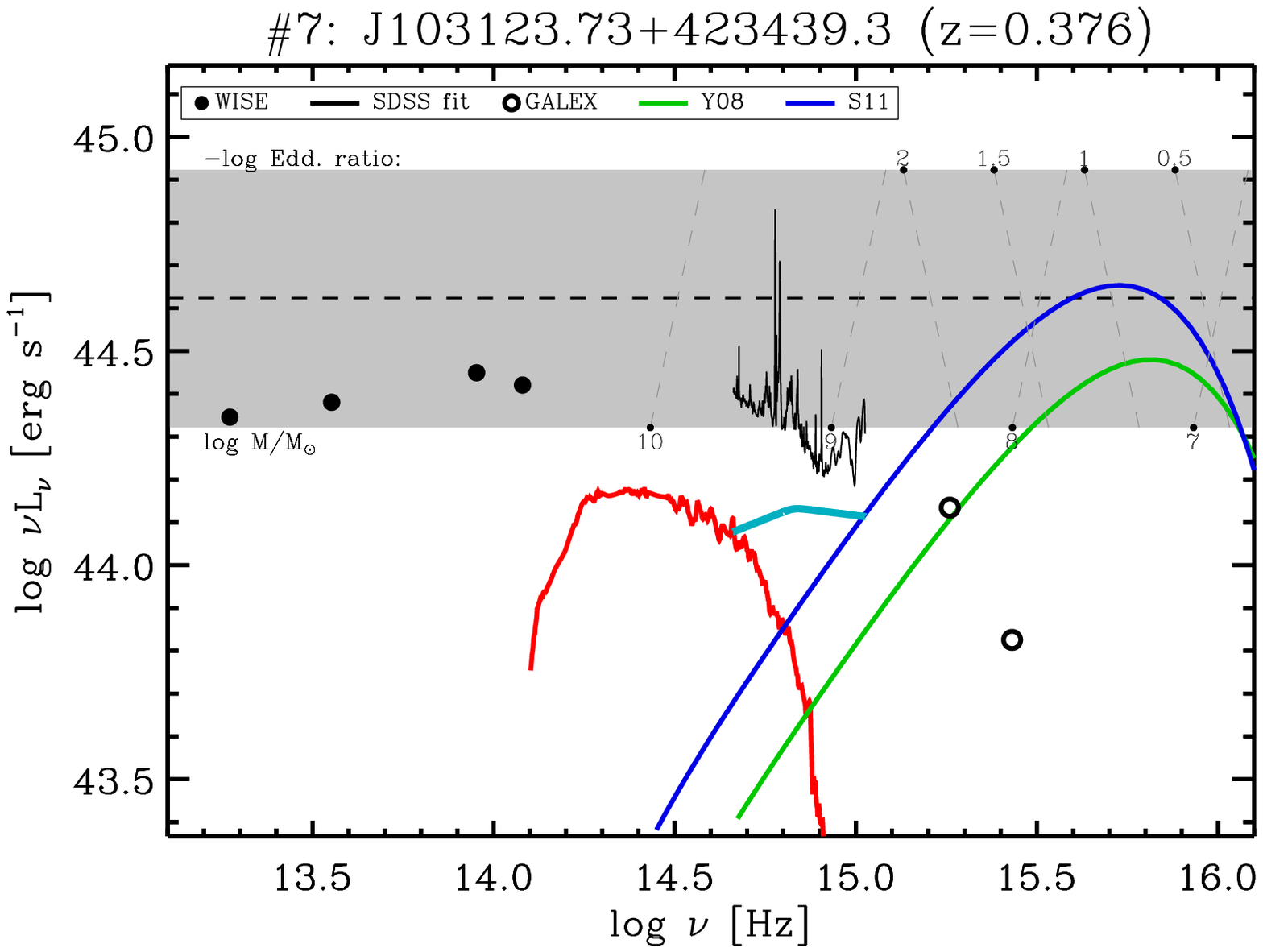}
    \includegraphics[width=.5\textwidth]{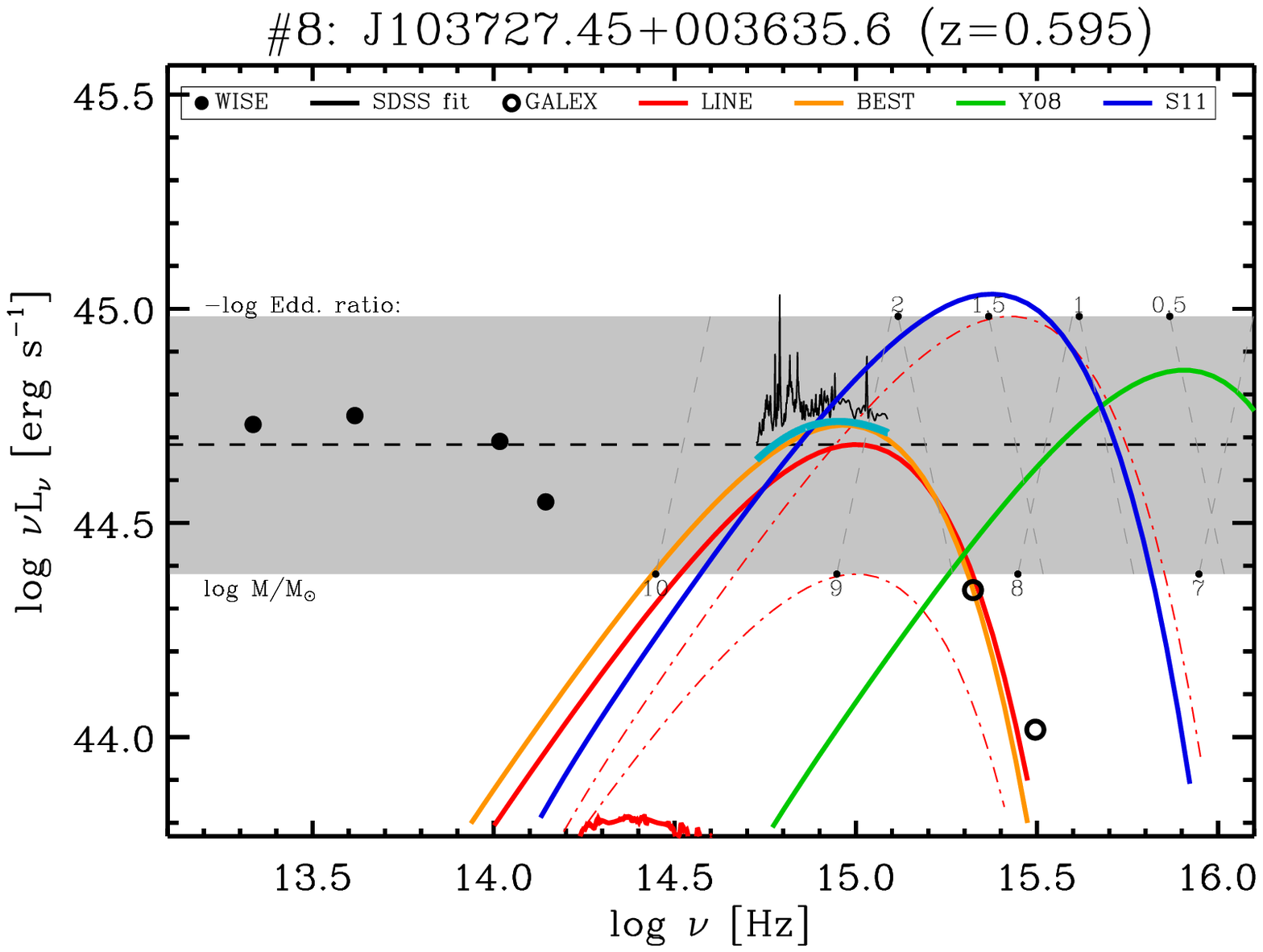}}\\
  \vspace{-0.7cm} \subfloat[][]{
    \includegraphics[width=.5\textwidth]{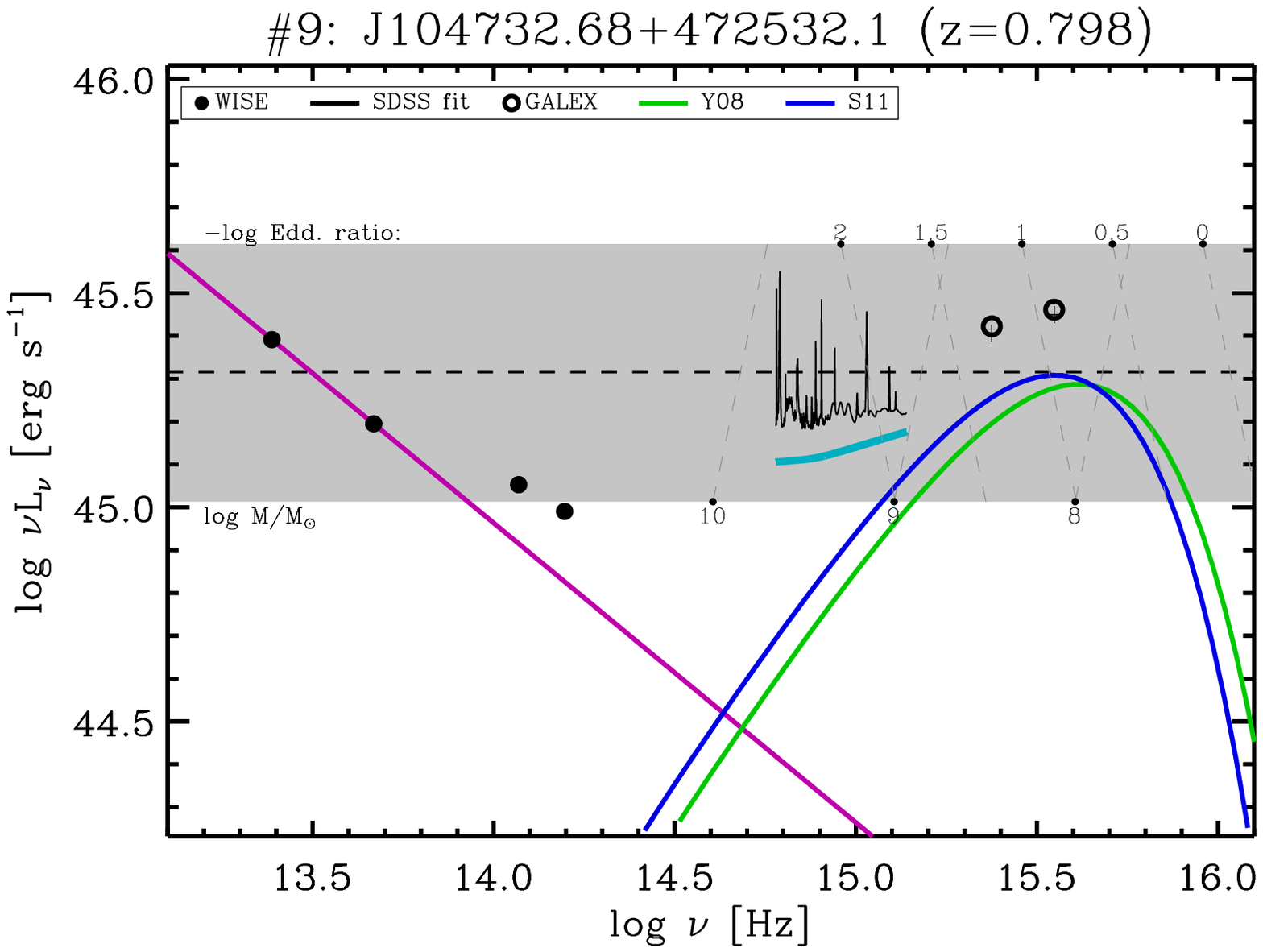}
    \includegraphics[width=.5\textwidth]{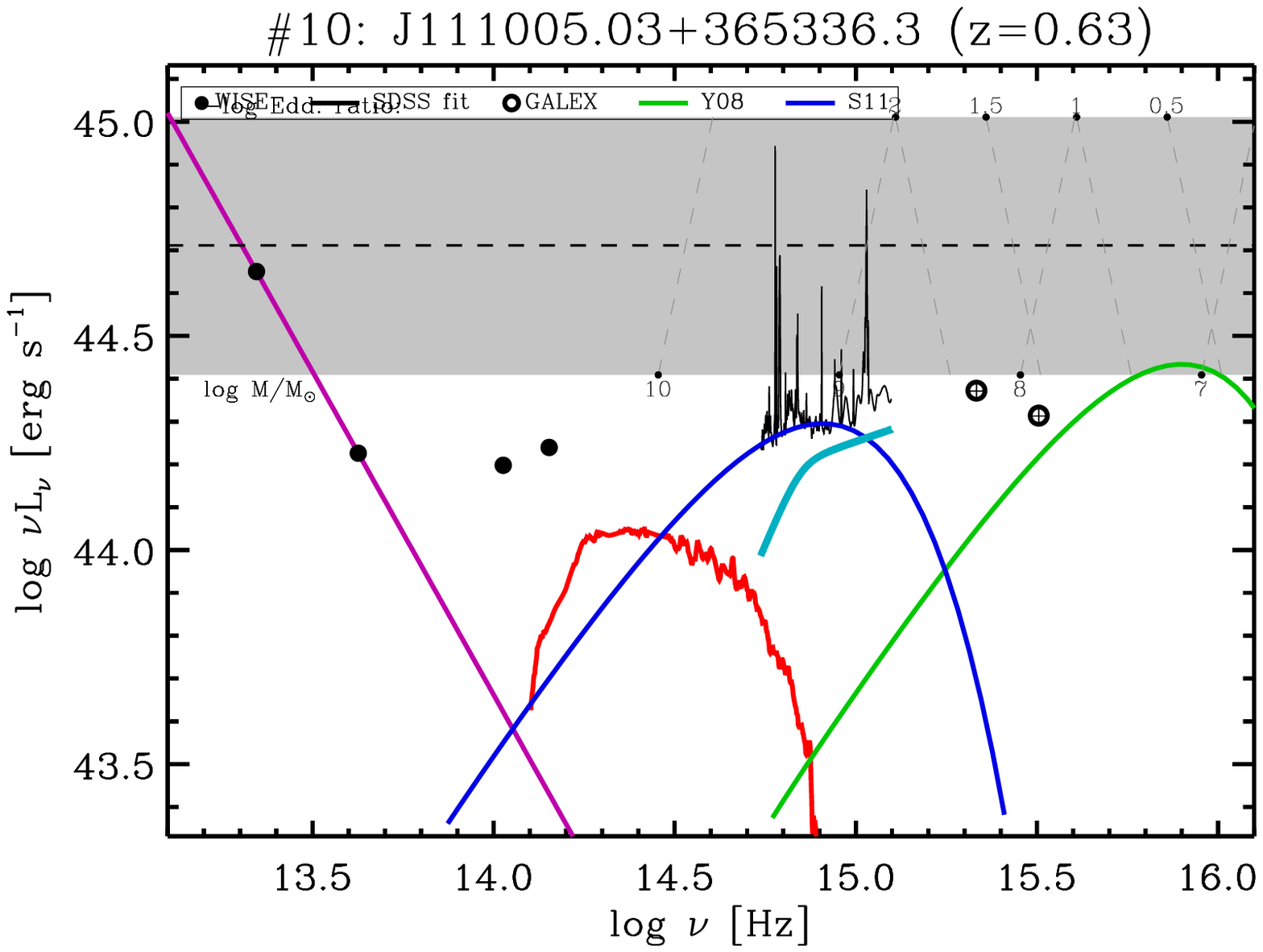}}\\
  \vspace{-0.7cm} \subfloat[][]{
    \includegraphics[width=.5\textwidth]{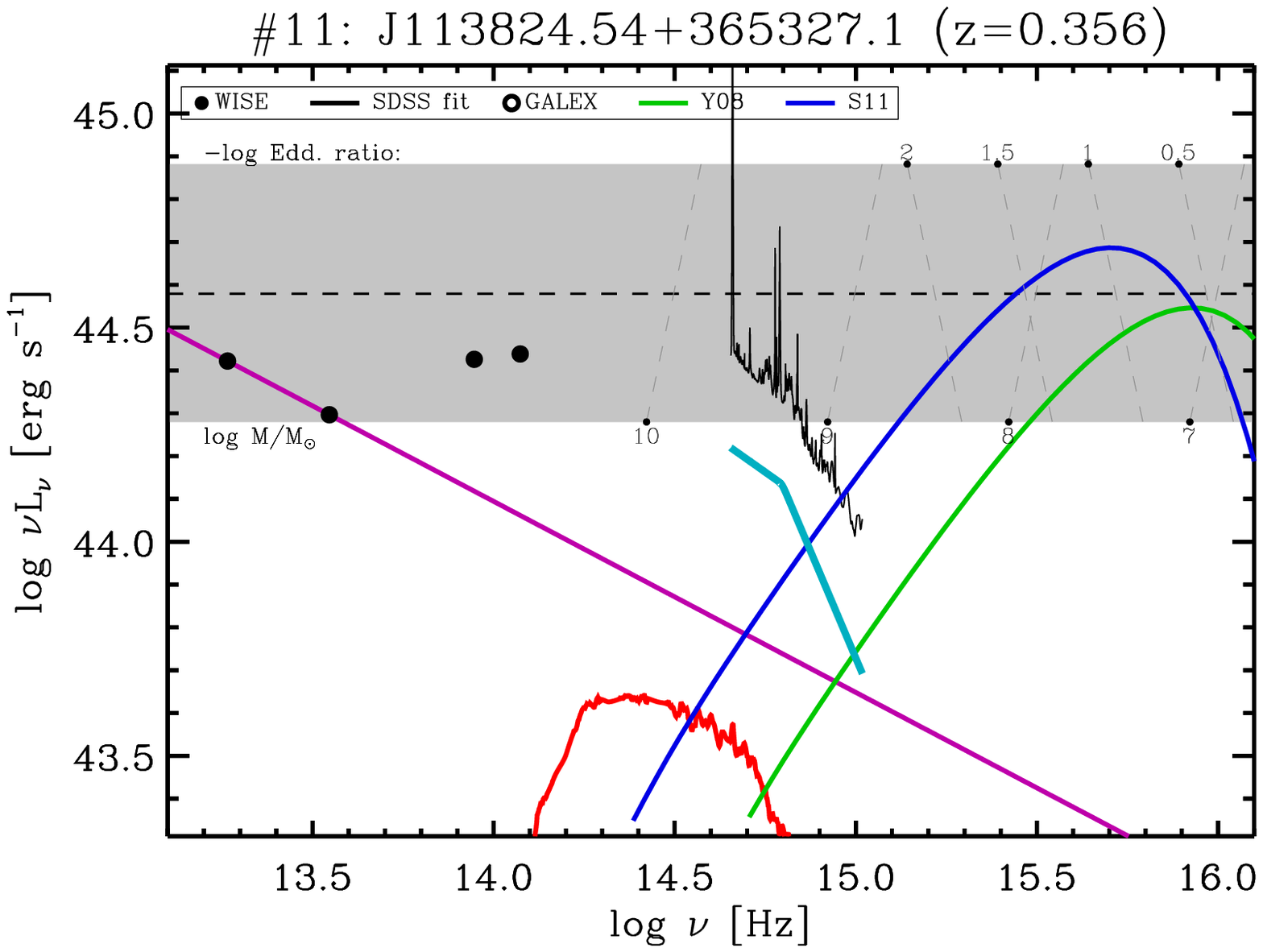}
    \includegraphics[width=.5\textwidth]{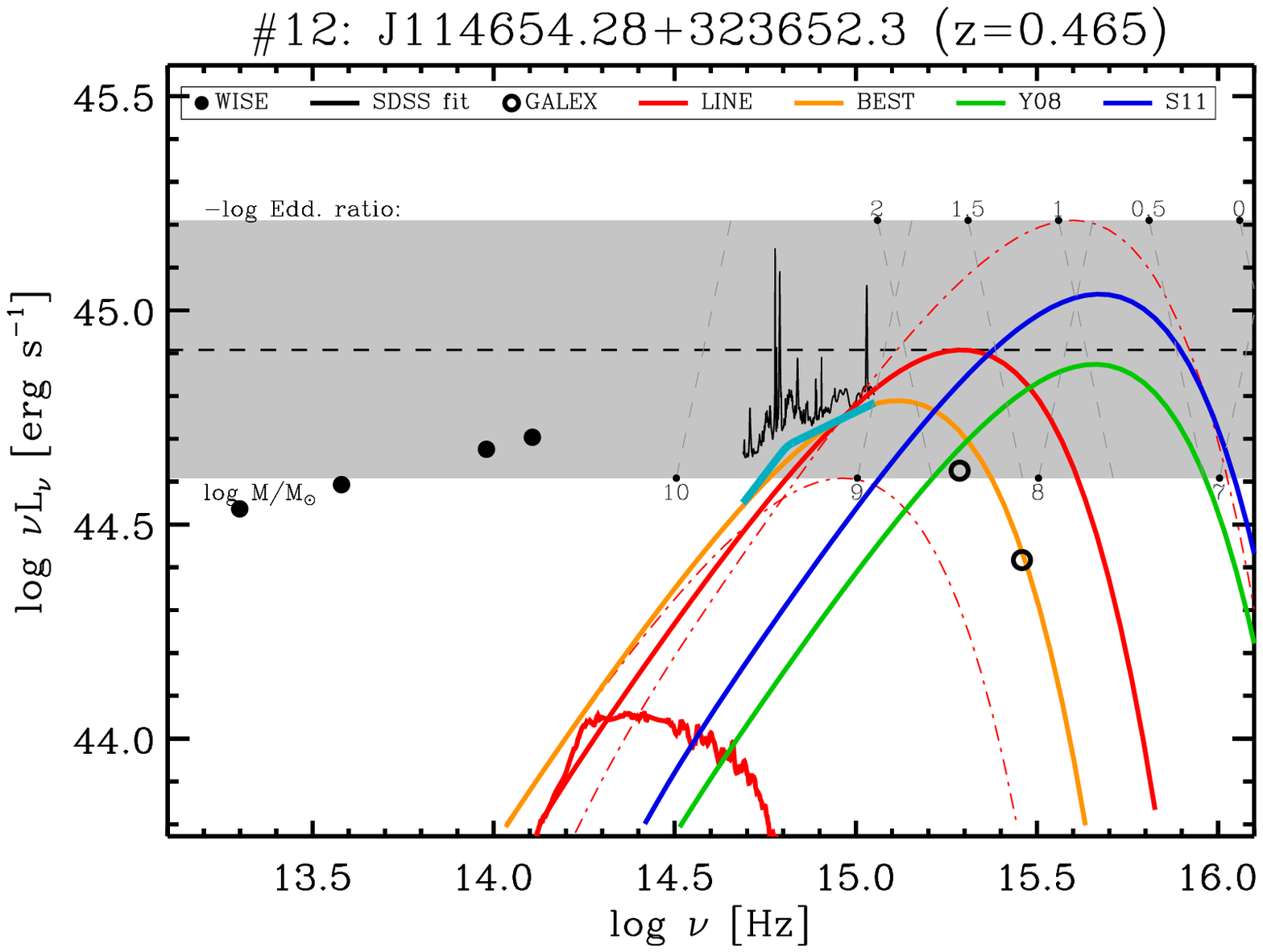}}\\
  \caption{(continued)}%
\end{figure*}

\begin{figure*}%
  \ContinuedFloat
  \centering
  \subfloat[][]{
    \includegraphics[width=.5\textwidth]{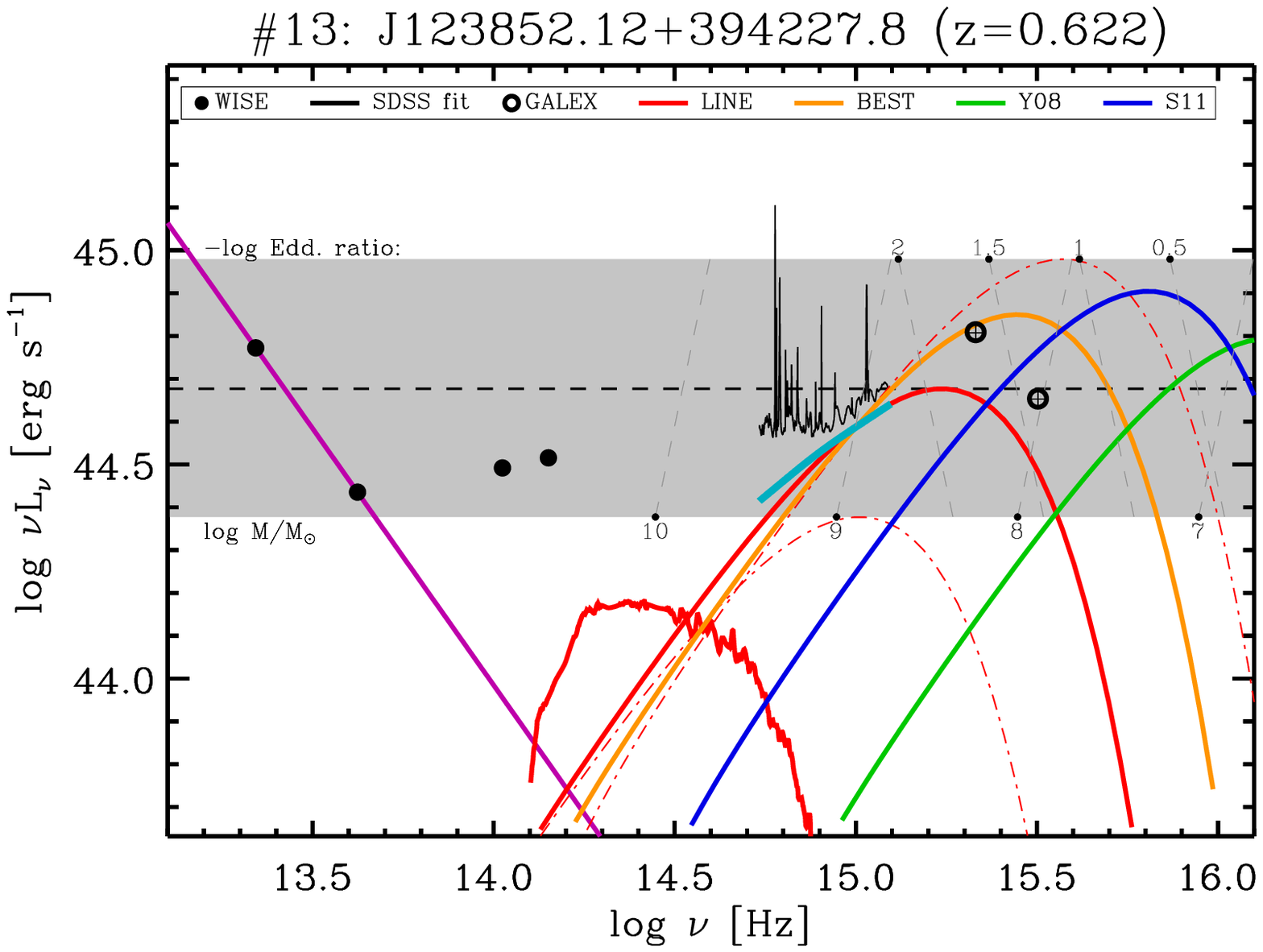}
    \includegraphics[width=.5\textwidth]{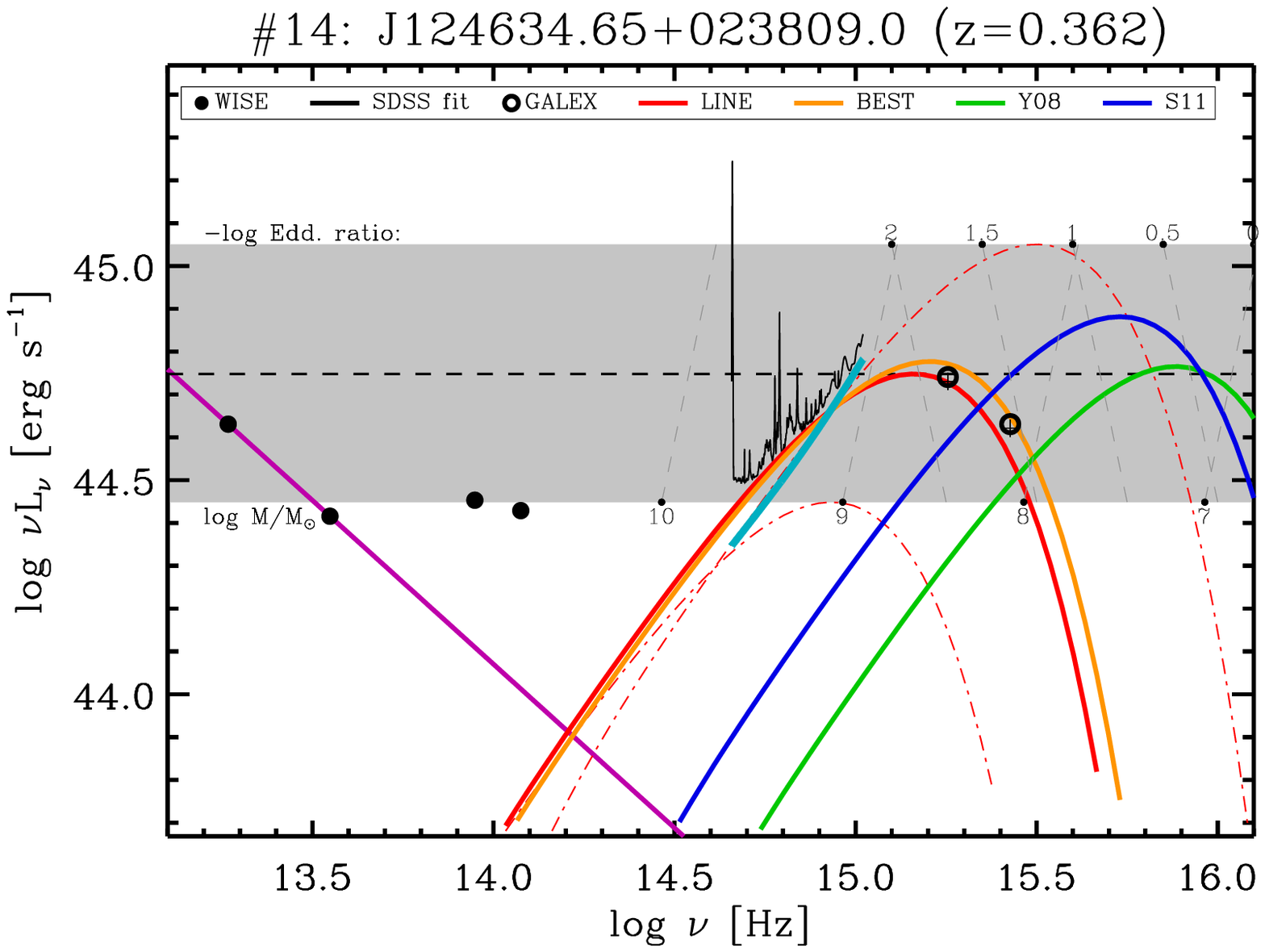}}\\
  \vspace{-0.7cm} \subfloat[][]{
    \includegraphics[width=.5\textwidth]{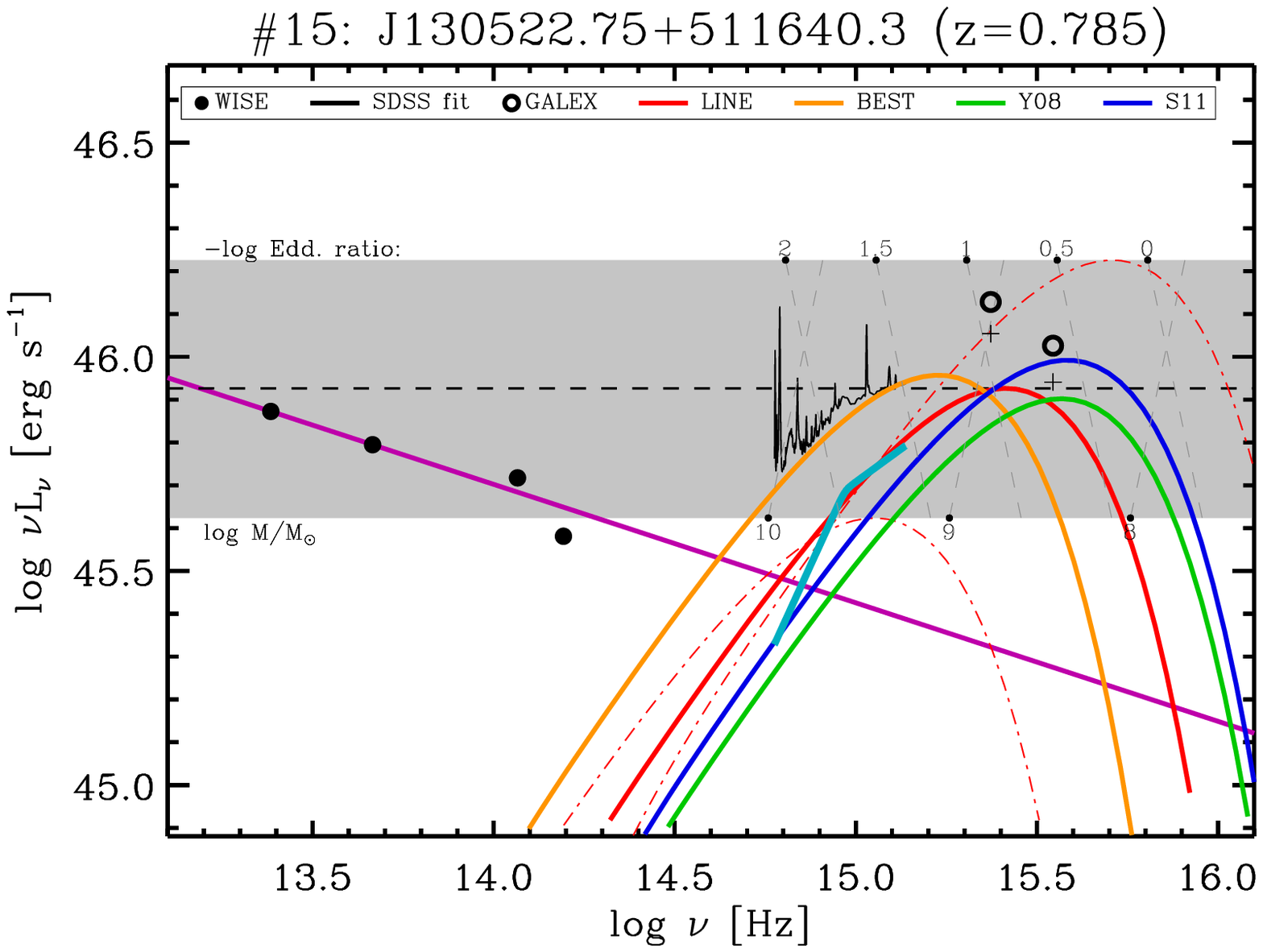}
    \includegraphics[width=.5\textwidth]{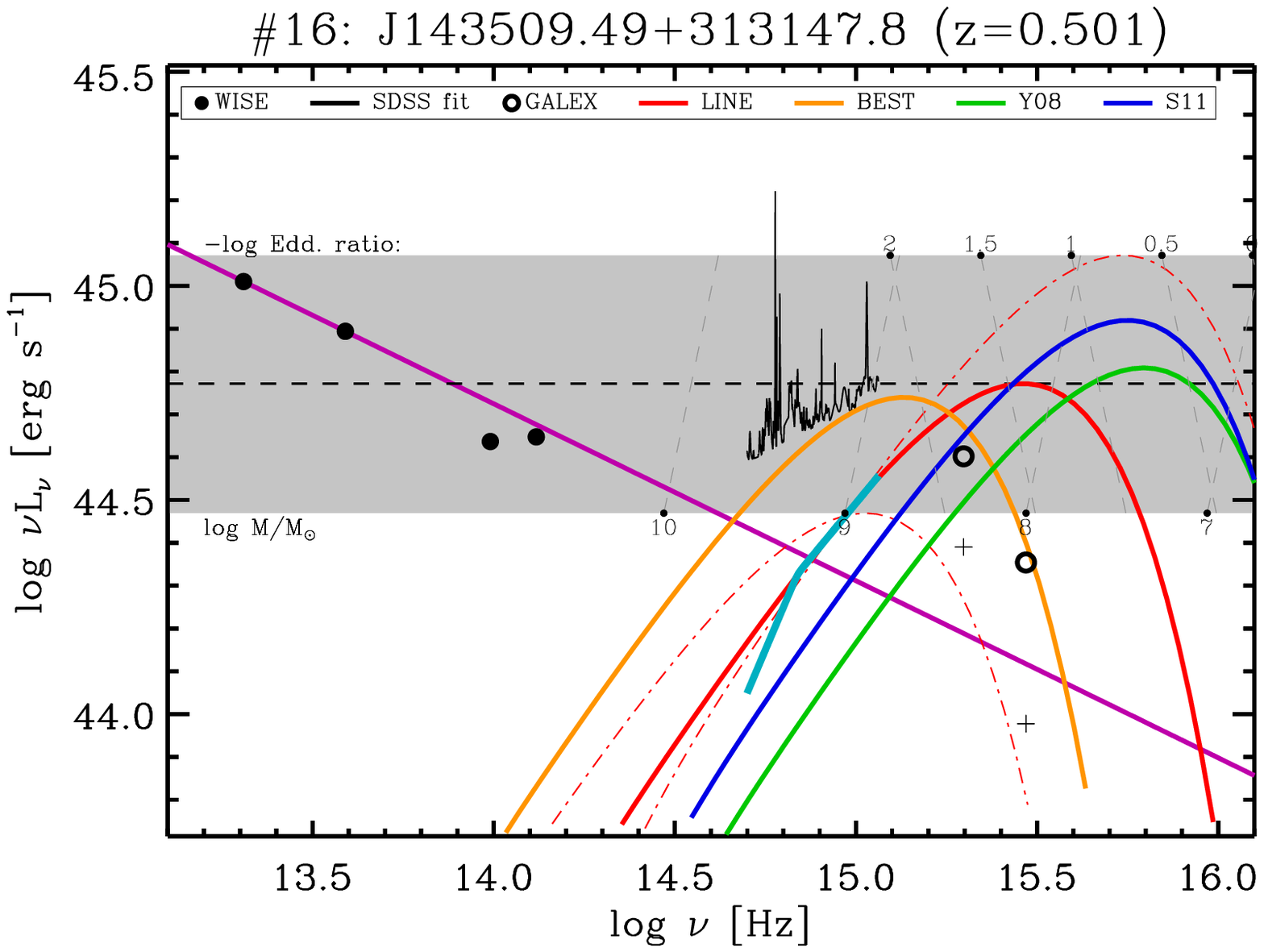}}\\
  \vspace{-0.7cm} \subfloat[][]{
    \includegraphics[width=.5\textwidth]{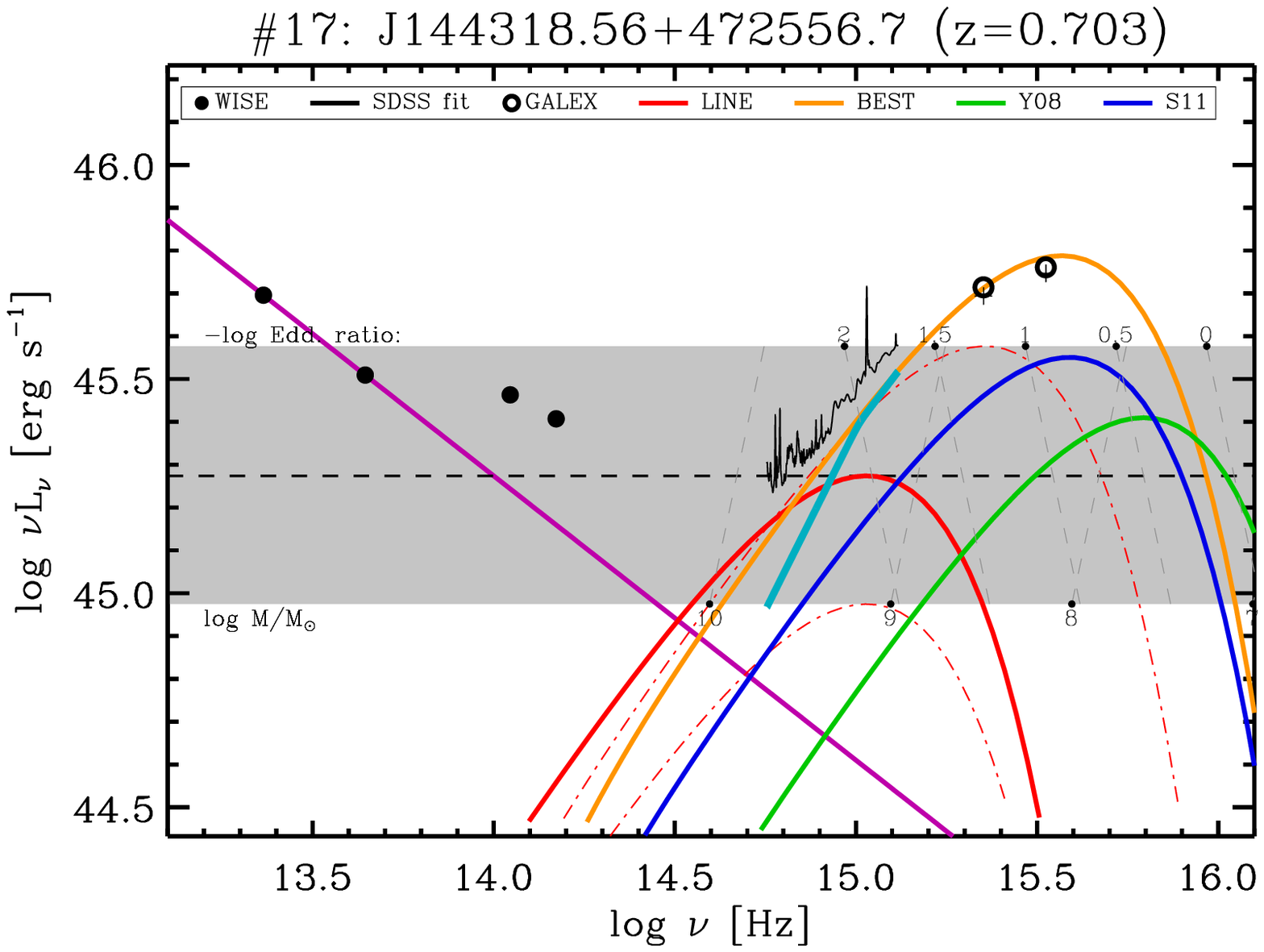}
    \includegraphics[width=.5\textwidth]{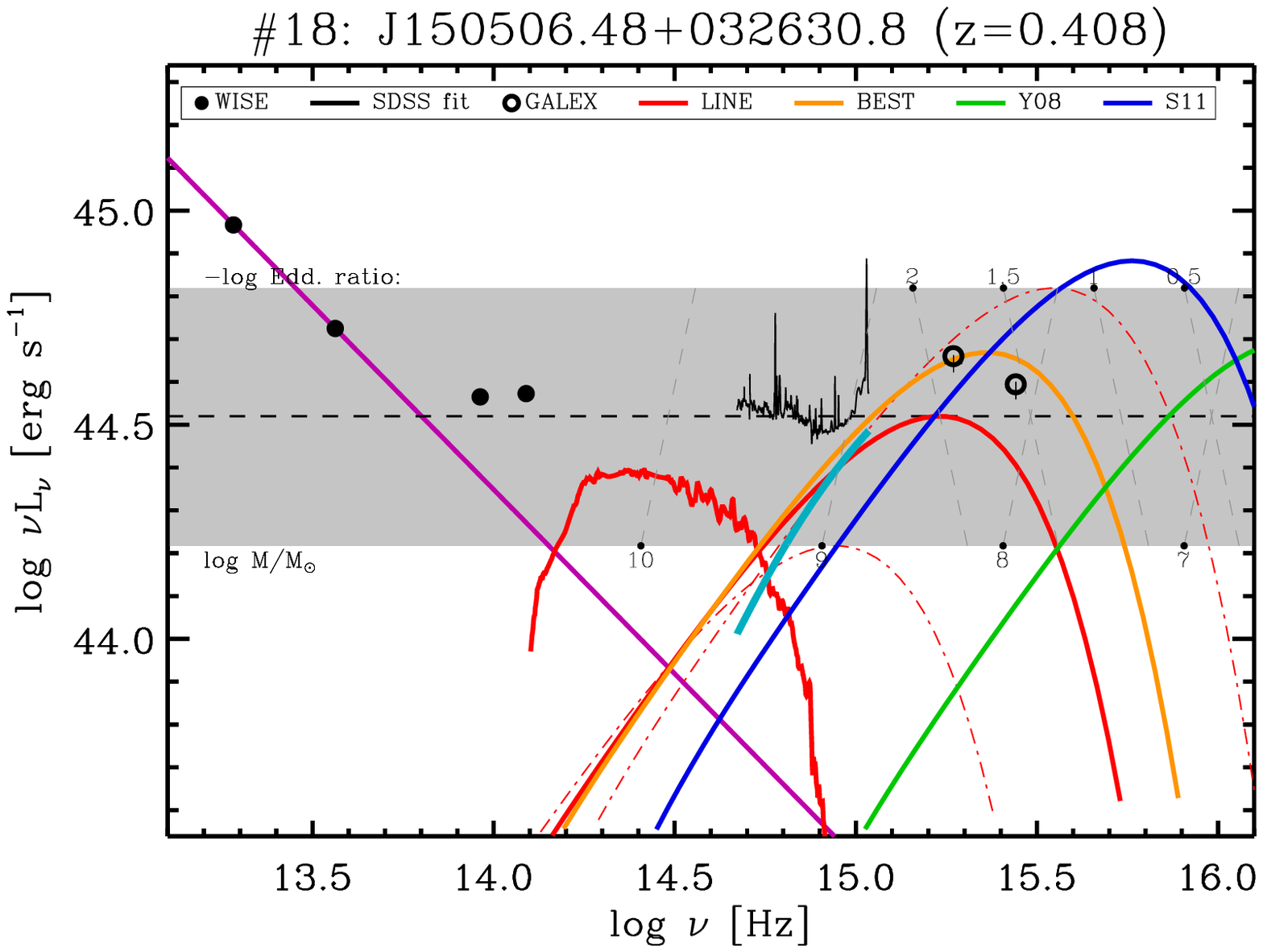}}\\
  \caption{(continued)}%
\end{figure*}

\begin{figure*}%
  \ContinuedFloat
  \centering
  \subfloat[][]{
    \includegraphics[width=.5\textwidth]{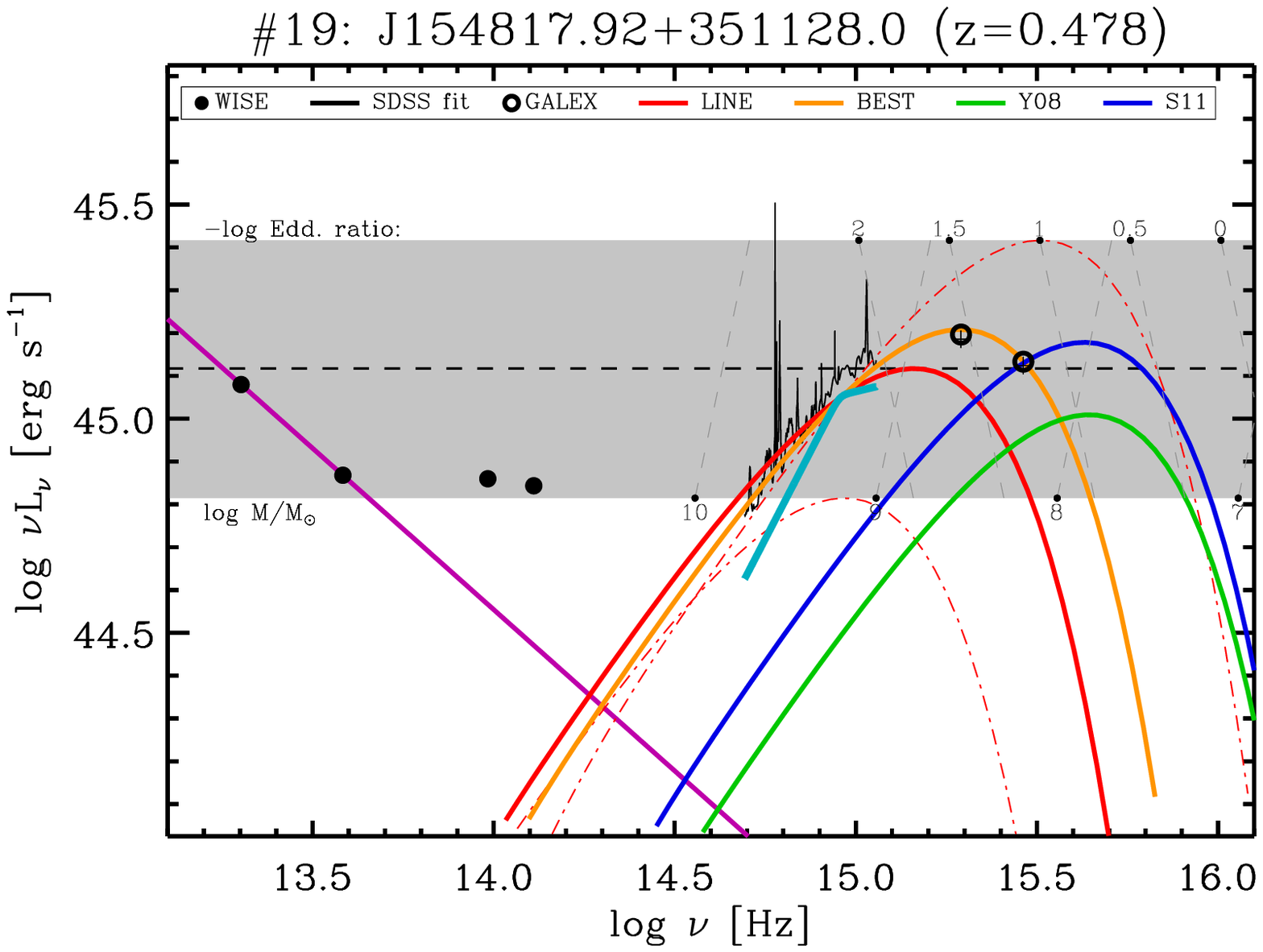}
    \includegraphics[width=.5\textwidth]{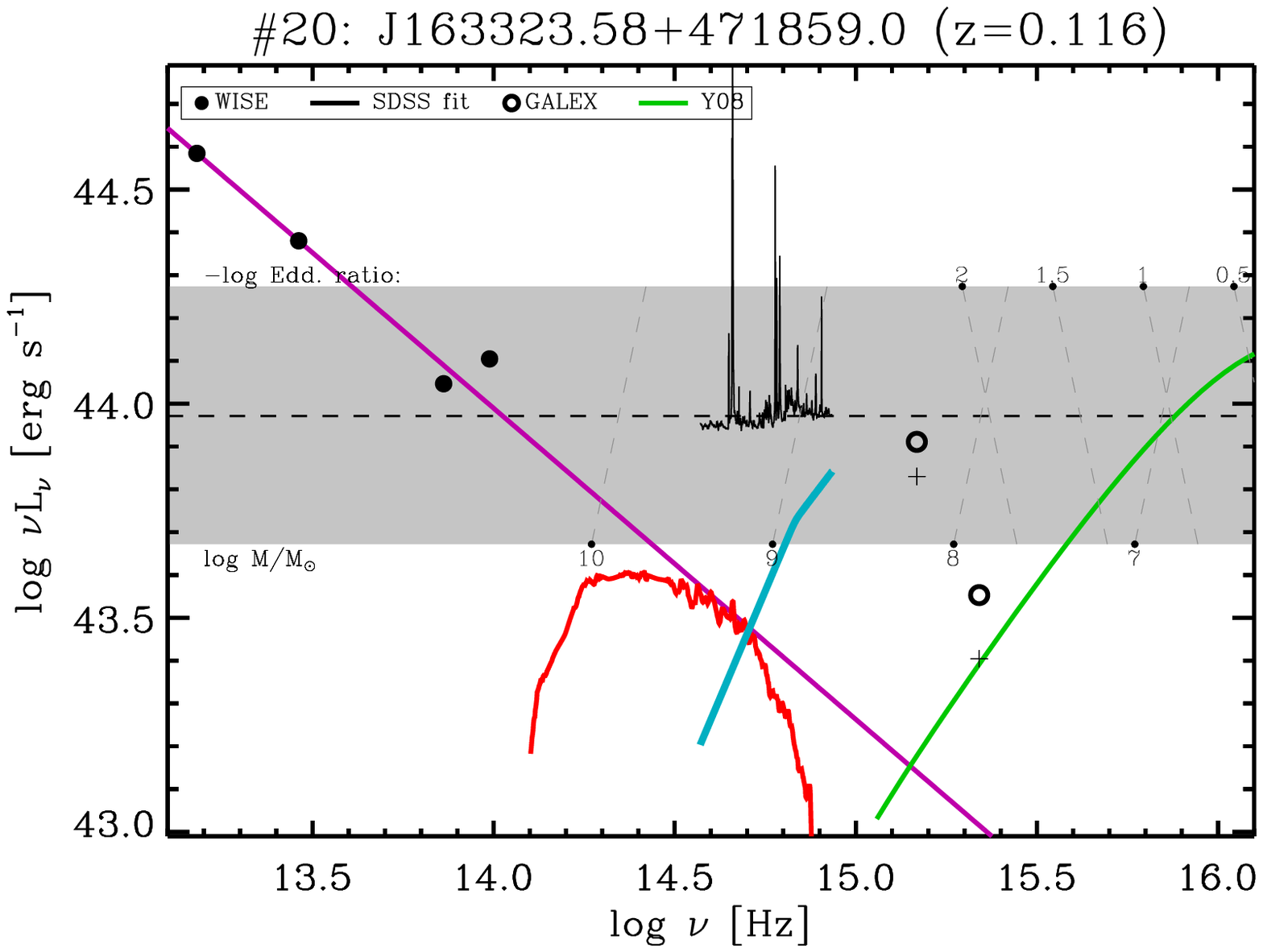}}\\
  \vspace{-0.7cm} \subfloat[][]{
    \includegraphics[width=.5\textwidth]{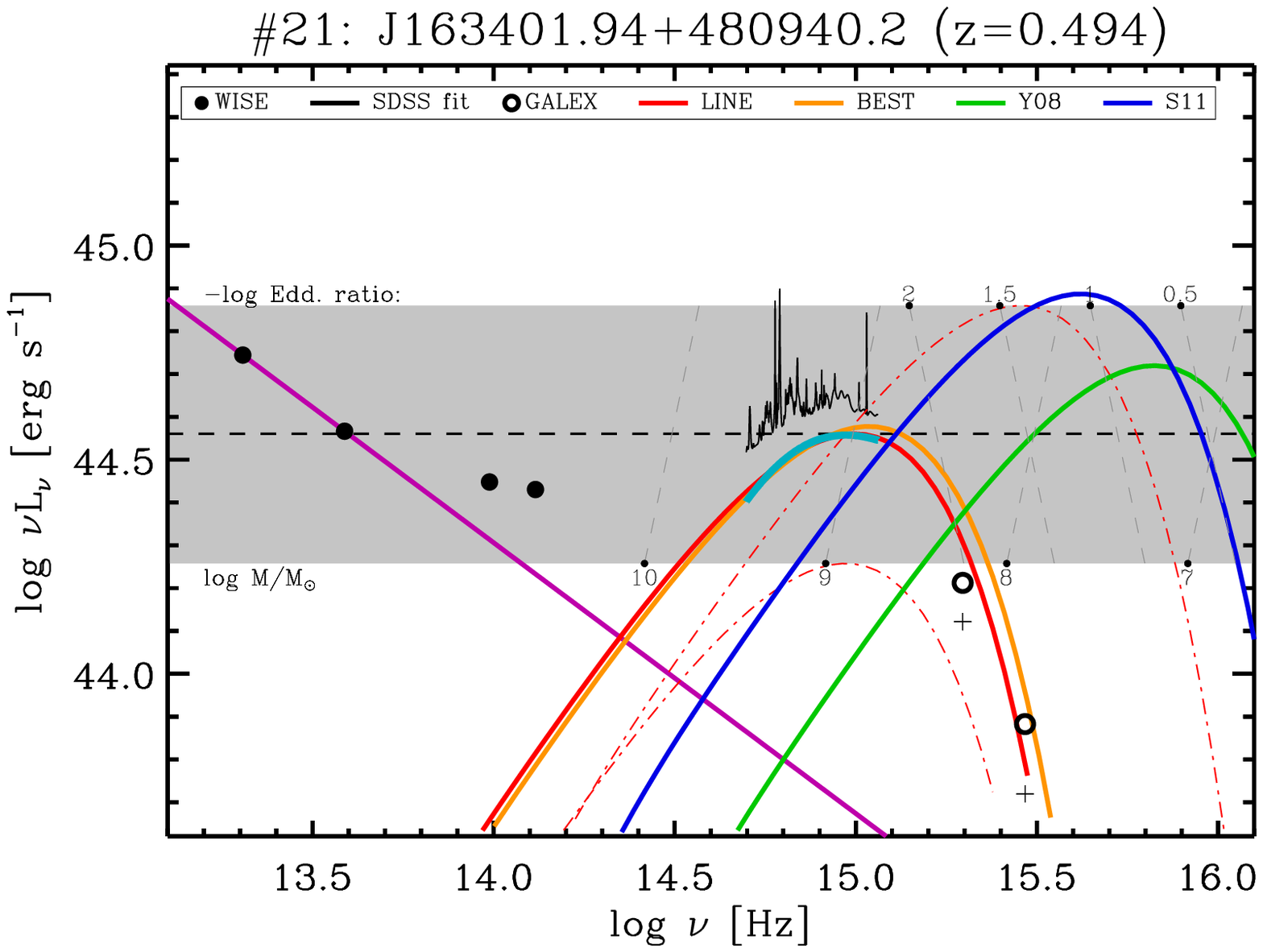}
    \includegraphics[width=.5\textwidth]{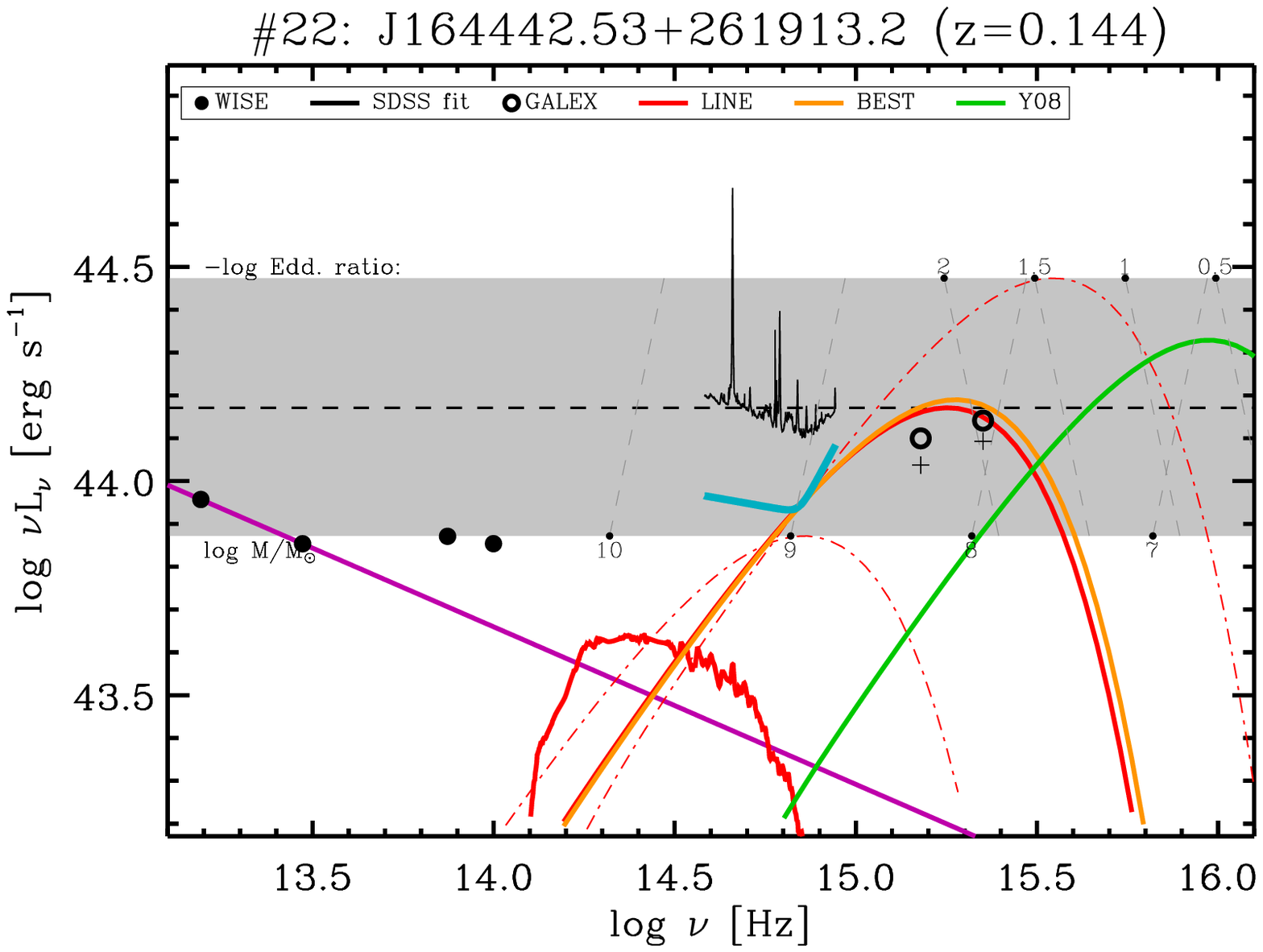}}\\
  \vspace{-0.7cm} \subfloat[][]{
    \includegraphics[width=.5\textwidth]{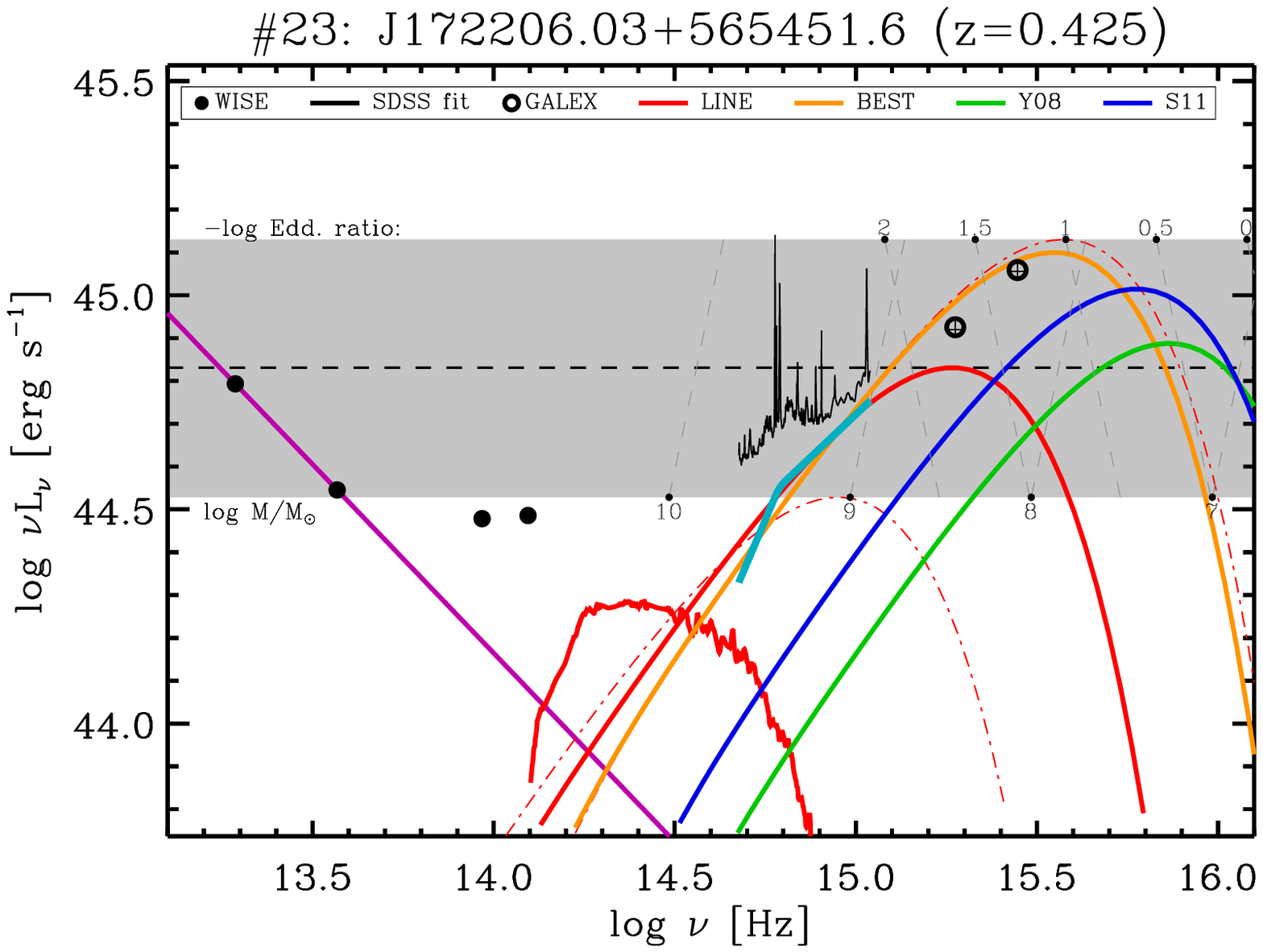}}
  \caption{(continued)}%
\end{figure*}

\label{lastpage}
\end{document}